\documentclass[11pt,onecolumn]{IEEEtran}
\usepackage[tbtags]{amsmath}
\usepackage{amsfonts}
\usepackage{amssymb}
\usepackage{amsthm}
\usepackage{algorithmicx}
\usepackage{algpseudocode}
\usepackage{algorithm}
\usepackage{subfigure}
\usepackage{graphicx}
\usepackage{cite}
\usepackage{calc}
\usepackage{color}
\usepackage{epsfig}
\usepackage{enumitem}
\usepackage{setspace}
\usepackage{multirow}
\usepackage{hyperref}
\usepackage{mathtools, cuted}
\usepackage{epstopdf}
\usepackage{lipsum}
\usepackage{mathtools}
\usepackage{cuted}

\makeatletter
\def\endthebibliography{%
	\def\@noitemerr{\@latex@warning{Empty `thebibliography' environment}}%
	\endlist
}
\makeatother

\makeatletter
\def\footnoterule{\kern-3\p@
	\hrule \@width 2in \kern 2.6\p@} 
\makeatother

\newtheorem{defn}{Definition}
\newtheorem{thm}{{\cal T}heorem}[section]
\newtheorem{cor}[thm]{Corollary}
\newtheorem{prop}{Proposition}
\newtheorem{lem}[thm]{Lemma}
\newtheorem{conj}[thm]{Conjecture}
\newtheorem{constr}[thm]{Construction}
\newtheorem{note}{Remark}
\newcommand{\bit}{\begin{itemize}}
\newcommand{\eit}{\end{itemize}}
\newcommand{\bcor}{\begin{cor}}
\newcommand{\ecor}{\end{cor}}
\newcommand{\beq}{\begin{equation}}
\newcommand{\eeq}{\end{equation}}
\newcommand{\beqn}{\begin{equation}}
\newcommand{\eeqn}{\end{equation}}
\newcommand{\bea}{\begin{eqnarray}}
\newcommand{\eea}{\end{eqnarray}}
\newcommand{\bean}{\begin{eqnarray*}}
\newcommand{\eean}{\end{eqnarray*}}
\newcommand{\ben}{\begin{enumerate}}
\newcommand{\een}{\end{enumerate}}
\newcommand{\bdefn}{\begin{defn}}
\newcommand{\edefn}{\end{defn}}
\newcommand{\bnote}{\begin{note}}
\newcommand{\enote}{\end{note}}
\newcommand{\bprop}{\begin{prop}}
\newcommand{\eprop}{\end{prop}}
\newcommand{\blem}{\begin{lem}}
\newcommand{\elem}{\end{lem}}
\newcommand{\bthm}{\begin{thm}}
\newcommand{\ethm}{\end{thm}}
\newcommand{\bconj}{\begin{conj}}
\newcommand{\econj}{\end{conj}}
\newcommand{\bconstr}{\begin{constr}}
\newcommand{\econstr}{\end{constr}}
\newcommand{\bpf}{\begin{proof}}
\newcommand{\epf}{\end{proof}}
\newcommand{\bprf}{{\em Proof: }}
\newcommand{\eprf}{\hfill $\Box$}

\newcommand{\cxyz}{\mbox{$C(x,y; \underline{z})$}}
\newcommand{\cxyzc}{\mbox{$C(z_y,y; \underline{z}(x \rightarrow z_y))$}}
\newcommand{\cxyznot}{\mbox{$C(x_0,y_0; \underline{z})$}}
\newcommand{\cxyzcnot}{\mbox{$C(x_0,y_0; \underline{z}(x \rightarrow z_{y_0}))$}}

\newcommand{\fQ}{\mbox{$\mathbb{F}_q$}}

\newcommand{\uu}{\mbox{$\underline{u}$}}

\newcommand{\uv}{\mbox{$\underline{v}$}}

\newcommand{\uz}{\mbox{$\underline{z}$}}
\newcommand{\uf}{\mbox{$\underline{f}$}}
\newcommand{\uw}{\mbox{$\underline{w}$}}

\newcommand{\Zs}{\mbox{$[0:s-1]$}}
\newcommand{\Zt}{\mbox{$[0:t-1]$}}
\newcommand{\Zst}{\mbox{$\mathbb{Z}_s^t$}}

\newcommand{\bc}{\begin{center}}
	\newcommand{\ec}{\end{center}}
\newcommand\MyLBrace[2]{%
	\left.\rule{0pt}{#1}\right\}\text{#2}}

\newcommand{\fq}{\ensuremath{\mathbb{F}_q}}

\newcommand{\uc}[1]{\ensuremath{\underline{c}_{#1}}}

\newcommand{\Ztwot}{\mbox{$\mathbb{Z}_2^t$}}

\newcommand{\smalld}{\ensuremath{\text{Small-$\mathsf{d}$}}}

\begin{document}
\title{\smalld\ MSR Codes with Optimal Access, Optimal Sub-Packetization and Linear Field Size}
\author{
	\IEEEauthorblockN{ Myna Vajha, S. B. Balaji, P. Vijay Kumar} \ \\
	\IEEEauthorblockA{
		Department of Electrical Communication Engineering, IISc Bangalore \\ \{mynaramana, balaji.profess,  pvk1729\}@gmail.com}
\footnote{This paper was presented in part at IEEE, International Symposium on Information Theory (ISIT) 2018 in \cite{VajBalKum}.}	
}
\maketitle

\begin{abstract}
	This paper presents an explicit construction of a class of optimal-access, minimum storage regenerating (MSR) codes, for small values of the number $d$ of helper nodes. The construction is valid for any parameter set $(n,k,d)$ with $d \in \{k+1, k+2, k+3\}$ and employs a finite field \fq\ of size $q=O(n)$.  We will refer to the constructed codes as \smalld\ MSR codes.  The sub-packetization level $\alpha$ is given by $\alpha = s^{{\lceil\frac{n}{s}\rceil}}$, where $s=d-k+1$.  By an earlier result on the sub-packetization level for optimal-access MSR codes, this is the smallest value possible.
\end{abstract}


\section{Introduction}

%
Erasure codes are of strong interest in distributed storage systems as they offer reliability at lower values of storage overhead in comparison with replication.   In the setting of distributed storage, the $B$ symbols of a given data file ${\cal F}$ are stored in redundant fashion, across $n$ storage units (nodes), such as a hard disk or a flash memory unit.  Among the class of erasure codes, Maximum Distance Separable (MDS) codes are of particular interest as they offer reliability at lowest possible value of storage overhead.  
Apart from reliability and storage overhead, an additional important concern in a distributed storage system is that of efficient single node repair \cite{RashmiShahGuKuangBorthakur}.   Efficient repair could either call for the amount of data download needed to repair a failed node to be kept to a low level or else, the number of helper nodes contacted for repair to be kept small.  The focus in the present paper, is on the first criterion, i.e., on lowering the amount of data download needed for node repair, also termed as the repair bandwidth.  

\subsection{Minimum Storage Regenerating (MSR) Codes} 

Regenerating codes \cite{DimakisKannaWuChangho} are codes that protect against data loss as well as single node failure with less repair bandwidth.  These codes have a vector symbol alphabet, given by $\mathbb{F}_q^{\alpha}$ where $\alpha$ is termed the level of sub-packetization of the regenerating code. Thus each storage unit stores $\alpha$  symbols from $\mathbb{F}_q$ associated to the file ${\cal F}$.  Protection against data loss is ensured by requiring that the stored data file be retrievable even in the face of the loss of upto $r, 1 \leq r \leq n$ storage units.  Thus the minimum Hamming distance $d_{\min}$ of the regenerating code must satisfy $d_{\min} \ge r+1$.  We define the parameter $k=n-r$.  Node repair is ensured by requiring that a failed node be repaired by downloading $\beta$ symbols over $\mathbb{F}_q$ from each node within a set of $d$ nodes, where the $d$ nodes are arbitrarily selected from the surviving $(n-1)$ nodes. 
Within the class of regenerating codes, the subclass of Minimum Storage Regenerating (MSR) codes are of particular interest, as this subclass falls within the class of MDS codes, and hence incur least-possible storage overhead when required to recover from the failure of $r$ storage units. To qualify for being called an MDS code, the regenerating code must satisfy the Singleton bound
\bean
q^B  = & \text{size of the code} & = \ q^{\alpha(n-d_{\min}+1)} \ = \ q^{\alpha k}, 
\eean
i.e., it must be that 
\bean
B & = &  k\alpha. 
\eean
It turns out that the minimal number symbols downloaded $\beta$, from each helper node  in an MSR code is necessarily given by 
\bea
\label{eq:rep_band}
\beta = \frac{\alpha}{d-k+1} \text{  (see \cite{DimakisKannaWuChangho}).} 
\eea
This is obtained by quantifying the condition under which the repair bandwidth $d \beta$ for the repair of a failed node, in an MDS code over $\fq^{\alpha}$ for a fixed value of $\{n,k,d,\alpha\}$, is the least possible.


Thus, MSR codes are characterized by the parameter set 
\bean
\left\{ (n,k,d),\ (\alpha,\beta),\ B, \ \fq) \right\}, 
\eean
where 
\bit 
\item \fq\ is the underlying finite field, 
\item $n$ is the number of code symbols $\{\underline{c}_i\}_{i=1}^n$, each of which is stored on a distinct node and 
\item each code symbol $\underline{c}_i$ is an element of  $\fq^{\alpha}$. 
\eit
Since each code symbol \uc{i} is stored on a distinct node, it follows that the index $i$ of a code symbol is synonymous with the index of the node upon which that code symbol is stored. Throughout this paper, we will focus on a linear MSR code i.e., on MSR codes where, the mapping from symbols comprising the data file and the symbols stored in the storage network takes on the linear form
\bean
\underline{m}^T G & = &  [\uc{1}^T,\hdots,\uc{n}^T], 
\eean
where $G$ is an $(k\alpha \times n\alpha)$ generator matrix over \fq\ and where $\underline{m}$ is a $(k\alpha \times 1)$ message vector over \fq\ corresponding to the $B=k\alpha$ message symbols of the data file, encoded by the MSR code.

\subsection{Desirable Properties of an MSR Code}

While MSR codes are MDS codes that need smallest repair bandwidth possible for single node repair, there is still scope for optimization with the class of MSR codes.  The additional features of interest are listed below.

\bit
\item {\em Optimal-Access:\ }
Optimal-access MSR codes \cite{TamoWangBruck} are a subclass of MSR codes having the property that during repair, the $\beta$ symbols that are transmitted by a helper node during repair, are simply a subset of the $\alpha$ symbols contained in the node.  This has two important and desirable, practical consequences.  Firstly, the number of symbols accessed in the node is as small as possible and secondly, no computations are required to generate the transmitted repair symbols.  
 \item {\em Low Values of Sub-Packetization:\ } Low values of sub-packetization level are desirable both to reduce complexity as well as to permit smaller file sizes $B=k\alpha$ to be encoded. 
\item {\em Low Field Size:\ } The need for a low field size is clear since the smaller the size of the finite field, the lesser is the implementation complexity.  
\eit

\subsection{Prior Work on MSR Codes}

Several constructions of an MSR code can be found in the literature.  In addition, there are constructions of systematic MDS codes in the literature where it is only the systematic nodes that can be recovered with minimal repair bandwidth, i.e., repaired by downloading $\frac{d\alpha}{d-k+1}$ symbols. We will refer to this latter class of codes as systematic MSR codes.  A detailed survey on MSR code constructions and sub-packetization level bounds can be found in \cite{BalNikVaj}. The product matrix construction in \cite{RasShaKum_pm} for any $2k-2 \le d \le n-1$ is one of the earliest constructions of an MSR code. These codes have smallest possible sub-packetization level possible of an MSR code, since $\beta=1$ in the product matrix construciton of an MSR code.   However, and possibly as a consequence of this, the rate of a product-matrix MSR code is bounded above by the quantity $\frac{1}{2}+\frac{1}{2n}$. 

In \cite{PapDimCad}, the authors provide a construction for a high-rate MSR code that makes use of Hadamard designs for any $(n, k, d)$ parameter set of the form $(n,k=n-2,d=n-1)$ with sub-packetization level $\alpha = 2^{k+1}$. In \cite{TamWanBru}, high-rate systematic MSR codes termed as Zigzag codes were constructed for $d=n-1$. These codes however have large field size and sub-packetization level that is exponential in the parameter $k$. This construction was subsequently extended in \cite{WanTamBru_allerton} to enable the repair of parity nodes. The existence of MSR codes for any value of $(n,k,d)$  as $\alpha$ tends to infinity is shown in \cite{CadJafMalRamSuh}. 

\subsubsection{Sub-Packetization Level} An open problem in the literature on regenerating codes is that of determining the smallest possible sub-packetization level $\alpha$ of an MSR code, for given parameters $\{(n,k,d=(n-1)\}$. 
This question is addressed in \cite{TamWanBru_access}, where a lower bound on $\alpha$ for MSR codes is presented by showing that $k \le \alpha {\alpha \choose \alpha / (n-k)}$.  In \cite{GopTamCal} it is established that:
\bean
k \leq 2 \log_2(\alpha) (\lfloor \log_{\frac{r}{r-1}} (\alpha) \rfloor + 1),
\eean
while in \cite{HuangParamXian} the authors prove that:
\bean
k \leq 2 \log_r(\alpha) (\lfloor \log_{\frac{r}{r-1}} (\alpha) \rfloor + 1). 
\eean
Most recently, in \cite{OmarGur} the authors prove that $ \alpha \ge e^{\Omega(\frac{k}{r})}$ for any general MSR code.

\subsubsection{Optimal-Access MSR Codes}  For the special case of an optimal-access MSR code, it was shown in \cite{TamWanBru_access} that:
\bean
\alpha \geq r^{\frac{k-1}{r}}.
\eean
The constructions presented in \cite{WanTamBru_allerton,TamWanBru} satisfy the optimal access property. However, they have sub-packetization level exponential in $k$ for a fixed rate. In \cite{AgaSasKum}, an optimal-access systematic MSR code is constructed for the case $d=n-1$ with $\alpha = r^{\frac{k}{r}}$. This was followed by in \cite{SasAgaKum}, by the construction of an optimal-access MSR for the case $d=n-1$ with $\alpha = r^{\lceil \frac{n}{r}\rceil}$. The construction in \cite{SasAgaKum} was extended to any $d \le n-1$ in \cite{RawKoyVis_msr} with $\alpha = s^{\lceil \frac{n}{s} \rceil}$ where $s=d-k+1$. The constructions in \cite{AgaSasKum, SasAgaKum, RawKoyVis_msr} are not explicit and need large field size. In \cite{YeBar_1} explicit MSR constructions for any $(n, k, d)$ with field size $O(n)$ and $\alpha = s^n$ are provided. In \cite{BalKum}, the authors improve upon the lower bound for optimal-access MSR case to $\alpha \ge s^{\frac{n}{s}}$ where $s = d-k+1$.  This turns out to settle the problem of determining the smallest sub-packetization level of an optimal-access MSR code as the optimal-access MSR code constructions in \cite{SasAgaKum, RawKoyVis_msr, SasVajKum, YeBarg_2017, LiTangTian} achieve this lower bound on $\alpha$ with equality.

\subsubsection{The Coupled-Layer MSR Code} 

In \cite{YeBarg_2017}, Ye-Barg presented an explicit construction of optimal-access MSR codes having parameters $(n, k, d=n-1)$ with sub-packetization level $\alpha = r^{\lceil \frac{n}{r} \rceil}$ and field size $q \ge n-1$.  In independent work, that followed shortly after, the authors of \cite{SasVajKum}, came up with essentially the same construction, but one that was presented from a coupled-layer perspective that involved the application of a pairwise coupling transform applied to a data cube wherein each horizontal layer was an MDS code.  Flexibility in selecting this MDS code meant that it was possible to construct a binary coupled layer MSR code by starting from a MDS code built over a binary vector alphabet.  Unknown to the authors of \cite{SasVajKum}, in \cite{LiTangTian}, the authors had employed the same coupling transform to transform an MDS code to one in which certain symbols could be optimally repaired.  This was later extended by the authors of \cite{LiTangTian}, after the appearance of \cite{YeBarg_2017} and \cite{SasVajKum} to show how an MDS code could be transformed to yield an MSR code through iterated application of the pairwise coupling transform. 

We will refer to the MSR code resulting form the constructions presented in \cite{YeBarg_2017,SasVajKum, LiTangTian} as the Clay code (Clay for Coupled LAYer), following the nomenclature introduced in \cite{VajRamPur_Clay}, where a detailed performance evaluation of the Clay code was conducted.  

\subsection{Our Contributions}

As discussed above, the Clay code is an optimal-access MSR code that is optimal with respect to sub-packetization level and has linear field size. However, the Clay code construction applies only to the case when the number of helper nodes $d$ contacted equals $(n-1)$ which is the largest possible.  As pointed out in the literature on locally recoverable codes, there is practical interest in minimizing the number of helper nodes that are contacted.   While the prior literature contains optimal-access MSR constructions for $d < n-1$, the resultant codes are either non-explicit with large field size \cite{RawKoyVis_msr, GopFazVar} or else have large sub-packetization level \cite{YeBar_1}.

In the present paper, we present explicit construction of MSR codes that are also optimal access, have optimal sub-packetization level, linear field size, but  where this time, $d$ is as small as possible. For this reason, we term these codes as \smalld\ MSR codes.  Specifically, we provide constructions for the cases $d\in\{k+1,k+2,k+3\}$.  The case $d=k$ is uninteresting since setting $d=k$ results in $\beta = \alpha$ from equation~\eqref{eq:rep_band} and thus there is no saving in repair bandwidth to be had in this case.   The parameters of the \smalld\ MSR codes constructed over a finite field $\mathbb{F}_q$ are given by: 
\bean
(n, k, d \in \{k+1, k+2, k+3\}), \ \ (\alpha=s^t, \beta=s^{t-1}), \ \ B=k\alpha, 
\eean
where
\bean
r=n-k, s = d-k+1, \ \ t= \left\lceil \frac{n}{s} \right\rceil, \ \ q=O(n).
\eean
It follows that the union of the \smalld\ and Clay code MSR constructions provide optimal-access, optimal sub-packetization level and linear-size code constructions for all $(n,k,d)$ parameter sets with $n- k \leq 5$.  Given the emphasis within industry on small block lengths and low values of storage overhead, this range is of practical interest.    The parameters of some example \smalld\ MSR codes is presented in Table~\ref{tab:example_codes}. 

	
\begin{table}[ht!]
	\caption{Parameters of some example \smalld\ MSR codes. Here, $r=n-k$, $s=d-k+1$, $t = \lceil \frac{n}{s}\rceil$, $\alpha=s^t$ and $\beta=s^{t-1}$. An $^{*}$ attached to the parameter $d$ identifies instances where the \smalld\ and Clay-code constructions yield codes with identical parameters. \label{tab:example_codes}}
		\bc
	\begin{tabular}{||c|c|c||c|c|c||c|c||c|c||}
	\hline \hline
	& & & & & & & & & \\
	$n$ & $k$ & $d$ & $r$ & $s$ & $t$ & $\alpha$ & $\beta$ & $B=k\alpha$ & $d \beta$\\  
	& & & & & & & & & \\ \hline
	10 & 8 & 9* & 2 & 2 & 5 & 32 & 16 & 256 & 144\\ \hline
	9 & 6 & 7 & 3 & 2 & 5 & 32 & 16 & 192 & 112\\ \hline
	9 & 6 & 8* & 3 & 3 & 3 & 27 & 9 & 162 & 72\\ \hline
	14 & 10 & 11 & 4 & 2 & 7 & 128 & 64 & 1280 & 704\\ \hline
	14 & 10 & 12 & 4 & 3 & 5 & 243 & 81 & 2430 & 972\\ \hline
	14 & 10 & 13* & 4 & 4 & 4 & 256 & 64 & 2560 & 832\\ 
	\hline \hline
	\end{tabular}
	\ec

\end{table}

\subsection{Outline}
We start by presenting the description of \smalld\ MSR code in Section~\ref{sec:smalld} and then introduce the notation and terminology in Section~\ref{sec:termi} that will be used to prove the MDS and optimal-access repair properties of the \smalld\ MSR code. In Section~\ref{sec:example}, we show that for an example case of $s = (d-k+1)=2$ and $r = n-k=3$, the \smalld\ MSR code is an optimal-access MSR code.
In Section~\ref{sec:mds_redn} we show that the MDS property of \smalld\ MSR code can be reduced to proving invertibility of a reduced matrix that is a sub-matrix of parity check matrix. Similar to Section~\ref{sec:mds_redn}, in Section~\ref{sec:oar_redn} we show that the optimal-access property of \smalld\ MSR code can also be reduced to proving invertibility of a reduced matrix. Finally, in Section~\ref{sec:inv_red} we show that the reduced matrix is invertible thereby proving that \smalld\ MSR code is an optimal-access MSR code for any $s = d-k+1 \in \{2,3,4\}$ and $r \ge s$.


We will adopt the following notation throughout the paper. 
\ben
\item $[a: b] = \{a, a+1, \cdots, b\}, [a] = [1:a]$ and $\mathbb{Z}_s = \Zs$.
\item Let $\uz = (z_0, z_1, \cdots, z_{t-1}) \in \Zst$ and $x \in \Zs$.  We define $\uz(x \rightarrow z_y)$ to be the vector obtained by replacing the $y$th component of \uz\ by $x$: 
\bean
\uz(x \rightarrow z_y) = (z_0, \cdots, z_{y-1}, x, z_{y+1}, \cdots, z_{t-1})
\eean
\een

\section{\smalld\ Construction \label{sec:smalld}}

A description of the \smalld\ MSR code construction is provided in this section.   This description includes associating a datacube structure with a codeword in a \smalld\ MSR code and identifying parity-checks that are imposed on this data structure.  This is the same datacube structure that appears in the description of the Coupled-Layer MSR code in \cite{SasVajKum}.  However, the parity-check equations take on a different form and this difference is explained in Section~\ref{sec:connections}.  Proof of the data collection and node repair properties of the \smalld\ MSR codes is deferred to Sections~\ref{sec:mds_redn} and \ref{sec:oar_redn}. 

\smalld\ MSR codes are constructed over a finite field \fq\ of size $q$ and have parameters given by 
\bean
(n=st, k=n-r, d=k+s-1), \ (\alpha=s^t, \beta=s^{t-1}), s \in \{2, 3, 4\}, 
\eean
where $r, t$ are integers such that $r \ge s \ge 2$, $t > 1$.  The field size is linear in the length $n$ of the code, i.e., $q = O(n)$, with the precise relationship (see Theorem~\ref{thm:main}) dependent on the value of $d$ within the set $\{k+1,k+2,k+3\}$. 

\subsection{Extension to General Parameter Sets} 

We note that through shortening, we can obtain codes for any $(n,k,d \in \{k+1,k+2, k+3\})$. In particular if $s \nmid n$ where $s=d-k+1$, then we can first set $t = \lceil \frac{n}{s}\rceil$ and $\delta =  n - st$ and proceed to construct a \smalld\ MSR code ${\cal C}'$ having parameters $(n+\delta, k+\delta, d+\delta)$.  We can shorten ${\cal C}'$ thereafter, to obtain the MSR code ${\cal C}$ having the desired parameters $(n,k,d)$.

\subsection{Data Cube Representation of the Codeword}

As in the case of the Coupled-Layer MSR code, each codeword in a \smalld\ MSR code is associated to a datacube of dimension $\big( s \times t \times s^t\big)$ (see Figure~\ref{fig:cube}).  It will be found convenient to view the datacube as a collection of $s^t$ planes, each of size $n = \big( s \times t \big)$.   Thus the data cube contains in all $n\alpha = ts^{t+1}$ points.  Each point in the data cube is indexed by the three tuple $(x,y,\uz)$ where $x \in \Zs$, $y \in \Zt$ and $\uz \in \Zst$ and is associated to a unique code symbol $\cxyz$. The collection of $n \alpha$ code symbols are given by:
\bean
\{\cxyz \ | \ x \in \Zs, y \in \Zt, \uz \in \Zst \}.
\eean
Each $(x,y)$ $2$-tuple is associated to a node or storage unit in the distributed data storage network comprising of $n=st$ nodes.  The vector $\uz$ is used to index the $\alpha=s^t$ planes and also serves as an index for the $\alpha$ code symbols contained within a node.  

\subsection{Pictorial Identification of the Planes in the Datacube} 

\begin{figure}[ht!]
	\begin{center}
		\subfigure[The data cube containing $( s \times t \times s^t ) $ symbols over the finite field \fQ. In this example, $s=2, t=3$. For presentation we have only shown $3$ out of the $s^t = 8$ planes each containing $s \times t = 6$ symbols. ]{\label{fig:cube}\includegraphics[height=2.0in]{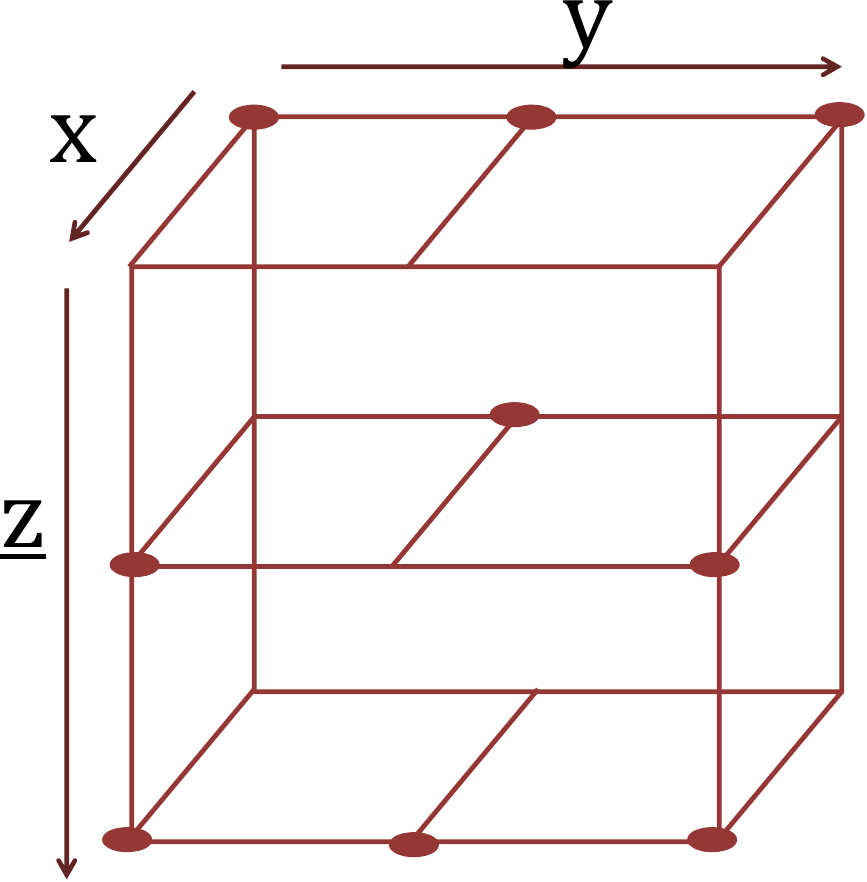}}
		\hspace{0.5in}
		\subfigure[We employ a dot notation to identify a plane. Red dots are placed at points of the form $(z_y, y)$. The example indicates the plane $\uz\ = (1, 0, 1)$.]{\label{fig:dot_rep}\includegraphics[width=2.0in]{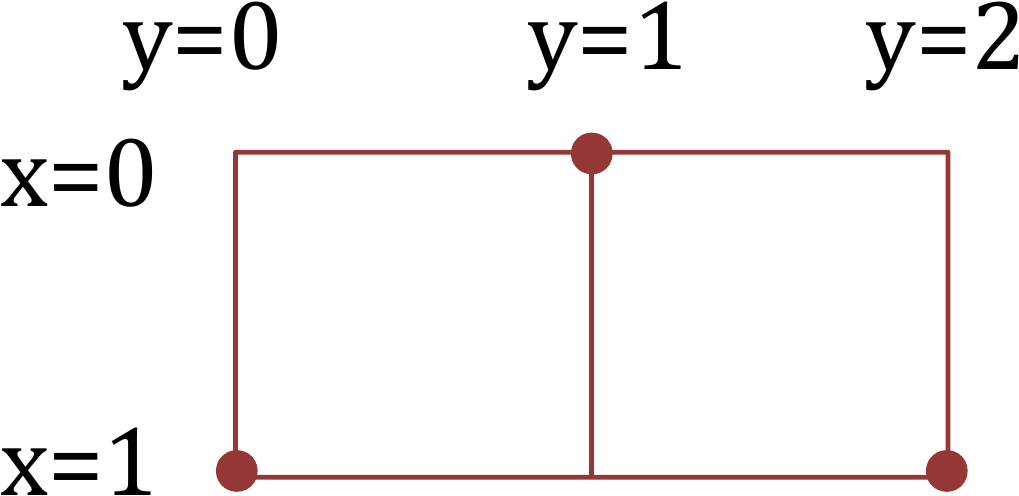}}
		\caption{Illustration of the data cube.\label{fig:data_cube}}
	\end{center}
\end{figure}

We associate a plane-dot-representation to each $(s \times t)$ plane indexed by $\underline{z}$ where a (red) dot is inserted in a location $(x,y)$ iff $z_y = x$. See Figure~\ref{fig:dot_rep} for an example where for a plane $\underline{z} = ({\color{red}1}, {\color{red}0}, {\color{red}1})$, the dots are inserted at locations $({\color{red}1},0),({\color{red}0},1),({\color{red}1},2)$.   Thus the location of the dot within the plane serves to uniquely identify the plane. 

\subsection{Parity Check Matrix\label{subsec:pcmatrix}}
%

The \smalld\ MSR code will be identified via an $(r \alpha \times n \alpha)$ parity check (p-c) matrix $H$ that imposes $r \alpha$ p-c equations on the code symbols associated with the datacube.  We associate $r$ parity checks to each plane $\uz$ and thus index a parity check equation or equivalently, a row of p-c matrix, using the pair $[\ell, \uz]$ where $\ell \in [r]$ and $\uz \in \Zst$. The $n\alpha = s\times t\times s^t$ columns of the parity check matrix $H$ are indexed using the three tuple $[x, y, \uu]$ with $x \in \Zs, y \in \Zt$ and $\uu \in \Zst$. The entries in the p-c matrix of the \smalld\ code are given by:
\bea
\label{eq:pcsup} H([\ell, \uz], [x,y,\uu]) &=& \begin{cases}
	\theta_{(x,y,u_y)}^{\ell-1}, & \uz = \uu, \\
	\gamma_{u_y, x} \theta_{(x,y,u_y)}^{\ell-1}, & \uz = \uu(x \rightarrow u_y), \ x \ne u_y, \\
	0, & \text{else},
\end{cases}
\eea
where $H([\ell, \uz], [x,y,\uu])$ is the element in the $[\ell,\uz]$-th row and $[x,y,\uu]$-th column of the parity check matrix. Also, 
\bea
\label{eq:gamma}
\gamma_{x, x'} = \begin{cases}
	\gamma & x < x'\\
	1 & x > x'\\
	0 & \text{Otherwise}
\end{cases}
\eea
such that $\gamma \in \fq \setminus \{0,1\}$.  Additionally, the element $\theta_{x,y,x'}$ is the entry in the $x$-th row and $x'$-th column of the matrix $\Lambda_{s,y}$, where
\bea
\label{eq:thetaassign}
\Lambda_{2,y} = \left[ \begin{array}{cc}
	\lambda_{0,y} &  \lambda_{1,y}\\
	\gamma \lambda_{1,y} & \lambda_{0,y}
\end{array} \right], \
\Lambda_{3,y} = \left[ \begin{array}{ccc}
	\lambda_{0,y} &  \lambda_{1,y} &  \lambda_{2,y}\\
	\gamma \lambda_{1,y} & \lambda_{0,y} &  \lambda_{3,y}\\
	\gamma \lambda_{2,y}& \gamma \lambda_{3,y} & \lambda_{0,y}
\end{array} \right], \		
\Lambda_{4,y} = \left[ \begin{array}{cccc}
	\lambda_{0,y} &  \lambda_{1,y} &  \lambda_{2,y} &  \lambda_{3,y}\\
	\gamma \lambda_{1,y} & \lambda_{0,y} &  \lambda_{3,y} &  \lambda_{2,y}\\
	\gamma \lambda_{2,y}& \gamma \lambda_{3,y} & \lambda_{0,y} &  \lambda_{1,y}\\
	\gamma \lambda_{3,y} & \gamma \lambda_{2,y} & \gamma \lambda_{1,y} & \lambda_{0,y}
\end{array} \right]. 
\eea
Further, the entries of the matrices $\Lambda_{s,y}$ are selected in such a way that for $s=2$, all the elements in the set $\{\lambda_{0,y},\lambda_{1,y}, \gamma \lambda_{1,y} \mid y \in \Zt\}$ form a set of $3t$ distinct nonzero elements of $\fq \setminus \{0\}$.  For $s \in \{3,4\}$, the analogous requirement is that all the elements in the set $\{\lambda_{0,y},\lambda_{i,y}, \gamma \lambda_{i,y} \mid i \in [3], y \in \Zt\}$ form a set of $7t$ distinct nonzero elements of $\fq \setminus \{0\}$.  Finally, $\fq$ is a field of characteristic $2$. In Theorem~\ref{thm:main} we show how to pick the elements $\{\lambda_{i,j}\}$ given that $q \ge 6t+2$ for $s=2$ and $q \ge 18t+2$ for $s \in \{3, 4\}$.

\begin{figure}[ht!]
	\begin{center}
		\subfigure[in-plane symbols in $\uz = (1,0,1)$]{\label{fig:inplane}\includegraphics[height=1.0in]{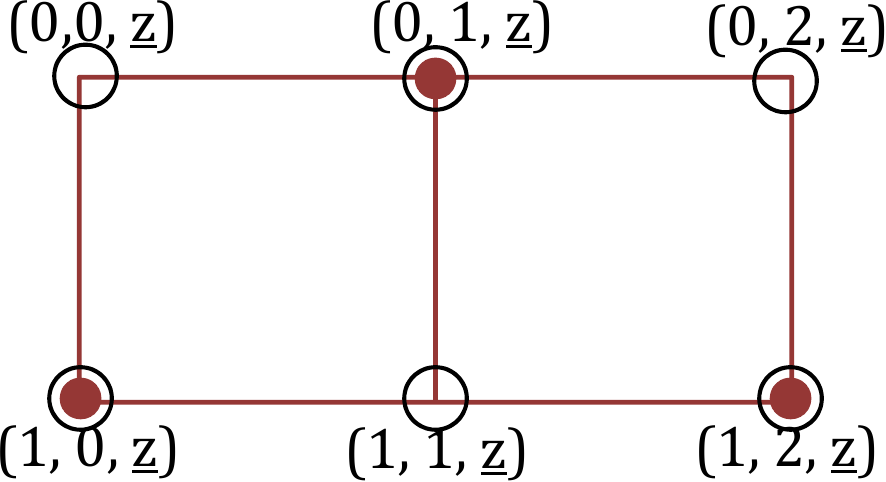}}
		\hspace{0.5in}
		\subfigure[out-of-plane symbols in plane $\uz_1 = \uz( 0 \rightarrow z_0) = (0, 0, 1)$ shown in black circle]{\label{fig:outplane0}\includegraphics[height=1.0in]{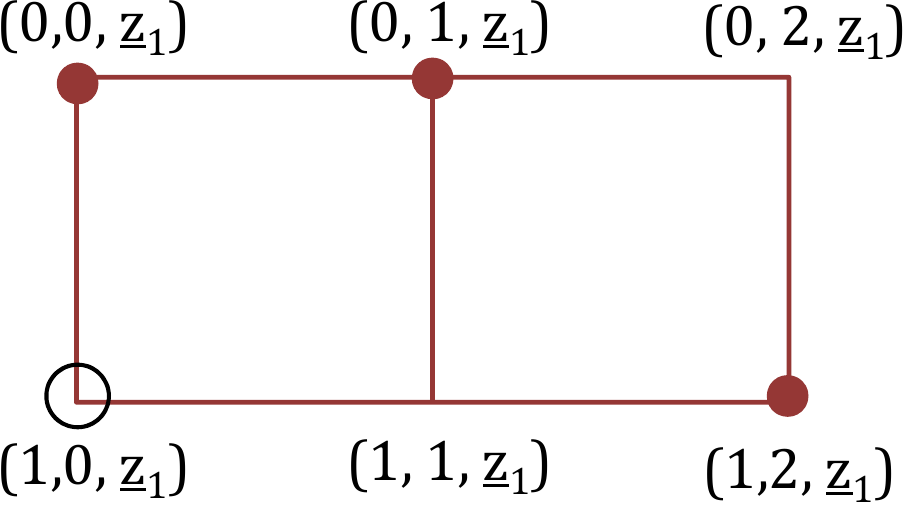}}
		\\
		\subfigure[out-of-plane symbols in  plane $\uz_2 = \uz(1 \rightarrow z_1) = (1, 1, 1)$ shown in black circle]{\label{fig:outplane1}\includegraphics[height=1.0in]{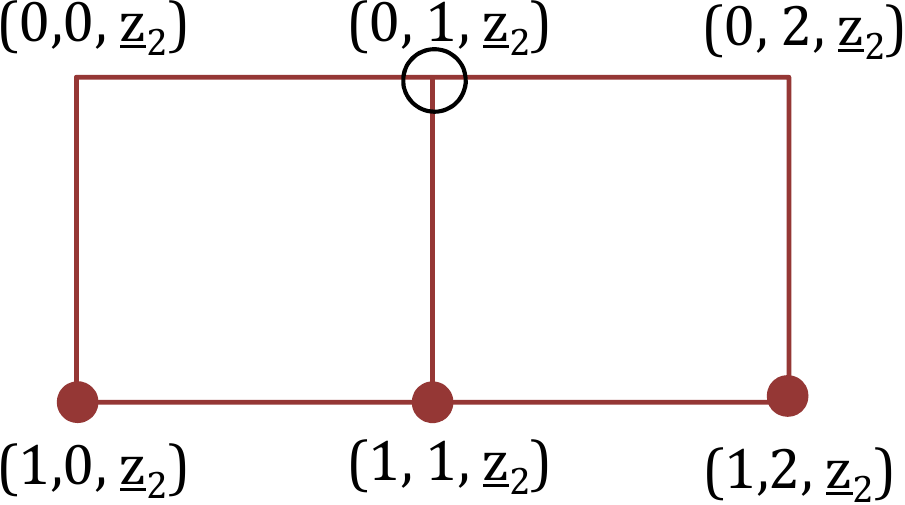}}
		\hspace{0.5in}
		\subfigure[out-of-plane symbols in plane $\uz_3 = \uz(0 \rightarrow z_2) = (1, 0, 0)$ shown in black circle]{\label{fig:outplane2}\includegraphics[height=1.0in]{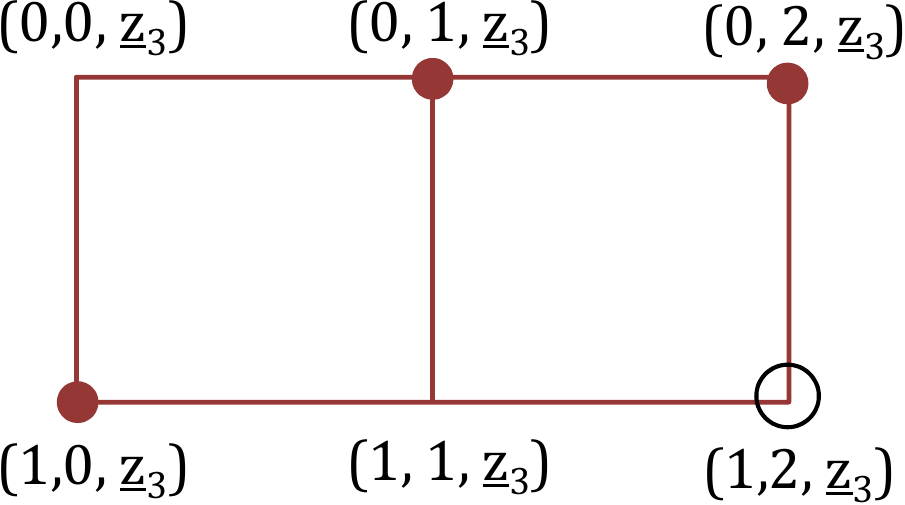}}
		\caption{Illustration of the symbols participating in parity check equation of plane $\uz=(1,0,1)$ shown circled in black.\label{fig:inoutsymbols}}
	\end{center}
\end{figure}

 The $[\ell,\uz]$-th parity check equation is given by:
\bea
\nonumber \sum\limits_{y=0}^{t-1} \sum\limits_{x=0}^{s-1} \sum\limits_{\uu \in \Zst} H([\ell,\uz],[x,y,\uu]) C(x, y; \uu) &=& 0.
\eea
By applying the equation \eqref{eq:pcsup} we get:
\bea
\label{eq:pc_msr_2}
\sum\limits_{y=0}^{t-1}\sum\limits_{x = 0}^{s-1} \left( \theta_{x,y,z_y}^{\ell-1}  \underbrace{\cxyz}_{\text{in-plane}} + \mathbf{1}_{x \ne z_y} \gamma_{x,z_y}\theta_{z_y,y,x}^{\ell-1} \underbrace{\cxyzc}_{\text{out-of-plane}} \right) &=& 0.
\eea
By parity check equations associated to the plane $\uz$, we will mean the p-c equations resulting from fixing $\uz \in \Zst$ and varying $\ell \in [r]$. The symbols participating within a p-c equation can be differentiated as in-plane symbols and out-of-plane symbols as indicated in equation \eqref{eq:pc_msr_2} and as illustrated using circles, in Fig.\ref{fig:inoutsymbols}.

Notice that there are $n=st$ in-plane symbols and $t(s-1)$ out-of-plane symbols participating in parity check equation shown in \eqref{eq:pc_msr_2}. We will now show in Lemma~\ref{lem:thetadistinct} that these $(2s-1)t$ symbols together are elements of an $[(2s-1)t, (2s-1)t-r]$ Generalized Reed Solomon (GRS) code by showing that the $(2s-1)t$ \ p-c variables  $\cup_{y=0}^{t-1} \left\lbrace \theta_{x, y, z_y}, \theta_{z_y, y, x} \mid x \in \Zs \right\rbrace$ that appear in equation \eqref{eq:pc_msr_2}, are all distinct. 

\blem\label{lem:thetadistinct}
The collection of $\theta$'s shown below correspond to $(2s-1)t$ distinct symbols in \fq\ for any $\uz \in \Zst$.
\bean
 \cup_{y=0}^{t-1} \underbrace{ \left\lbrace \theta_{x, y, z_y}, \theta_{z_y, y, x} \mid x \in \Zs \right\rbrace}_{A_y} 
\eean
\elem
\bprf We will first show that there are $(2s-1)$ distinct elements in $A_y$. $\{\theta_{x, y, z_y}, \theta_{z_y, y, x} \mid x \in [0:s-1] \}$ are $(2s-1)$ elements in $(s \times s)$ matrix $\Lambda_{s, y}$ where $s$ elements are from column $z_y$ of $\Lambda_{s, y}$ and remaining $s-1$ elements are non-diagonal elements from row indexed by $z_y$ of matrix . From equation~\eqref{eq:thetaassign} it can be verified that these elements are distinct for any $y$. It is clear to see that $A_{y} \cap A_{y'} = \phi$ for $y \ne y'$ as the elements in $A_y$, $A_{y'}$ are picked from matrices $\Lambda_{s, y}$, $\Lambda_{s, y'}$ respectively and by definition these matrices have distinct symbols from $\fq$.
\eprf

\subsection{Making a Connection with the Clay Code} \label{sec:connections}
An $(n=st, k=n-r, d=k+s-1)$ Clay code can be defined using the $(r\alpha \times n\alpha)$ p-c matrix $H_{\tiny \text{Clay}}$ shown below:
\bea
\label{eq:pcsup_clay} H_{\tiny \text{Clay}}([\ell, \uz], [x,y,\uu]) &=& \begin{cases}
	\theta_{(x,y)}^{\ell-1} & \uz = \uu\\
	\gamma \theta_{(u_y,y)}^{\ell-1} & \uz = \uu(x \rightarrow u_y), x \ne u_y\\
	0 & \text{otherwise},
\end{cases}
\eea
where the pair $[\ell, \uz]$ indexes the rows with $\ell \in [r]$, $\uz \in \Zst$ and the triple $[x, y, \uu] \in \Zs \times \Zt \times \Zst$ indexes the columns.    Here, we impose the condition that $\gamma^2 \ne 1$ and $\{\theta_{x, y} \mid x \in \Zs, y \in \Zt \}$ are a collection of $n = st$ distinct elements in $\fq$ where $q \ge n$.   If the symbols $\{ \cxyz \mid x \in \Zs, y \in \Zt, \uz \in \Zst\}$ are the $n\alpha$ code symbols of the Clay code, the parity check equations are then given by:
\bea
\label{eq:pc_msr_clay}
\sum\limits_{y=0}^{t-1}\sum\limits_{x = 0}^{s-1} \left( \theta_{(x,y)}^{\ell-1}  \underbrace{\cxyz}_{\text{in-plane}} + \mathbf{1}_{x \ne z_y} \gamma \theta_{(x,y)}^{\ell-1} \underbrace{\cxyzc}_{\text{out-of-plane}} \right) &=& 0 \text{ for any } \ell \in [r], \uz \in \Zst.
\eea
Notice that $(2s-1)t$ code symbols appear in the p-c equations associated to a plane $\uz$. However, they do not form a GRS code unlike in the case of the \smalld\ code.  As a result of this Clay code structure, when one attempts to carry out single-node repair using a collection of $d < n-1$ helper nodes, during repair of failed node $(x_0, y_0)$ the $(s-1)$ nodes $\{ (x, y_0) \mid x \in \Zs \setminus \{x_0 \}\}$ must necessarily be part of the $d$ helper nodes. The remaining $d-s+1=k$ helper nodes can be chosen arbitrarily.  Thus, one cannot choose any $d$ nodes to aid in node reapir as is required of a regenerating code.    This problem is circumvented in the case of the \smalld\ MSR code construction by ensuring that all $(2s-1)t$ p-c variables  appearing in the p-c equation~\eqref{eq:pc_msr_2}, are distinct.


\section{Partitioning of Erasure Patterns and the Equivalence Classes of Planes\label{sec:termi}}


We will now introduce the notation and terminology that will be used to show that the \smalld\ MSR code is indeed an optimal-access MSR code. For this, we have to show that the \smalld\ MSR code is an MDS code and that it possesses the optimal-access repair property.

\subsection{Steps Involved in Establishing the MDS and Optimal-Access Repair Properties} 

In order to prove the MDS property it is enough to show that the code is able to recover from the erasure of the code symbols associated to any $(n-k)=r$ nodes. This implies recovering $r$ symbols from each of the $\alpha=s^t$ planes. We provide a sequential decoding algorithm where the planes are first associated with an intersection score (see Definition~\ref{defn:intscore}) and are ordered by that score. The erased symbols corresponding to planes with lower intersection score are decoded first followed by planes having larger intersection score.  The planes having the same intersection score, are partitioned into equivalence classes and all the planes within the same equivalence class are decoded together.  The partitioning into equivalence classes is introduced in Definition~\ref{defn:decplanes}.  It will be shown in the subsequent section, Section~\ref{sec:mds_redn}, how recovery of erased symbols reduces to the problem of proving the invertibility of certain sub-matrices of the p-c matrix that are introduced within the present section, in Definitions~\ref{defn:Hsubdefn},\ref{defn:Hsubred}.  Specifying these sub-matrices calls for a partitioning of the erasure pattern into three distinct subsets (see Definition~\ref{defn:eraspat}) of erasures, a partitioning that is dependent on the plane index $\underline{z}$.

To establish the optimal-access repair property, we provide a sequential repair algorithm in which the $(n-1-d)=r-s$ nodes that do not participate in the repair process are regarded as nodes that have been erased.  We use the term {\emph aloof nodes} to refer to these nodes.   
Only a subset $\beta$ of the $\alpha$ planes within the datacube participate in the repair process and are hereby referred to as repair planes. We associate an intersection score with each of the repair planes and as was the case with the sequential decoding algorithm described above, we partition the repair planes into equivalence classes and simultaneously repair planes lying within the same equivalence class.   In Section~\ref{sec:oar_redn} we will show how recovery of failed node can be reduced to showing the invertibility of certain sub-matrices of the p-c matrix.  

\begin{note}
In this section, we will use the symbol $E$ in two different ways.  During the proof of the MDS property of the \smalld\ MSR code, $E$ will denote the set of $r$ erased nodes.  During the proof of the optimal-access repair property of the \smalld\ MSR code, $E$ will refer to the $(r-s)$ aloof nodes that do not participate in the repair process and thus may be regarded as being erased. 
\end{note}


	
\begin{figure}[ht!]
	\begin{center}
		\subfigure[$\uz = (1,1,1)$, $\sigma(E, \uz) = 0$]{\label{fig:is0}\includegraphics[height=1.35in]{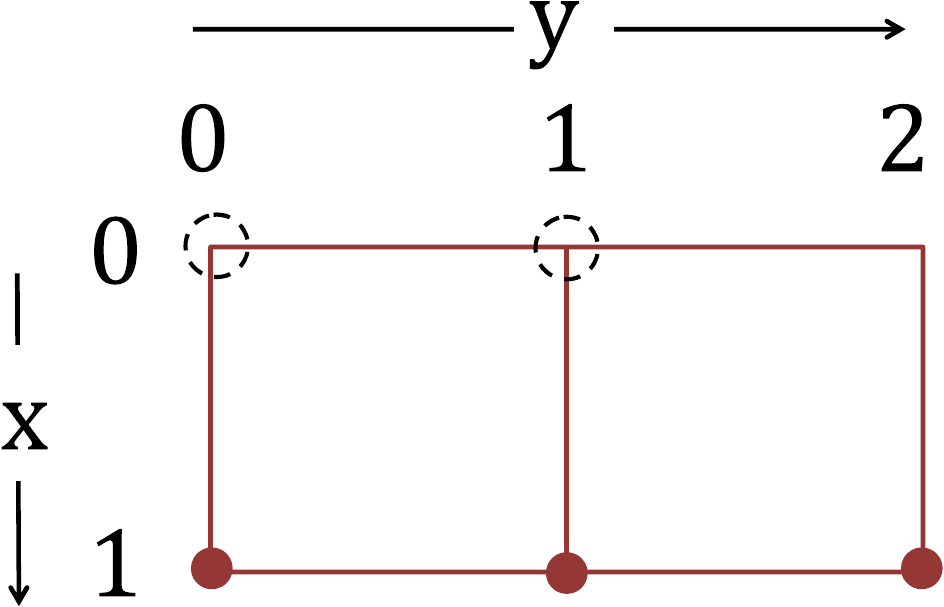}}
		\hspace{0.1in}
		\subfigure[$\uz = (1,0,1)$, $\sigma(E, \uz) = 1$]{\label{fig:is1}\includegraphics[height=0.9in]{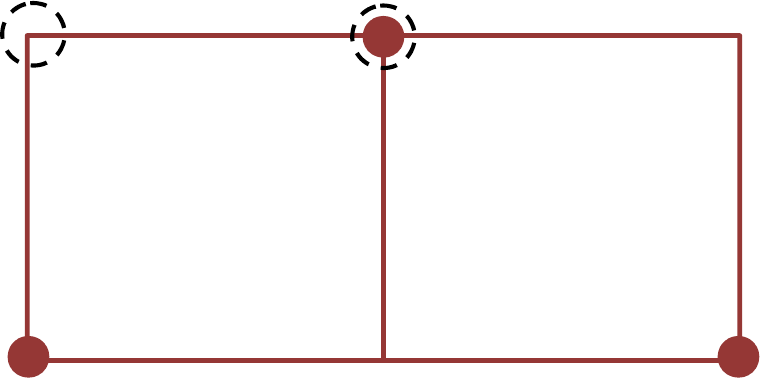}}
		\hspace{0.1in}
		\subfigure[$\uz = (0,0,1)$, $\sigma(E, \uz) = 2$]{\label{fig:is2}\includegraphics[height=0.9in]{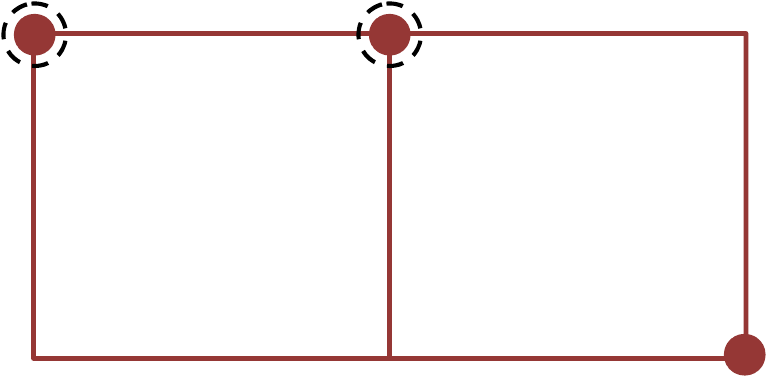}}
		\caption{Illustration of the Intersection Scores of erasure (aloof-node) pattern $E = \{(0,0),(0,1)\}$ over various planes and erasures (aloof-nodes) are indicated as holes (the dotted circles).\label{fig:is}}
	\end{center}
\end{figure}

\bdefn\label{defn:intscore}
Intersection Score corresponding a plane $\uz$ and an erasure pattern (aloof node) set $E$ is given by:
\bea
\label{eq:is}\sigma(E, \uz) = |\{ (z_y, y) \in E | y \in \Zt \}| = |E_{0, \uz}|.
\eea
This can also be seen as the hole-dot count in the plane-dot representation, where holes (dotted-circles) correspond to the erasure (aloof-node) pattern and dots indicate the plane index. See Fig.\ref{fig:is} for an illustration.
\edefn

In the sequential decoding (repair) algorithm, the planes within an equivalence class are decoded (repaired) together. 
Given an erasure (aloof node) pattern $E$ and a plane $\uz$, we define the equivalence class of planes $Q(E, \uz)$ below. 

\bdefn[Equivalence Class of $\uz$]\label{defn:decplanes}
Given a plane \uz, we use $Q(E, \uz)$ to denote the collection of planes 
\bean
Q(E, \uz) & = & S_0 \times S_1 \cdots S_{t-1}, 
\eean
where the $S_y$, $y \in \Zt$ are given by:
\bea
\label{eq:qezcomp}S_y = \begin{cases}
	\{ x | (x,y) \in E \} & (z_y, y) \in E\\
	\{z_y\} & \text{otherwise.}
\end{cases}
\eea
The collection $Q(E, \uz)$ contains \uz\ and will turn out to represent the set of planes that are decoded together during the process of recovering the erased symbols $E$ contained within the plane $\uz$.  The collection $Q(E, \uz)$ can be verified to satisfy the following closure property:
\bean
\uw \in Q(E,\uz) & \Leftrightarrow & Q(E,\uz) \ = \ Q(E,\uw) .
\eean

\begin{figure}[ht!]
	\begin{center}
		\subfigure[$\uz = (1,0,1)$, $E_{0,\uz} = \{(1,0)\}$, $E_{1,\uz}=\{(1,1)\}$, $E_{2,\uz}=\{(0,0)\}$  ]{\label{fig:eraspat1}\includegraphics[height=1.35in]{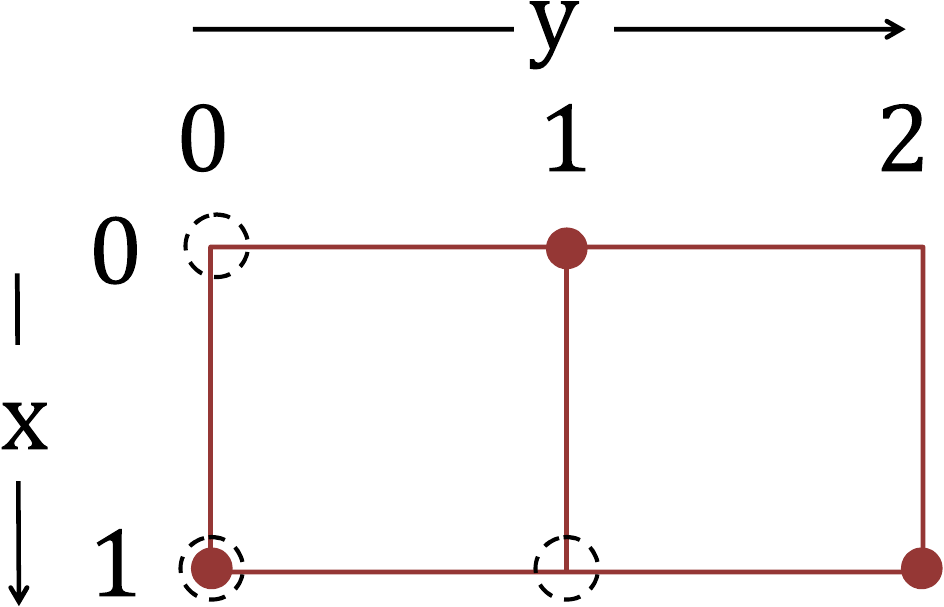}}
		\hspace{0.1in}
		\subfigure[$\uz = (0,0,1)$, $E_{0,\uz} = \{(0,0)\}$, $E_{1,\uz}=\{(1,1)\}$, $E_{2,\uz}=\{(1,0)\}$]{\label{fig:eraspat2}\includegraphics[height=0.9in]{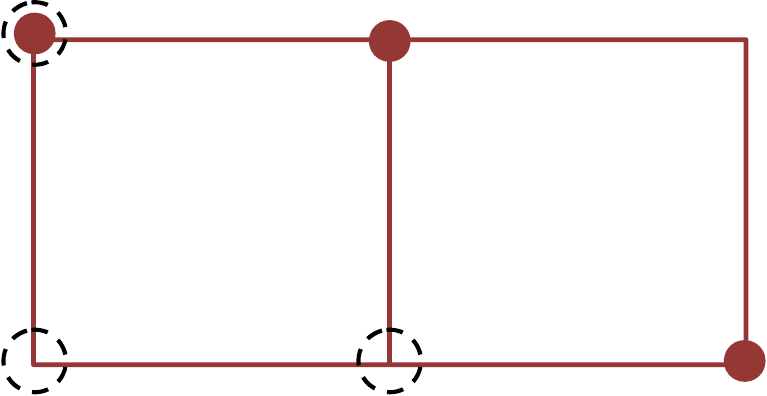}}
		\hspace{0.1in}
		\subfigure[$\uz = (1,1,1)$, $E_{0,\uz} = \{(1,0), (1,1)\}$, $E_{2,\uz}=\{(0,0)\}$, $E_{1,\uz}=\phi$]{\label{fig:eraspat3}\includegraphics[height=0.9in]{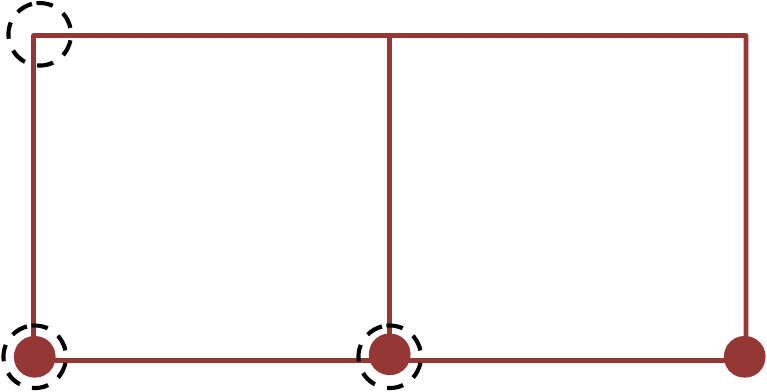}}
		\caption{Illustration of the erasure pattern partitioning for $E = \{(0,0),(1,0),(1,1)\}$ over various planes.\label{fig:eraspat}}
	\end{center}
\end{figure}

\edefn
\begin{figure}[ht!]
	\begin{center}
		\subfigure[$\uz_1=(1,0,1)$]{\label{fig:deceraspat1}\includegraphics[width=1.75in]{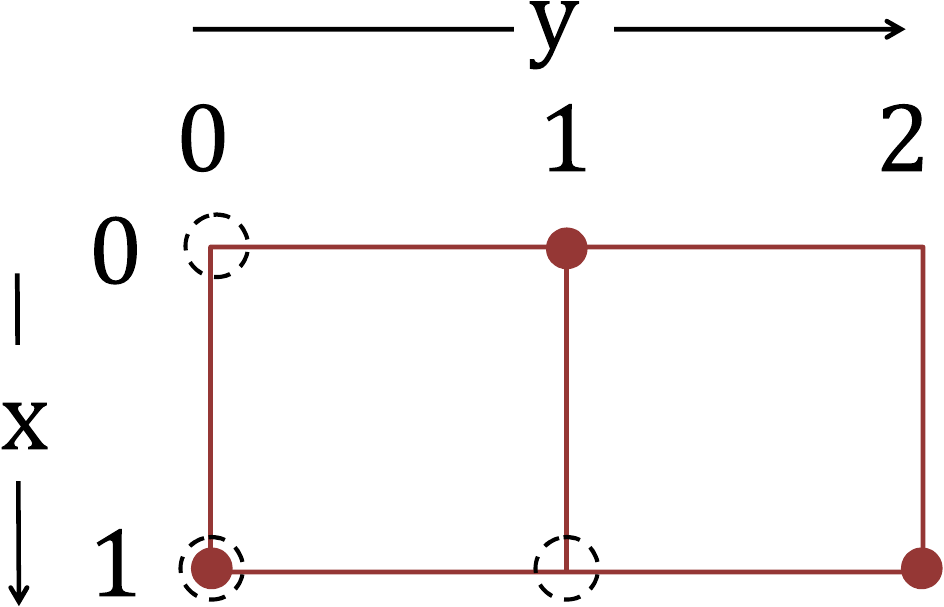}}
		\subfigure[$\uz_2 = (0,0,1)$]{\label{fig:deceraspat2}\includegraphics[width=1.5in]{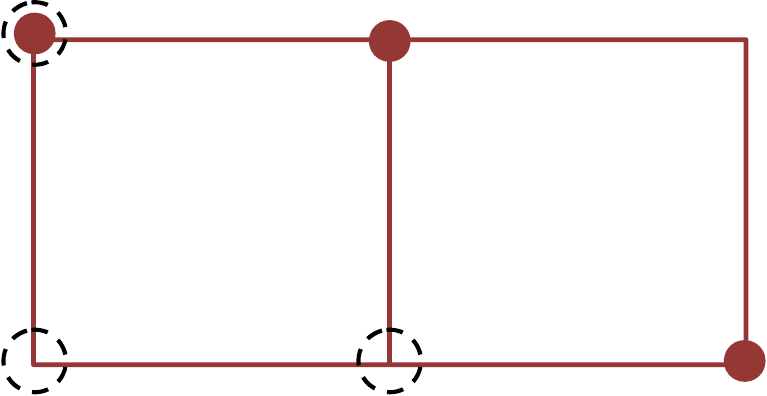}}		
		\subfigure[$\uz_3 = (1,1,1)$]{\label{fig:deceraspat3}\includegraphics[width=1.5in]{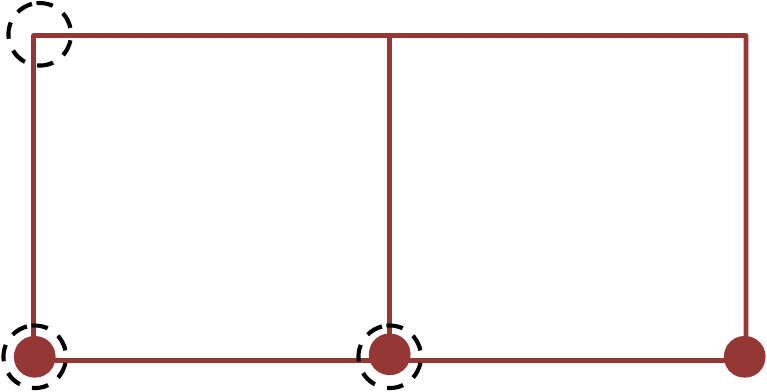}}
		\subfigure[$\uz_4 = (1,0,1)$]{\label{fig:deceraspat4}\includegraphics[width=1.5in]{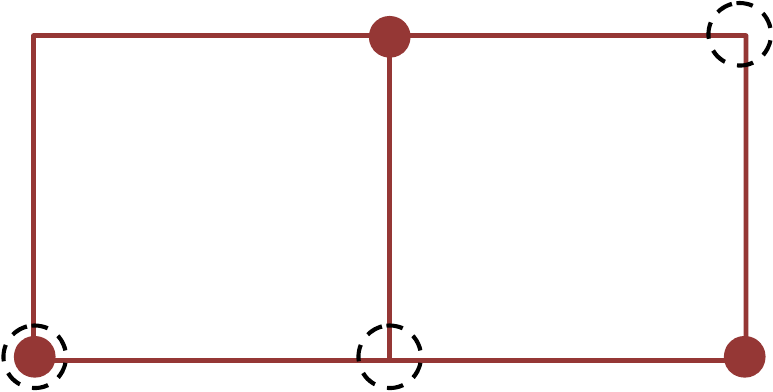}}
		\caption{Illustrating the definition of the equivalence class of planes that are decoded together for erasure pattern $E = \{(0,0),(1,0),(1,1)\}$:  $Q(E,\uz_1) =  Q(E,\uz_2) = \{\uz_1,\uz_2\}$, \ $Q(E,\uz_3) =  Q(E,\uz_5)  =  \{\uz_3,\uz_5\}$ where $\uz_5=(0,1,1)$ and $Q(E,\uz_4)=\{\uz_4\}$.  \label{fig:deceraspat}}
	\end{center}
\end{figure}
\begin{note}
	It is clear to see that $Q(E, \uz)$ is an equivalence class of $\uz$ as for any $\uw \in Q(E, \uz)$, $Q(E,\uw) = Q(E,\uz)$ and for any $\uw \notin Q(E, \uz)$, $Q(E, \uz) \cap Q(E, \uw) = \phi$.
\end{note}
We define below a partitioning of erasure pattern set $E$ into three subsets given a plane $\uz$. This will be used in defining the p-c sub-matrices whose invertibility will imply the MDS and optimal access properties. In Fig.~\ref{fig:eraspat}, erasures are indicated as holes (the dotted circles).

\bdefn[Erasure Patterns] \label{defn:eraspat} Given an erasure pattern $E$ and a plane $\uz \in \Zst$ we define a partitioning of the erasure patterns as following:
\bean
E_{0,\uz} &=& \{(z_y, y) \in E \} \ \ \text{(hole-dot pairs)} \\
E_{1,\uz} &=& \{(x,y) \in E \mid (z_y, y) \notin E \} \ \ \text{(holes without hole-dot pair in their column)}\\
E_{2,\uz} &=& \{(x,y) \in E \mid x \ne z_y, \ (z_y, y) \in E \} \ \ \text{(holes with hole-dot pair in their column)}
\eean
\edefn

We will refer to the subsets in the partition $E_{0,\uz}, E_{1, \uz}$ and $E_{2, \uz}$  as mild, moderate and serious erasures respectively.
\blem\label{lem:decplane_prop}
For any plane $\uw$ in the equivalence class of $\uz$, $\uw \in Q(E, \uz)$ it follows that:
\ben
\item The subset of moderate erasures remains the same i.e., $E_{1, \uw} = E_{1, \uz}$,
\item The cardinality of the mild and serious erasures remains the same i.e., $|E_{0, \uw}| = |E_{0, \uz}|$, $|E_{2, \uw}| = |E_{2, \uz}|$,
\item The intersection score is the same i.e., $\sigma(E, \uw) = \sigma(E, \uz)$,
\item $|Q(E,\uz)| = \prod\limits_{y = 0}^{t-1}(e_y+1)$, $|E_{2, \uz}| = \sum\limits_{y=0}^{t-1} e_y$, $S_y = \{z_y\} \cup E_{2, \uz}(y)$ where:
\bean
E_{2,\uz}(y) = \{ x' \mid (x', y') \in E_{2,\uz},  y' = y \} \text{ and } e_y = |E_{2, \uz}(y)|.
\eean
\item If $|E_{2,\uz}| = 0$ then $Q(E, \uz) = \{\uz\}$.
\een
\elem
\bprf
The proof follows directly from Definitions \ref{defn:eraspat}, \ref{defn:decplanes} and \ref{defn:intscore}.
\eprf

\subsection{MDS Sub-Matrix and The Reduced Matrix}
Let $H$ denote the overall p-c matrix of the MSR code appearing in equation \ref{eq:pcsup}.  We will now define sub-matrices of the p-c matrix, $H_{E,{\tiny \uz}}$ for any $\uz \in \Zst$, $E \subseteq \Zs \times \Zt$ such that $|E|=r$, whose invertibility would imply the MDS property. The proof of this appears in Theorem~\ref{thm:reduction1}.

\bdefn[MDS Sub-Matrix]\label{defn:Hsubdefn}
Given an erasure pattern $E$ such that $|E| = r$ and plane $\uz \in \Zst$, we set $p = |Q(E, \uz)|$.   We use $H_{E, {\tiny \uz}}$ to denote the $(r p \times r p)$ sub-matrix of $H$ obtained by restricting attention to p-c equations indexed by planes in $Q(E,\uz)$ and erased symbols $E$ within planes in $Q(E, \uz)$ i.e., 
\bea
\label{eq:Hsubdefn}H_{E, {\tiny \uz}}([\ell,\uv], [x,y,\uu]) = H([\ell, \uv], [x,y,\uu]) , \ \ \uu, \uv \in Q(E,\uz), \ (x,y) \in E, \ \ \ell \in [r]. 
\eea
\edefn

We now define a further small sub-matrix of p-c matrix whose invertibility implies invertibility of $H_{E, {\tiny \uz}}$. This will be shown in Theorem~\ref{thm:reduction2}.

\bdefn[The Reduced Matrix] \label{defn:Hsubred}
Given an erasure or aloof-node pattern $E$ and plane $\uz \in \Zst$, we set $p = |Q(E, \uz)|$ and $\mu = |E_{2,\uz}|$.   We use $H^{\text{\tiny Red}}_{E, \tiny{\uz}}$ to denote the $(\mu p \times \mu p)$ sub-matrix of $H$ obtained by restricting attention to $\mu$ p-c equations indexed by planes in $Q(E,\uz)$ and erased symbols from set $E_{2, \uu}$ for planes $\uu \in Q(E, \uz)$ i.e., 
\bea
\label{eq:Hsubdefn2}
H^{\text{\tiny Red}}_{E, {\tiny \uz}}([\ell, \uv], [x,y,\uu]) &=& H([\ell, \uv], [x,y,\uu]), \ \ \uu, \uv \in Q(E,\uz), \ (x,y) \in E_{2,\uu}, \ \ \ell \in [\mu]. 
\eea
\edefn

\section{An Example Code $s=2$, $r=3$ \label{sec:example}}
Before proving the MDS and optimal-access repair properties for the general case of any $s \in \{2, 3, 4\}$ and $r \ge s$, we present the proof for example case of  $s=2$ and $r=3$. The ideas used to prove the lemmas in this section will help understand the reduction proofs for MDS property in Theorems~\ref{thm:reduction1}, \ref{thm:reduction2} and the reduction proofs for optimal-access repair property presented in Theorems~\ref{thm:reduction1_rep}, \ref{thm:reduction2_rep}. The parameters of \smalld\ code for this example are as follows:
\bean
(n=2t, k=n-3, d=n-2), \ \ (\alpha = 2^t, \beta=2^{t-1}).
\eean
Note that by using the idea of shortening described in Section~\ref{sec:smalld} one can construct optimal-access MSR code for any $(n, k=n-3, d=n-2)$.
We will start by showing the MDS property for the example in Lemma~\ref{lem:example_mds}.

\blem[MDS property for $s=2$, $r=3$ ] \label{lem:example_mds} \smalld\ construction for $s=2, r=3$ is an MDS code. 
\elem
\bprf
MDS property can be shown by proving that we can recover from any $r=3$ erasures. The type of erasure patterns $E$ can be classified in to two cases, (1) where all the three erasures have different $y$ and (2) where two erasures have same $y$.
In both the cases erased symbols are recovered by arranging the planes sequentially in increasing order of intersection score and decoding erased symbols plane by plane.

\subsubsection{Case 1: Three erasures with different $y$}
Let $E = \{(x_1, y_1), (x_2, y_2), (x_3, y_3) \}$ be the set of erasures where $y_1, y_2, y_3$ are distinct. In this case, for any $\uz \in \Ztwot$, the equivalence class of $\uz$ contains $\uz$ alone i.e., $Q(E, \uz) = \{\uz\}$.
\ben
\item[0]-th Step: Consider planes $\uz \in \Ztwot$ with intersection score $\sigma(E, \uz) = 0$. In this case for all $y \in \Zt$, $(z_y, y) \notin E$. Therefore all the out-of-plane symbols participating in the $[\ell, \uz]$-th p-c equation are known. Hence the $[\ell,\uz]$-th p-c equation~\eqref{eq:pc_msr_2} reduces to:
\bea
\label{eq:exmds1}
\sum\limits_{(x,y) \in E} \theta_{x, y, z_y}^{\ell -1} \cxyz\ &=& \kappa^*,
\eea
where $\kappa^*$ can be computed from the unerased symbols. Therefore by varying $\ell \in [3]$, erased symbols corresponding to this plane $\{ \cxyz \mid (x,y)\in E \}$ can be recovered as $\theta_{x, y, z_y}$'s are distinct for $(x, y) \in E$.
\item[j]-th Step: Consider planes $\uz \in \Ztwot$ with intersection score $\sigma(E, \uz) = j$. Then the $[\ell, \uz]$-th p-c equation can be written as:
\bean
\sum\limits_{(x,y) \in E} \theta_{x, y, z_y}^{\ell-1} \cxyz + \sum\limits_{\substack{y : (z_y, y) \in E \\ x \ne z_y}} 
\gamma_{x, z_y} \theta_{z_y, y, x}^{\ell-1} \cxyzc\ &=& \kappa^*,
\eean
 where $\kappa^*$ can be computed from the unerased symbols. We will now make an observation that the out-of-plane symbols appearing in the above equation are known. For $y$ such that $(z_y, y) \in E$, by the choice of erasure pattern, there are no more erasures in with same $y$ ie., for any $x \ne z_y$, $(x, y) \not \in E$ and therefore $\sigma(\uz(x \rightarrow z_y)) = j-1$. Hence the out-of-plane symbol \cxyzc\ is recovered in $(j-1)$-th step and is available during the $j$-th step. Therefore the $[\ell, \uz]$-th p-c equation reduces to equation~\eqref{eq:exmds1} and the erased symbols in this plane can be recovered due to distinctness of $\theta_{x,y,z_y}$'s for $(x,y) \in E$.
\een
By end of all steps we have recovered all the erased symbols $\{\cxyz \mid (x,y) \in E, \uz \in \Ztwot \}$.

\subsubsection{Case 2: Two erasures with same $y$}
Let $E = \{(0, y_1), (1, y_1), (x_2, y_2) \}$ be the set of erasures where $y_1 \ne y_2$. The intersection scores that are possible in this case are $1, 2$ with plane $\uz$ having intersection score $1$ when $z_{y_2} \ne x_2$ and intersection score $2$ when $z_{y_2} = x_2$.
\ben
\item[1]-st Step: Consider planes $\uz \in \Ztwot$ such that $z_{y_1} = 0$, $z_{y_2} \ne x_2$. These planes have intersection score $\sigma(E, \uz) = 1$. The $[\ell, \uz]$-th p-c equation~\eqref{eq:pc_msr_2} reduces to:
\bea
\nonumber
\sum\limits_{(x,y) \in E}\theta_{x, y, z_y}^{\ell-1} \cxyz\ + \sum\limits_{\substack{y:(z_y, y) \in E \\ x \ne z_y}}\gamma_{x, z_y} \theta_{z_y, y, x}^{\ell-1} \cxyzc\  &=& \kappa_*\\
\label{eq:exmds2} \implies \sum\limits_{(x,y) \in E} \theta_{x, y, z_y}^{\ell-1} \cxyz\ + \gamma_{1, 0} \theta_{0, y_1, 1}^{\ell-1} C(0, y_1; \uz( 1 \rightarrow z_{y_1}) )  &=& \kappa_*.
\eea
Here the out-of-plane symbol $C(0, y_1; \uz( 1 \rightarrow z_{y_1}) )$ is unknown as the intersection score of the plane $\uw = \uz( 1 \rightarrow z_{y_1})$ is $\sigma(E,\uw)=1$. Therefore there are $4$ unknowns and $3$ p-c equations by varying $\ell \in [3]$. The equivalence class of $\uz$ in this case is given by $Q(E, \uz) = \{ \uz, \uw\}$. We will therefore also consider the $[\ell, \uw]$-th p-c equations for $\ell \in [3]$:
\bea
\label{eq:exmds3} \sum\limits_{(x,y) \in E} \theta_{x, y, w_y}^{\ell-1} C(x,y; \uw) + \gamma_{0, 1} \theta_{1, y_1, 0}^{\ell-1} C(1, y_1; \underbrace{\uw(0 \rightarrow w_{y_1})}_{\uz})  &=& \kappa_*
\eea
Together the $6$ equations in \eqref{eq:exmds2} and \eqref{eq:exmds3} have $6$ unknowns and the equations are as shown below.
\bean
\left[\begin{array}{ccc|ccc}
	1 & 1 & 1 & 1 & &\\
	\theta_{0, y_1, 0} & \theta_{1, y_1, 0} & \theta_{x_2, y_2, z_{y_2}} & \theta_{0, y_1, 1} & &\\
	\theta_{0, y_1, 0}^2 & \theta_{1, y_1, 0}^2 & \theta_{x_2, y_2, z_{y_2}}^2 &  \theta_{0, y_1, 1}^2 & &\\ \hline
	& \gamma & & 1 & 1 & 1\\
	& \gamma \theta_{1, y_1, 0} & & \theta_{0, y_1, 1} & \theta_{1, y_1, 1} & \theta_{x_2, y_2, z_{y_2}}\\
	& \gamma \theta_{1, y_1, 0}^2 & & \theta_{0, y_1, 1}^2 & \theta_{1, y_1, 1}^2 & \theta_{x_2, y_2, z_{y_2}}^2
\end{array}\right] \left[ \begin{array}{c}
	C(0, y_1, \uz) \\
	C(1, y_1, \uz) \\
	C(x_2, y_2, \uz) \\
	C(0, y_1, \uw) \\
	C(1, y_1, \uw) \\
	C(x_2, y_2, \uw)
\end{array} \right] &=& \kappa_*.
\eean
Note that $\gamma_{1,0}=1$ and $\gamma_{0,1}=\gamma$ from equation~\eqref{eq:gamma}. Therefore the erased symbols corresponding to the planes $\uz$, $\uw$ can be recovered given the following matrix is invertible.
\bea
\label{eq:hezexa}
H_{E, {\tiny \uz}} &=& \left[\begin{array}{ccc|ccc}
	1 & 1 & 1 & 1 & &\\
	\theta_{0, y_1, 0} & \theta_{1, y_1, 0} & \theta_{x_2, y_2, z_{y_2}} & \theta_{0, y_1, 1} & &\\
	\theta_{0, y_1, 0}^2 & \theta_{1, y_1, 0}^2 & \theta_{x_2, y_2, z_{y_2}}^2 &  \theta_{0, y_1, 1}^2 & &\\ \hline
	& \gamma & & 1 & 1 & 1\\
	& \gamma \theta_{1, y_1, 0} & & \theta_{0, y_1, 1} & \theta_{1, y_1, 1} & \theta_{x_2, y_2, z_{y_2}}\\
	& \gamma \theta_{1, y_1, 0}^2 & & \theta_{0, y_1, 1}^2 & \theta_{1, y_1, 1}^2 & \theta_{x_2, y_2, z_{y_2}}^2
\end{array}\right].
\eea
Let the vector $\uf = (f_{0,0}, f_{0,1}, f_{0,2}, f_{1,0}, f_{1,1}, f_{1,2})^T$ be in the left null space of $H_{E, {\tiny \uz}}$ i.e., $\uf^T H_{E, {\tiny \uz}}=0$ and let $f_i(x) = \sum\limits_{\ell=1}^3 f_{i,\ell} \theta^{\ell-1}$ for $i=0,1$. It is clear to see that:
\bea
\nonumber f_0(\theta_{0, y_1, 0}) = f_0(\theta_{x_2, y_2, z_{y_2}}) = 0\\
\nonumber f_1(\theta_{1, y_1, 1}) = f_1(\theta_{x_2, y_2, z_{y_2}}) = 0\\
\label{eq:exmds4}f_0(\theta_{0, y_1, 1}) + f_1(\theta_{0, y_1, 1}) = 0\\
\label{eq:exmds5}f_0(\theta_{1, y_1, 0}) + \gamma f_1(\theta_{1, y_1, 0}) = 0.
\eea
By the assignment of coefficients shown in equation \eqref{eq:thetaassign}, $\theta_{0, y_1, 0} = \theta_{1, y_1, 1} = \lambda_{0, y_1}$. Therefore, both the polynomials $f_0, f_1$ have $\lambda_{0, y_1}$ and $\theta_{x_2, y_2, z_{y_2}}$ as roots and hence can be expressed as $f_i(\theta) = f_i^{\text{\tiny Red}} (\theta - \lambda_{0, y_1})(\theta - \theta_{x_2, y_2, z_{y_2}})$ for $i = 0, 1$ where $f_0^{\text{\tiny Red}}, f_1^{\text{\tiny Red}}$ are constants. Substituting this expression in equations \eqref{eq:exmds4} and \eqref{eq:exmds5} we get:
\bean
\underbrace{\left[\begin{array}{cc}
		1 & 1\\
		1 & \gamma
	\end{array}\right]}_{H_{E, {\tiny \uz}}^{\text{\tiny Red}}} \left[\begin{array}{c}
	f_0^{\text{\tiny Red}}\\
	f_1^{\text{\tiny Red}}
\end{array}\right] &=& 0.
\eean
Therefore, $f_0^{\text{\tiny Red}}=f_1^{\text{\tiny Red}}=0$ as $\gamma \ne 1$ implying that the polynomials $f_0, f_1$ are zeroes and that $\uf = \underline{0}$. Therefore the erased symbols corresponding to the planes $\uz$, $\uz(1 \rightarrow z_{y_1})$ given by $\{ \cxyz, C(x,y,\uz(1 \rightarrow z_{y_1})) \mid (x,y)\in E \}$ can be recovered.
\item[2]-nd Step: Consider planes $\uz \in \Ztwot$ such that $z_{y_1} = 0$ and $z_{y_2} = x_2$. These planes have intersection score $\sigma(E, \uz) = 2$ and the $[\ell, \uz]$-th p-c equation can be written as:
\bean
\sum\limits_{(x,y) \in E} \theta_{x, y, z_y}^{\ell-1} \cxyz\ + \sum\limits_{\substack{y:(z_y, y) \in E \\ x \ne z_y}}\gamma_{x, z_y} \theta_{z_y, y, x}^{\ell-1} \cxyzc\  =\kappa_*\\
\scalebox{0.9}{$\sum\limits_{(x,y) \in E} \theta_{x, y, z_y}^{\ell-1} \cxyz\ + \gamma_{1, 0} \theta_{0, y_1, 1}^{\ell-1} C(0, y_1; \uz(1 \rightarrow z_{y_1})) + \gamma_{x_2\oplus 1, x_2} \theta_{x_2, y_2, x_2 \oplus 1}^{\ell-1}C(x_2, y_2; \uz((x_2 \oplus 1) \rightarrow z_{y_2}))=\kappa_*$}.
\eean
The plane  $\uz((x_2 \oplus 1) \rightarrow z_{y_2})$ has intersection score $\sigma(E, \uz((x_2 \oplus 1) \rightarrow z_{y_2})) = 1$, therefore the symbol $C(x_2, y_2; \uz((x_2 \oplus 1) \rightarrow z_{y_2}))$ is already recovered in the first step. Hence the $[\ell,\uz]$-th p-c equation reduces to equation~\eqref{eq:exmds2}. The equivalence class of $\uz$, $Q(E, \uz) = \{\uz, \uz(1 \rightarrow z_{y_1})\}$. Therefore, we look at p-c equations corresponding to plane $\uw = \uz(1 \rightarrow z_{y_1})$. The $[\ell, \uw]$-th p-c equation reduces to equation~\eqref{eq:exmds3}. Therefore erased symbols corresponding to planes $\uz, \uz(1 \rightarrow z_{y_1})$, $\{C(x,y,\uz), C(x,y,\uz(1 \rightarrow z_{y_1})) \mid (x,y) \in E \}$ can be recovered due to invertibility of $H_{E, {\tiny \uz}}$ shown in equation \eqref{eq:hezexa}. 
\een
By the end of the two steps all the erased symbols are recovered.
\eprf
%
\  \\ 

We will now prove the optimal-access property which along with previous lemma proves that the  \smalld\ code is an MSR code for $s=2$, $r=3$.
\blem[Optimal Access Property for $s=2$, $r=3$] \smalld\
code for $s=2$, $r=3$ satisfies the optimal-access repair property.
\elem
\bprf
Let $(x_0, y_0)$ be the failed node and $E = \{ (x_1, y_1) \}$ be the set of aloof nodes. The number of aloof nodes in this case is $(n-1-d) = (r-s)=1$. We consider two cases for aloof nodes (1) when aloof node has same $y$ as the failed node (2) when aloof node has different $y$.

A helper node $(x,y)$ sends symbols $\{\cxyz\ \mid \uz \in R\}$ from the repair planes $R = \{\uz \in \Ztwot \mid z_{y_0} = x_0\}$. Therefore, the number of symbols downloaded from each helper node is $\beta = 2^{t-1}$.
\ben 
\item Case 1: Aloof node is $(x_1, y_1)$ where $y_0 = y_1$ and $x_1 = x_0 \oplus 1$
The $[\ell,\uz]$-th p-c equation corresponding to a repair plane $\uz \in R$ reduces to:
\bea
\label{eq:exmds6}
\theta_{x_0, y_0, x_0}^{\ell-1}C(x_0, y_0; \uz) + \theta_{x_1, y_1, z_{y_1}}^{\ell-1} C(x_1, y_1; \uz) + \gamma_{x_0 \oplus 1, x_0}\theta_{x_0, y_0, x_0 \oplus 1}^{\ell-1}C(x_0, y_0; \uz((x_0 \oplus 1) \rightarrow z_{y_0}) = \kappa_*.
\eea
This is because all the in-plane symbols of $\uz$ other than the failed node and aloof node symbol given by $\{C(x_0, y_0, \uz), C(x_1, y_1, \uz)\}$ are known. Among the out-of-plane symbols given by $\{\cxyzc \mid y \in \Zt, x \ne z_y \}$ the symbol $C(z_{y_0}, y_0, \uz( (x_0 \oplus 1) \rightarrow z_{y_0})) = C(x_0, y_0, \uz( (x_0 \oplus 1) \rightarrow z_{y_0}))$ is a failed node symbol and is hence unkown. For $y \ne y_0$, $(z_y, y) \notin E$  and therefore it is a helper node and the out-of-plane symbol $\cxyzc$ belongs to plane $\uz(x \rightarrow z_{y})$ which belong to the repair planes set $R$. Therefore the symbol $\cxyzc$ is available as helper information. Now by varying $\ell \in [3]$ in equation~\eqref{eq:exmds6} there are $3$ equations and $3$ unknowns and by Lemma~\ref{lem:thetadistinct} the symbols corresponding to failed node $\{C(x_0, y_0, \uz), C(x_0, y_0, \uz((x_0 \oplus 1) \rightarrow z_{y_0}) \}$ can be recovered along with one aloof node symbol $C(x_1, y_1; \uz)$. By varying $\uz \in R$ we can recover
\bean
\{\cxyzcnot \mid \uz \in R, x \in \mathbb{Z}_2 \} = \{\cxyznot \mid \uz \in \Ztwot \}. \eean

\item Case 2: Aloof node set is $E = \{(x_1 , y_1)\}$ where $y_1 \ne y_0$. The repair in this case is sequential. First the repair planes $\uz \in R$ such that $z_{y_1} \ne x_1$ are repaired as the intersection score for such planes is $\sigma(E, \uz) = 0$ and then the repair planes with $z_{y_1} = x_1$ are looked at as they have intersection score $1$.
\ben
\item[0]-th Step: Let $\uz \in R$ such that $z_{y_1} = x_1 \oplus 1$. Then the $[\ell, \uz]$-th p-c equation reduces to equation \eqref{eq:exmds6} as in this case too the only unknown symbols are the in-plane symbols $\{C(x_0, y_0, \uz), C(x_1, y_1, \uz)\}$ and  the out-of-plane symbol $C(x_0, y_0; \uz((x_0\oplus 1) \rightarrow z_{y_0})$. Therefore the failed node symbols $C(x_0, y_0; \uz)$, $C(x_0, y_0; \uz((x_0\oplus 1) \rightarrow z_{y_0})$ and the aloof node symbol $C(x_1, y_1; \uz)$ can be recovered. By the end of Step $0$, we would have recovered all the aloof node symbols in plane $\uz \in R$ such that $z_{y_1} = x_1 \oplus 1$.
\item[1]-th Step: Let $\uz \in R$ such that $z_{y_1} = x_1$, the $[\ell, \uz]$-th p-c equation in this case is given by:
\bean
\theta_{x_0, y_0, x_0}^{\ell-1}C(x_0, y_0; \uz) + \gamma_{x_0 \oplus 1, x_0}\theta_{x_0, y_0, x_0 \oplus 1}^{\ell-1}C(x_0, y_0; \uz((x_0 \oplus 1) \rightarrow z_{y_0}))  \ \ \ \ \ \ \ \ \ \ \ \ \ \ \ \ \ \ \ \ \\ 
+ \theta_{x_1, y_1, z_{y_1}}^{\ell-1} C(x_1, y_1; \uz) + \gamma_{x_1 \oplus 1, x_1}\theta_{x_1, y_1, x_1 \oplus 1}^{\ell-1} C(x_1, y_1, \uz((x_1 \oplus 1) \rightarrow z_{y_1})) = \kappa_*.
\eean
This is because the only unknown in-plane symbols are $\{C(x_0, y_0, \uz), C(x_1, y_1, \uz) \}$ and out-of-plane symbols are $\{C(x_0, y_0, \uz((x_0 \oplus 1) \rightarrow z_{y_0})), C(x_1, y_1; \uz((x_1 \oplus 1) \rightarrow z_{y_1}))\}$. However the aloof-node symbol $C(x_1, y_1; \uz((x_1 \oplus 1) \rightarrow z_{y_1}))$ is already recovered in the first step. Therefore $[\ell,\uz]$-th p-c equation reduces to equation~\eqref{eq:exmds6} and hence the failed node symbols $C(x_0, y_0, \uz)$, $C(x_0, y_0, \uz((x_0 \oplus 1) \rightarrow z_{y_0}))$ and aloof node symbol $C(x_1, y_1, \uz)$ can be recovered.
\een
Therefore by the end of the algorithm all the $\alpha$ symbols corresponding to the failed node $\{C(x_0, y_0, \uz) \mid \uz \in \Zst \}$ are recovered.
\een
\eprf

\section{MDS Property: The Reductions \label{sec:mds_redn}}

In this section we first start by showing in Theorem~\ref{thm:reduction1} that invertibility of the MDS Sub-Matrix $H_{E, {\tiny \uz}}$ (see Definition~\ref{defn:Hsubdefn}) for any erasure pattern $E \subseteq \Zs \times \Zt$ such that $|E|=r$ and any plane $\uz \in \Zst$,  implies that the \smalld\ code satisfies the MDS property. We follow this up with Theorem~\ref{thm:reduction2} where we show that invertibility of further reduced matrix $H_{E,{\tiny \uz}}^{\tiny \text{Red}}$ implies invertibility of $H_{E,{\tiny \uz}}$.

\bthm[The Reduction I: MDS property]\label{thm:reduction1}
To show that \smalld\ construction yields an MDS code, it suffices to show that for any erasure pattern $E$ such that $|E| = r$, and for any plane $\uz \in \Zst$, the matrix $H_{E, {\tiny \uz}}$ is invertible, where $H_{E, {\tiny \uz}}$ is as shown in Definition~\ref{defn:Hsubdefn}.
\ethm
\bprf
To show that \smalld\ code is an MDS code, it is enough to show that the code can recover from any erasure pattern $E$ such that $|E| = r$.

Given an erasure pattern we recover the erased symbols sequentially by ordering the planes in increasing order of their intersection scores, starting from $0$ and recovering erased symbols lying in planes $\uz$ having intersection score $\sigma(E, \uz)=0$, then $\sigma(E, \uz)=1$ and so on. Among the planes that have same intersection score, say for plane $\uz$ such that $\sigma(E, \uz) = j$,  we look at the planes in equivalence class of $\uz$, $Q(E, \uz)$ and decode them together. This can be done because all the planes in $Q(E, \uz)$ have same intersection score by Lemma~\ref{lem:decplane_prop}.

\begin{enumerate}[wide, labelwidth=!, labelindent=0pt]
\item[Step 0:]  Let $\uz \in \Zst$ with intersection score $\sigma(E, \uz) = 0$. Then in the $[\ell,\uz]$-th p-c equation shown in \eqref{eq:pc_msr_2}, all the out-of-plane symbols $\{\cxyzc \mid y \in \Zt, x \in \Zs \setminus \{z_y\}\}$ are known as for any $y \in \Zt, (z_y, y) \notin E$. Therefore the $[\ell, \uz]$-th p-c equation~\eqref{eq:pc_msr_2} reduces to:
\bean
\sum\limits_{(x,y) \in E} \theta_{x,y,z_y}^{\ell-1} \cxyz &=& \kappa^* \text{ for all } \ell \in [r],
\eean
where $\kappa^*$ indicates the quantity that can be computed from known symbols. The equivalence class of $\uz$, $Q(E, \uz)$ consists of just the single plane $\{\uz\}$, therefore $H_{E, {\tiny \uz}}$ is an $(r \times r)$ matrix and $H_{E, {\tiny \uz}}([\ell,\uz],[x,y,\uz]) = \theta_{x,y,z_y}^{\ell-1}$ for any $\ell \in [r]$ and $(x,y) \in E$. It follows that if $H_{E, {\tiny \uz}}$ is invertible, the erased symbols corresponding to this plane $\uz$ can be recovered.
\item[Step $j$:] Let $\uz \in \Zst$ be such that $\sigma(E, \uz) = j$ and let us assume that the erased symbols corresponding to planes $\underline{w}$ having intersection score $\sigma(E,\underline{w}) \leq (j-1)$ have already been recovered, then the $[\ell, \uz]$-th p-c equation~\eqref{eq:pc_msr_2} reduces to:
\bean
\sum\limits_{(x,y) \in E} \theta_{x,y,z_y}^{\ell-1} \cxyz + \sum\limits_{y: (z_y, y) \in E}\sum\limits_{x = 0}^{s-1} \gamma_{x,z_y}\theta_{z_y,y,x}^{\ell-1} \cxyzc &=& \kappa^* \text{ for all } \ell \in [r].
\eean
Let $(z_y, y) \in E$, then $\sigma(E, \uz(x \rightarrow z_y)) \le j$ with $\sigma(E, \uz(x \rightarrow z_y)) = j$ iff $(x,y) \in E$. This implies that the symbols $\cxyzc$ are recovered in the $(j-1)$ step if $(x, y) \notin E$, whereas they are unknown if $(x, y) \in E$. Therefore the $[\ell,\uz]$-th p-c equation further simplifies to:
\bea
\label{eq:red1}\sum\limits_{(x,y) \in E} \theta_{x,y,z_y}^{\ell-1} \cxyz + \sum\limits_{ (x, y) \in E_{2, {\tiny \uz}}} \gamma_{x,z_y}\theta_{z_y,y,x}^{\ell-1} \cxyzc &=& \kappa^* \text{ for all } \ell \in [r].
\eea
This follows as $\gamma_{x, z_y} = 0$ if $x = z_y$ and by the Definition~\ref{defn:eraspat} of the set $E_{2, \uz}$. Notice that for the case when $|E_{2,\uz}|=0$ there are $r$ equations and $r$ unknowns in the above equation. From Lemma~\ref{lem:decplane_prop}, $Q(E, \uz) = \{\uz\}$ and $H_{E,{\tiny \uz}}$ is an $r \times r$ matrix with $H_{E,{\tiny \uz}}([\ell, \uz], [x, y, \uz]) = \theta_{x,y,z_y}^{\ell-1}$ for all $\ell \in [r]$ and $(x,y) \in E$. Invertibility of $H_{E, {\tiny \uz}}$ implies that we can recover the erased symbols $\{\cxyz \mid (x,y) \in E\}$.

In the case when $|E_{2,\uz}|>0$, the number of erased symbols appearing in the p-c equation \eqref{eq:red1} associated to plane \uz\ is greater than the number of equations $r$. It turns out that in this case, if we consider the p-c equations corresponding to all the planes in the equivalence class of $\uz$ 
 then we have a situation where the number of unknowns $rp$ equals the number of equations $rp$ where $p = |Q(E,\uz)|$.  The p-c matrix associated to this set of equations is precisely $H_{E,{\tiny \uz}}$ and hence if this p-c matrix is invertible, then all such erasures can be recovered. We will now go ahead and show that the unknowns in the p-c equations defined by indices $\{[\ell, \uw] \mid \uw \in Q(E,\uz), \ell \in [r]\}$ correspond to the erased symbols in planes $Q(E, \uz)$. This will imply that the number of unknowns and number of equations is $r p$.


For any $\uw \in Q(E, \uz)$, the p-c equations are given by:
\bean
\sum\limits_{(x,y) \in E} \theta_{x,y,w_y}^{\ell-1} C(x,y;\uw) + \sum\limits_{ (x, y) \in E_{2,{\tiny \uw}}} \gamma_{x,w_y}\theta_{w_y,y,x}^{\ell-1} C(w_y,y; \uw(x \rightarrow w_y)) &=& \kappa^* \text{ for all } \ell \in [r],
\eean
and the symbol $C(w_y, y, \uw(x \rightarrow w_y))$ corresponds to the plane $\uw(x \rightarrow w_y)$ where $(x,y) \in E_{2,  \uw}$. Therefore from Lemma~\ref{lem:decplane_prop} it follows that $\uw(x \rightarrow w_y) \in Q(E, \uw)$ and from the definition of equivalence class of planes in Definition~\ref{defn:decplanes} it is clear that $Q(E, \uw) = Q(E, \uz)$ implying $\uw(x \rightarrow w_y) \in Q(E, \uz)$. 
%
This would mean that the parity checks corresponding to planes in $Q(E, \uz)$ involve erased symbols corresponding to planes in $Q(E, \uz)$ alone and therefore invertibility of sub matrix $H_{E, {\tiny \uz}}$ would imply recoverability of erased symbols in planes $Q(E, \uz)$.
\end{enumerate}
\eprf

\bthm[The Reduction II: MDS property]\label{thm:reduction2} Let $E$ be an erasure pattern of size $r$ and let $\uz$ be a plane. For the case when $|E_{2,\uz}| = 0$, $H_{E, {\tiny \uz}}$ is invertible. Otherwise, $H_{E, {\tiny \uz}}$ is invertible if $H^{\text{\tiny Red}}_{E, {\tiny \uz}}$ is invertible.
\ethm
\bprf
Let $p = |Q(E, \uz)|$ and $\uf$ be a vector in $\mathbb{F}_q^{rp}$ such that $\uf^T H_{E,{\tiny \uz}} = 0$ and $\uf^T = (f_{\ell,\uv} \mid \ell \in [r], \uv \in Q(E, \uz))$. Let $f_{\uv}$ be a polynomial defined as:
\bean
f_{\uv}(x) = \sum\limits_{\ell=1}^r f_{\ell,\uv}x^{\ell-1}.
\eean
 Given $\uf^T H_{E,{\tiny \uz}} = \underline{0}$ we want to show that $\uf = \underline{0}$ to prove the invertibility of $H_{E,{\tiny \uz}}$. $\uf^T H_{E,{\tiny \uz}} = \underline{0}$ implies that:
 \bea
 \label{eq:red2eq1} \sum\limits_{\uv \in Q(E, \uz)} \sum\limits_{\ell =1}^r f_{\ell,\uv} H([\ell,\uv], [x,y,\uu]) &=& 0  \text{ for any } (x,y) \in E \text{ and } \uu \in Q(E, \uz).
 \eea
By definition of \smalld\ construction, the assignment of $H([\ell,\uv], [x,y,\uu])$ is given by:
\bean
H([\ell,\uv], [x,y,\uu]) &=& \begin{cases}
	\theta_{x,y,u_y}^{\ell-1} & \uv = \uu\\
	\gamma_{ u_y, x} \theta_{x, y, u_y}^{\ell-1} & \uv = \uu( x \rightarrow u_y), x \ne u_y\\
	0 & \text{otherwise}.
\end{cases}
\eean
$H([\ell,\uv], [x,y,\uu])$ is non-zero only when $\uv =\uu$ and $\uv = \uu(x \rightarrow u_y)$. For any $y$ such that $(z_y,y) \in E_{0,\uz}$, and for any $\uu \in Q(E, \uz)$ it is implied that $(u_y, y) \in E$ by the definition of equivalence class of planes in Definition~\ref{defn:decplanes}. By considering $[x,y,\uu]=[u_y, y, \uu]$, equation~\eqref{eq:red2eq1} reduces to:
\bea
\label{eq:red2eq2}\sum\limits_{\ell = 1}^r f_{\ell, \uu} \theta_{u_y,y,u_y}^{\ell-1} = 0 \implies f_{\uu}(\theta_{u_y,y,u_y}) = 0 \overset{\text{from equation } \eqref{eq:thetaassign}}\Rightarrow f_{\uu}(\lambda_{0,y}) = 0 \text{ for all } \uu \in Q(E, \uz), (z_y,y) \in E_{0,\uz}.
\eea
For $(x,y) \in E_{1,\uz}$ it implies that $(z_y, y) \notin E$ and therefore $S_y = \{z_y\}$ (in definition shown in equation \eqref{eq:qezcomp}) and for any $\uu \in Q(E, \uz)$, $u_y = z_y$ and $\uu(x \rightarrow u_y) \notin Q(E, \uz)$. Equation~\eqref{eq:red2eq1} in this case reduces to:
\bea
\label{eq:red2eq3}\sum\limits_{\ell =1}^r f_{\ell, \uu} \theta_{x,y,u_y}^{\ell-1} &=& 0 \implies f_{\uu}(\theta_{x,y,u_y}) = f_{\uu}(\theta_{x,y,z_y}) = 0 \text{ for all } \uu \in Q(E, \uz), (x,y) \in E_{1,\uz}.
\eea
For any $\uu \in Q(E, \uz)$ and $(x,y)\in E_{2,\uu}$, it is implied that $(u_y, y) \in E$, $ x \ne u_y$ and $\uu(x \rightarrow u_y) \in Q(E, \uz)$, therefore  equation~\eqref{eq:red2eq1} in this case reduces to:
\bea
\label{eq:red2eq4}\sum\limits_{\ell =1}^r \left( f_{\ell,\uu} \theta_{x,y,u_y}^{\ell-1} + \gamma_{ u_y, x} f_{\ell,\uu(x \rightarrow u_y)} \theta_{x, y, u_y}^{\ell-1} \right)= 0 \implies f_{\uu}(\theta_{x,y,u_y})+\gamma_{u_y, x}f_{\uu(x \rightarrow u_y)}(\theta_{x, y, u_y}) = 0,
\eea
for all $\uu \in Q(E, \uz)$ and $(x,y) \in E_{2,\uu}$. 
For the case when $|E_{2,\uz}| = 0$, equations~\eqref{eq:red2eq2} and \eqref{eq:red2eq3} imply that there are $|E_{0,\uz}| + |E_{1,\uz}|=r$ roots for $f_{\uu}(x)$ for any $\uu \in Q(E, \uz)$ given by:
\bean
\{ \theta_{x, y, u_y} \mid (x, y) \in E \}.
\eean
By Lemma~\ref{lem:thetadistinct}, all these $r$ roots are distinct. But $f_{\uu}(x)$ is an $(r-1)$ degree polynomial implying that $f_{\uu}(x) = 0$ for all $\uu \in Q(E, \uz)$. This also implies that $\uf = 0$ and hence $H_{E, {\tiny \uz}}$ is invertible.\\ \ \\
For the case when $\mu = |E_{2,\uz}| >0$, from equations~\eqref{eq:red2eq2} and \eqref{eq:red2eq3} it is implied that:
\bea
\label{eq:red2eq5}f_{\uu}(x) = \left(\prod\limits_{(z_y,y) \in E_{0,{\tiny \uz}}} (x -\lambda_{0,y}) \right) \left(\prod\limits_{(\hat{x},y) \in E_{1,{\tiny \uz}}} (x - \theta_{\hat{x},y,z_y})\right) f^{\text{\tiny Red}}_{\uu}(x),
\eea
where $f^{\text{\tiny Red}}_{\uu}(x)$ is a polynomial of degree $\mu-1$.\\
By substituting equation~\eqref{eq:red2eq5} in \eqref{eq:red2eq4} we get that for any  $\uu \in Q(E, \uz), (x,y) \in E_{2, \uu}$:
\bea
\label{eq:red2eq60} \underbrace{\left(\prod\limits_{(z_{\hat{y}},\hat{y}) \in E_{0, {\tiny \uz}}} (\theta_{x,y,u_y} -\lambda_{0,\hat{y}}) \right)}_{P_1} \underbrace{\left(\prod\limits_{(\hat{x},\hat{y}) \in E_{1,{\tiny \uz}}} (\theta_{x,y,u_y} - \theta_{\hat{x},\hat{y},z_{\hat{y}}})\right)}_{P_2} \left( f^{\text{\tiny Red}}_{\uu}(\theta_{x,y,u_y})+\gamma_{u_y, x}f^{\text{\tiny Red}}_{\uu(x \rightarrow u_y)}(\theta_{x, y, u_y}) \right) = 0.
\eea
The term $P_1$ is clearly non zero as $(x, y) \in E_{2, \uu}$ it is implied that $x \ne u_y$ therefore by the assignment in equation \eqref{eq:thetaassign}, $\theta_{x, y, u_y} \ne \lambda_{0, \hat{y}}$ for any $\hat{y} \in [0: t-1]$ as $\theta_{x, y, u_y}$ corresponds to a non-diagonal element of $\Lambda_{s, y}$. We will now look at term $P_2$. By the definition~\ref{defn:eraspat} of erasure partitions if $(\hat{x}, \hat{y}) \in E_{1, \uz}$ it is implied that $(z_{\hat{y}}, \hat{y}) \notin E$. It follows from equation \eqref{eq:qezcomp} that $S_{\hat{y}} = \{z_{\hat{y}}\}$ and therefore for any $\uu \in Q(E, \uz)$, $u_{\hat{y}} = z_{\hat{y}}$. From Lemma~\ref{lem:decplane_prop} $E_{1, \uz} = E_{1, \uu}$, hence the term $P_2$ can be written as:
\bean
P_2 = \left(\prod\limits_{(\hat{x},\hat{y}) \in E_{1,{\tiny \uu}}} (\theta_{x,y,u_y} - \theta_{\hat{x},\hat{y},u_{\hat{y}}})\right).
\eean
From Lemma~\ref{lem:thetadistinct} it is clear that $P_2 \ne 0$. Therefore it follows from equation~\eqref{eq:red2eq60} that for any $\uu \in Q(E, \uz)$, $(x, y) \in E_{2, \uu}$:
\bea
\label{eq:red2eq6}f^{\text{\tiny Red}}_{\uu}(\theta_{x,y,u_y})+\gamma_{u_y, x}f^{\text{\tiny Red}}_{\uu(x \rightarrow u_y)}(\theta_{x, y, u_y}) = 0.
\eea
Let $f^{\text{\tiny Red}}_{\uu}(x) = \sum\limits_{\ell=1}^{\mu} f^{\text{\tiny Red}}_{\ell,\uu}x^{\ell-1}$ and let $\uf^{\text{\tiny Red}}$ be a vector in $\mathbb{F}_q^{\mu p}$ and $\uf^{\text{\tiny Red}} = (f_{\ell,\uv}^{\text{\tiny Red}} \mid \ell \in [\mu], \uv \in Q(E, \uz))^T$. Equation~\eqref{eq:red2eq6} can be rewritten as:
\bean
\sum\limits_{\ell=1}^{\mu} f^{\text{\tiny Red}}_{\ell,\uu} H([\ell, \uu], [x,y,\uu]) + f^{\text{\tiny Red}}_{\ell, \uu(x \rightarrow u_y)} H([\ell, \uu(x \rightarrow u_y)], [x,y,\uu]) &=& 0 \text{ for all } \uu \in Q(E, \uz), \ (x,y) \in E_{2, \uu}\\
\sum\limits_{\ell=1}^{\mu} \sum\limits_{\uv \in Q(E, \uz)} f^{\text{\tiny Red}}_{\ell,\uv} H([\ell, \uv], [x,y,\uu]) &=& 0 \text{ for all } \uu \in Q(E, \uz), \ (x,y) \in E_{2, \uu} \\
{\uf^{\text{\tiny Red}}}^T H^{\text{\tiny Red}}_{E, {\tiny \uz}} &=& \underline{0} \text{ from the definition in equation~\eqref{eq:Hsubdefn2}}.
\eean
If $H^{\text{\tiny Red}}_{E, {\tiny \uz}}$ is invertible, this would imply that $\uf^{\text{\tiny Red}} = \underline{0}$. From equation~\eqref{eq:red2eq5}, if $\uf^{\text{\tiny Red}} = \underline{0}$, it follows that $\uf=\underline{0}$ implying $H_{E, {\tiny \uz}}$ is invertible.
\eprf

The Theorems~\ref{thm:reduction1} and \ref{thm:reduction2} together imply that it is enough to show invertibility of reduced matrix $H_{E, {\tiny \uz}}^{\tiny \text{Red}}$ to prove the MDS property. We prove that reduced matrix is invertible in Section~\ref{sec:inv_red}.

\section{Optimal Access Repair Property: The Reductions \label{sec:oar_redn}}
Recall that during a single node repair, $d$ helper nodes are contacted among the remaining $n-1$ nodes. Therefore $(n-1-d) = r-s$ nodes remain aloof in the repair process. To prove the optimal-access property we will first introduce a sub-matrix of the p-c matrix $H$ called Repair Sub-Matrix.  For any failed node $(x_0, y_0)$, aloof node set $E \subseteq (\Zs \times \Zt) \setminus \{(x_0, y_0)\}$ such that $|E|=r-s$ and repair plane $\uz \in R$, where $R = \{ \uz \in \Zst \mid z_{y_0} = x_0 \}$ we define the sub-matrix $H_{E,(x_0, y_0), {\tiny \uz}}$ in Definition~\ref{defn:Hsubdefn_rep}. We will later show in Theorem~\ref{thm:reduction1_rep} that the invertibility of this sub-matrix for any aloof node set $E$, repair plane $\uz$ would imply the optimal-access property.

We first show in the following lemma that for any repair plane $\uz \in R$, the planes in equivalence class of $\uz$ are in indeed repair planes. 
\blem\label{lem:qedrepair}
Let $(x_0, y_0)$ be a failed node and let $E$ be the set of $(r-s)$ aloof nodes such that $(x_0, y_0) \notin E$. Let $\uz \in R$ where $R = \{\uz \in \Zst \mid z_{y_0} = x_0\}$, then $Q(E, \uz) \subseteq R$.
\elem
\bprf
By the Definition~\ref{defn:decplanes}, the equivalence class of $\uz$ $Q(E,\uz)= S_0 \times S_1 \times \cdots \times S_{t-1}$ where $S_y$ is defined as shown in equation~\eqref{eq:qezcomp}. It is clear to see that $S_{y_0} = \{x_0\}$ as $(z_{y_0}, y_0) = (x_0, y_0) \notin E$. Therefore for any $\uu \in Q(E, \uz)$, $u_{y_0} = x_0$ i.e., $\uu \in R$.
\eprf

We now define the repair sub-matrix $H_{E, (x_0, y_0), {\tiny \uz}}$ by looking at p-c equations corresponding to the planes in equivalence class of $\uz$, $Q(E,\uz)$ and the failed node, aloof node symbols that participate in those equations.

\bdefn[Repair Sub-Matrix]\label{defn:Hsubdefn_rep}
Given a node $(x_0, y_0)$, an aloof node pattern $E$, such that $(x_0, y_0) \notin E$,  $|E| = r-s$, and plane $\uz \in R$, where $R = \{ \uz \in \Zst \mid z_{y_0} = x_0 \}$, $H_{E, (x_0, y_0), {\tiny \uz}}$ is defined as an $(r p \times r p)$ sub-matrix of the parity check matrix $H$ where $p = |Q(E, \uz)|$.\\
We can index the rows of the matrix by $[\ell, \uv]$ where $\ell \in [r]$ and $\uv \in Q(E, \uz)$ and columns by $[x,y,\uu]$ where,
\bean 
[x,y, \uu] \in \underbrace{(E \times Q(E, \uz))}_{(r-s)p \text{ aloof node symbols}} \cup \underbrace{\{(x_0, y_0, \uw( \hat{x} \rightarrow w_{y_0})) \mid  \hat{x} \in \Zs, \uw \in Q(E, \uz) \}}_{sp \text{ failed node symbols}},
\eean
\bea
\label{eq:Hsubdefn_rep}H_{E, (x_0, y_0), {\tiny \uz}}([\ell,\uv], [x,y,\uu]) = H([\ell, \uv]), [x,y,\uu]).
\eea
Columns of this matrix correspond to the $(r-s)p$ aloof node symbols within the planes $Q(E, \uz)$ and $sp$ failed node symbols that are not limited to planes in $Q(E, \uz)$.
\edefn

Using the repair sub-matrix definition we will show that its invertibility implies the optimal-access repair property in Theorem~\ref{thm:reduction1_rep}. 

\bthm[The Reduction I: Optimal-Access Property]\label{thm:reduction1_rep}
\smalld\ construction satisfies the optimal-access repair property, if for any node $(x_0, y_0)$, aloof node pattern $E \subseteq (\Zs \times \Zt) \setminus \{(x_0, y_0)\}$ such that $|E|=(r-s)$, and for any plane $\uz \in R$, where $R = \{\uz \in \Zst \mid z_{y_0} = x_0\}$ the repair sub-matrix $H_{E, (x_0, y_0), {\tiny \uz}}$ is invertible, where $H_{E, (x_0, y_0), {\tiny \uz}}$ is defined as shown in Definition~\ref{defn:Hsubdefn_rep}.
\ethm
\bprf
To show that \smalld\ construction satisfies the optimal-access property, we will show that it can recover any node $(x_0, y_0)$ with the help of $\beta = \frac{\alpha}{s} = s^{t-1}$ symbols from any $d$ helper nodes. Let $E$ denote the set of aloof nodes that do not participate in repair. Therefore, $|E| = (n-1-d) = r-s$.
The helper information sent by a node $(x,y) \notin E \cup \{(x_0, y_0)\}$ is given by:
\bean
\{ \cxyz \mid \uz \in R \}, R = \{ \uz \in \Zst | z_{y_0} = x_0 \}.
\eean
Given an aloof node pattern $E$ we recover the failed node symbols sequentially by first ordering the repair planes, $R$ by the intersection scores and then recovering failed node symbols and aloof node symbols within the repair plane. In this method, the repair planes belong to the same equivalence class are repaired together.

\begin{enumerate}[wide, labelwidth=!, labelindent=0pt]
	\item[Step 0:]  $\uz \in R$ such that $\sigma(E, \uz) = 0$, then $Q(E, \uz) = \{\uz\}$ and the $[\ell, \uz]$-th p-c equation~\ref{eq:pc_msr_2} reduces to:
	\bean
	\theta_{x_0,y_0,x_0}^{\ell-1} C(x_0,y_0;\uz) + \sum\limits_{(x,y) \in E} \theta_{x,y,z_y}^{\ell-1} \cxyz + \sum\limits_{x \ne x_0} \gamma_{x,x_0} \theta_{x_0, y_0, x}^{\ell-1} C(x_0, y_0; \uz(x \rightarrow z_{y_0})) &=& \kappa^* \text{ for all } \ell \in [r].
	\eean
	This is because the only unknown in-plane symbols are the $(r-s+1)$ symbols $\{ \cxyz \mid (x,y) \in E \} \cup \{C(x_0, y_0, \uz)\}$ and the unknown out-of-plane symbols are the $(s-1)$-symbols $\{ C(x_0,y_0,\uz(x \rightarrow z_{y_0})) \mid x \in \Zs \setminus \{x_0\}\}$. The remaining $(s-1)(t-1)$ out-of-plane symbols $\{ \cxyzc \mid y \ne \Zt \setminus \{y_0\}, x \in \Zs \setminus \{z_y\} \}$ belong to a helper node as $(z_y, y) \notin E$ and are part of repair planes as $\uz(x \rightarrow z_y) \in R$.
	
	Therefore, there are $|E|+s = r$ unknowns in the above equations corresponding to $s$ failed node symbols and $(r-s)$ aloof node symbols. Clearly, for this plane $\uz$, $H_{E, (x_0,y_0), {\tiny \uz}}$ is an $r \times r$ matrix and if it is invertible then we can recover the $s$ failed node symbols $\{\cxyzcnot \mid x \in [0:s-1]\}$ and $(r-s)$ aloof node symbols $\{\cxyz \mid (x, y) \in E\}$.
	\item[Step $j$:] Let $\uz \in R$ such that $\sigma(E, \uz) = j$ and let us assume by induction that the aloof node symbols corresponding to repair planes with intersection score $<j$ are already recovered, then the $[\ell, \uz]$-th p-c equation \ref{eq:pc_msr_2} reduces to:
	\bean
	\theta_{x_0,y_0,x_0}^{\ell-1} C(x_0,y_0;\uz) +  \sum\limits_{x \ne x_0} \gamma_{x,x_0} \theta_{x_0, y_0, x}^{\ell-1} C(x_0, y_0; \uz( x \rightarrow z_{y_0})) + \sum\limits_{(x,y) \in E} \theta_{x,y,z_y}^{\ell-1} \cxyz + \\
   \sum\limits_{y: (z_y, y) \in E}\sum\limits_{x=0}^{s-1} \gamma_{x,z_y}\theta_{z_y,y,x}^{\ell-1} \cxyzc &=& \kappa^* \text{ for all } \ell \in [r].
	\eean
	Let $(z_y, y) \in E$, then $\sigma(E, \uz(x \rightarrow z_y)) = j$ iff $(x,y) \in E$. This implies that the symbols $\cxyzc$ are recovered in the $(j-1)$ step if $(x, y) \notin E$, whereas they are unknown if $(x, y) \in E$. Therefore the $[\ell, \uz]$-th p-c equation further simplifies to:
	\bean
	\theta_{x_0,y_0,x_0}^{\ell-1} C(x_0,y_0;\uz) +  \sum\limits_{x \ne x_0} \gamma_{x,x_0} \theta_{x_0, y_0, x}^{\ell-1} C(x_0, y_0; \uz( x \rightarrow z_{y_0})) + \sum\limits_{(x,y) \in E} \theta_{x,y,z_y}^{\ell-1} \cxyz +\\
    \sum\limits_{ (x, y) \in E_{2,{\tiny \uz}}} \gamma_{x,z_y}\theta_{z_y,y,x}^{\ell-1} \cxyzc &=& \kappa^* \text{ for all } \ell \in [r],
	\eean
	by the Definition~\ref{defn:eraspat} of $E_{2,\uz}$. Clearly, when $|E_{2,\uz}| = 0$ there are $r$ equations and $r$ unknowns in the above equation. From Lemma~\ref{lem:decplane_prop}, $Q(E,\uz) = \{\uz\}$ when $|E_{2,\uz}| = 0$ and $H_{E, (x_0, y_0), {\tiny \uz}}$ is the $r \times r$ matrix corresponding to the $r$ p-c equations indexed by $\{[\ell, \uz] \mid \ell \in [r] \}$ and the $r$ unknowns participating. Therefore its invertibility implies recoverability of $s$ failed node symbols $\{C(x_0, y_0, \uz(x \rightarrow z_{y_0})) \mid x \in \Zs \}$ and $(r-s)$ aloof node symbols $\{\cxyz \mid (x,y) \in E\}$.
	
When $|E_{2,\uz}| >0$, the number of unknowns $r + |E_{2,\uz}|$ is greater than the number of equations $r$, therefore we need to use more parity checks in order to recover aloof node symbols corresponding to the plane $\uz$. 
It turns out that in this case, if we consider the p-c equations corresponding to planes in equivalence class of $\uz$ then we have a situation where the number of unknowns $rp$ equals the number of equations $rp$ where $p = |Q(E,\uz)|$.  The p-c matrix associated to this set of equations is precisely $H_{E,(x_0, y_0),{\tiny \uz}}$. We will now go ahead and show that the unknowns in the p-c equations indexed by $\{[\ell, \uw] \mid \uw \in Q(E,\uz), \ell \in [r]\}$ correspond to the $(r-s)p$ aloof symbols in planes $Q(E, \uz)$ and $sp$ failed node symbols given by $\{C(x_0, y_0; \uw(x \rightarrow z_{y_0})) \mid x \in [0:s-1], \uw \in Q(E, \uz)\}$. This will imply that the number of unknowns and number of equations is $r p$.

We therefore consider all the equations corresponding to the planes in $Q(E, \uz)$. From Lemma~\ref{lem:qedrepair}, $Q(E, \uz) \subseteq R$. 

For any $\uw \in Q(E, \uz)$, the $[\ell, \uw]$-th p-c equation is given by:
\bean
\theta_{x_0,y_0,x_0}^{\ell-1} C(x_0,y_0;\uw) +  \sum\limits_{x \ne x_0} \gamma_{x,x_0} \theta_{x_0, y_0, x}^{\ell-1} C(x_0, y_0; \uw(x \rightarrow w_{y_0})) + \sum\limits_{(x,y) \in E} \theta_{x,y, w_y}^{\ell-1} C(x,y; \uw) +\\
 + \sum\limits_{ (x, y) \in E_{2,{\tiny \uw}}} \gamma_{x,w_y}\theta_{w_y,y,x}^{\ell-1} C(w_y,y;\uw(x \rightarrow w_y)) = \kappa^* \text{ for all } \ell \in [r].
\eean
and the symbol $C(w_y, y; \uw(x \rightarrow w_y))$ corresponds to the plane $\uw(x \rightarrow w_y)$. For $(x, y) \in E_{2, \uw}$ it is clear from the Definition~\ref{defn:decplanes} of the equivalence class of planes that $\uw(x \rightarrow w_y) \in Q(E, \uw)$. It is also known that $Q(E, \uw) = Q(E, \uz)$. Therefore $\uw(x \rightarrow w_y) \in Q(E, \uz)$.
This would mean that the parity checks corresponding to planes in $Q(E, \uz)$ involve $(r-s)p$ aloof node symbols corresponding to planes within $Q(E, \uz)$ and $sp$ failed node symbols in planes $\{\uw( x \rightarrow w_{y_0}) \mid x \in \Zs, \uw \in Q(E, \uz) \}$, therefore invertibility of sub matrix $H_{E, (x_0,y_0), {\tiny \uz}}$ would imply recoverability of aloof node symbols in planes $Q(E, \uz)$ and failed node $(x_0,y_0)$'s symbols in planes  $\{\uw( x \rightarrow w_{y_0}) \mid x \in \Zs, \uw \in Q(E, \uz) \}$.
\end{enumerate}
Therefore, by the end of all the steps invertibility of $H_{E,(x_0,y_0),\uz}$ for all $\uz \in R$ implies recovery of all the $\alpha$ failed node symbols:
\bean
\{ C(x_0, y_0, \uz(x \rightarrow z_{y_0})) \mid \uz \in R \} = \{ C(x_0, y_0, \uz) \mid \uz \in \Zst \}
\eean
and recovery of $\beta (r-s)$ aloof node symbols $\{\cxyz \mid (x,y) \in E, \uz \in R\}$.
\eprf

Now, in Theorem~\ref{thm:reduction2_rep} we show that invertibility of reduced matrix introduced in Definition~\ref{defn:Hsubred} implies invertibility of the repair sub-matrix.

\blem[The Reduction II: Optimal Access Property]\label{thm:reduction2_rep}
Let $(x_0, y_0)$ be the failed node and $E$ be an aloof node pattern of size $(r-s)$ such that $(x_0, y_0) \notin E$ and let $\uz \in R$ where $R = \{\uu \in \Zst \mid u_{y_0} = x_0\}$. For the case when $|E_{2,\uz}| = 0$, $H_{E, (x_0, y_0),{\tiny \uz}}$ is invertible. Otherwise, $H_{E, (x_0, y_0), {\tiny \uz}}$ is invertible if $H_{E, {\tiny \uz}}^{\text{\tiny Red}}$ (see Definition~\ref{defn:Hsubred}) is invertible.
\elem
\bprf
Let $p = |Q(E, \uz)|$ and $\uf$ be a vector in $\mathbb{F}_q^{r p}$ such that $\uf^T H_{E, (x_0, y_0), {\tiny \uz}} = 0$ and $\uf^T = (f_{\ell, \uv} \mid \ell \in [r], \uv \in Q(E, \uz))$. Let $f_{\uv}$ be a polynomial defined as:
\bean
f_{\uv}(x) = \sum\limits_{\ell=1}^r f_{\ell,\uv}x^{\ell-1}.
\eean
Given $\uf^T H_{E,(x_0, y_0),{\tiny \uz}} = \underline{0}$ we want to show that $\uf = \underline{0}$. $\uf^T H_{E,(x_0,y_0),{\tiny \uz}} = \underline{0}$ implies that:
\bea
\label{eq:red2eq1_rep}\sum\limits_{\ell=1}^r \sum\limits_{\uv \in Q(E, \uz)} f_{\ell,\uv} H([\ell,\uv], [x,y,\uu]) = 0,  \ \ \ \ \ \ \ \ \ \  \ \ \ \ \ \ \ \ \ \ \\
\nonumber \text{for any } [x,y, \uu] \in (E \times Q(E, \uz)) \cup \{(x_0, y_0, \uw(\hat{x} \rightarrow w_{y_0})) \mid \hat{x} \in \Zs,  \uw \in Q(E, \uz) \}.
\eea

By definition of \smalld\ construction, $H([\ell,\uv], [x,y,\uu])$ is non-zero only for $\uv = \uu, \uu(x \rightarrow u_y)$. 
For any $y$ such that $(z_y,y) \in E_{0,\uz}$, and for any $\uu \in Q(E, \uz)$ it is implied that $(u_y, y) \in E$, by considering $[x,y,\uu]=[u_y, y, \uu]$, equation~\eqref{eq:red2eq1_rep} reduces to:
\bea
\label{eq:red2eq2_rep}\sum\limits_{\ell = 1}^r f_{\ell, \uu} \theta_{u_y,y,u_y}^{\ell-1} = 0 \implies f_{\uu}(\theta_{u_y,y,u_y}) = 0 \overset{\text{from equation } \eqref{eq:thetaassign}} \implies f_{\uu}(\lambda_{0,y}) = 0 \text{ for all } \uu \in Q(E, \uz), (z_y,y) \in E_{0,\uz}
\eea
For $(x,y) \in E_{1,\uz}$ it implies that $(z_y, y) \notin E$ and that $x \ne z_y$, therefore $S_y = \{z_y\}$ (see definition in equation \eqref{eq:qezcomp}) and given $\uu \in Q(E, \uz)$, $\uu(x \rightarrow u_y) \notin Q(E, \uz)$ and $u_y = z_y$. Equation~\eqref{eq:red2eq1_rep} in this case reduces to:
\bea
\label{eq:red2eq3_rep}\sum\limits_{\ell = 1}^r f_{\ell,\uu} \theta_{x,y,u_y}^{\ell-1} &=& 0 \implies f_{\uu}(\theta_{x,y,u_y}) = f_{\uu}(\theta_{x,y,z_y}) = 0 \text{ for all } \uu \in Q(E, \uz), (x,y) \in E_{1,\uz} = E_{1, \uu}.
\eea
For any $\uu \in Q(E, \uz)$ and $(x,y)\in E_{2,\uu}$, it is implied that $(u_y, y) \in E$, $ x \ne u_y$ and $\uu(x \rightarrow u_y) \in Q(E, \uz)$, therefore:
\bea
\label{eq:red2eq5_rep}\sum\limits_{\ell = 1}^r \left( f_{\ell,\uu} \theta_{x,y,u_y}^{\ell-1} + \gamma_{ u_y, x} f_{\ell,\uu(x \rightarrow u_y)} \theta_{x, y, u_y}^{\ell-1} \right)= 0 \implies f_{\uu}(\theta_{x,y,u_y})+\gamma_{u_y, x}f_{\uu(x \rightarrow u_y)}(\theta_{x, y, u_y}) = 0 .
\eea
For $(x,y, \uu) = (x_0, y_0, \uw(\hat{x} \rightarrow w_{y_0}))$ and $\uw \in Q(E, \uz)$, $\hat{x} \in \Zs$, the $\uv$ where $H([\ell,\uv],[x,y,\uu])$ is non-zero is $\uv = \uw, \uw(\hat{x} \rightarrow w_{y_0})$. However for $\hat{x} \ne w_{y_0} = x_0$,  $\uw(\hat{x} \rightarrow w_{y_0}) \notin R$. From Lemma~\ref{lem:qedrepair}, $Q(E, \uz) \subseteq R$. Therefore $\uw(\hat{x} \rightarrow w_{y_0}) \notin Q(E, \uz)$ and the only $\uv \in Q(E, \uz)$ for which $H([\ell,\uv],[x,y,\uu])$ is non-zero is $\uv = \uw$. Equation~\eqref{eq:red2eq1_rep} reduces to:
\bea
\label{eq:red2eq4_rep}\sum\limits_{\ell=1}^r f_{\ell,\uw} \theta_{x_0,y_0,\hat{x}}^{\ell-1} &=& 0 \implies f_{\uw}(\theta_{x_0,y_0,\hat{x}}) = 0 \text{ for all } \uw \in Q(E, \uz), \hat{x} \in \Zs.
\eea

For the case when $\mu = |E_{2,\uz}| = 0$, \eqref{eq:red2eq2_rep}, \eqref{eq:red2eq3_rep} and \eqref{eq:red2eq4_rep} imply that there are $|E_{0,\uz}|+|E_{1,\uz}|+s=|E|+s=r$ roots for $f_{\uu}(x)$ for any $\uu \in Q(E, \uz)$ given by:
\bean
\{ \theta_{x, y, z_y} \mid (x, y) \in E \} \cup \{\theta_{x_0, y_0, x} \mid x \in [0:s-1]\}.
\eean 
From  Lemma~\ref{lem:thetadistinct} that there are $r$ distinct roots for $f_{\uu}(x)$. 
But $f_{\uu}(x)$ is an $r-1$ degree polynomial implying that $f_{\uu}(x) = 0$ for all $\uu \in Q(E, \uz)$. This also implies that $\uf = 0$ and hence $H_{E, (x_0, y_0), {\tiny \uz}}$ is invertible.\\ \ \\
For the case when $\mu = |E_{2,\uz}| >0$, from \eqref{eq:red2eq2_rep}, \eqref{eq:red2eq3_rep} and \eqref{eq:red2eq4_rep} it is implied that:
\bea
\label{eq:red2eq6_rep}f_{\uu}(x) = \left(\prod\limits_{\hat{x}=0}^{s-1} (x -\theta_{x_0,y_0, \hat{x}}) \right) \left(\prod\limits_{(z_y,y) \in E_{0,{\tiny \uz}}} (x -\lambda_{0,y}) \right) \left(\prod\limits_{(\hat{x},y) \in E_{1,{\tiny \uz}}} (x - \theta_{\hat{x},y,u_y})\right) f^{\text{\tiny Red}}_{\uu}(x),
\eea
where $f^{\text{\tiny Red}}_{\uu}(x)$ is a polynomial of degree $\mu-1$.\\
By substituting \eqref{eq:red2eq6_rep} in \eqref{eq:red2eq5_rep} we get that for any  $\uu \in Q(E, \uz), (x,y) \in E_{2, \uu}$:
\bean
P_1 P_2 P_3  \left( f_{\uu}^{\text{\tiny Red}}(\theta_{x,y,u_y})+\gamma_{u_y, x} f_{\uu(x \rightarrow u_y)}^{\text{\tiny Red}}(\theta_{x, y, u_y}) \right) = 0,
\eean
\bean
\text{ where } P_1 =  \left(\prod\limits_{\hat{x}=0}^{s-1} (\theta_{x, y, u_y} -\theta_{x_0,y_0, \hat{x}}) \right), P_2 = \left(\prod\limits_{(z_{\hat{y}},\hat{y}) \in E_{0,{\tiny \uz}}} (\theta_{x, y, u_y} -\lambda_{0,\hat{y}}) \right), P_3 = \left(\prod\limits_{(\hat{x},\hat{y}) \in E_{1,{\tiny \uz}}} (\theta_{x,y,u_y} - \theta_{\hat{x},\hat{y},u_{\hat{y}}})\right).
\eean
It follows from Lemma~\ref{lem:thetadistinct} that $P_1, P_2, P_3$ are non zero and hence equation~\eqref{eq:red2eq6_rep} for any $\uu \in Q(E, \uz)$ and $(x, y) \in E_{2, \uu}$, reduces to:
\bea
\label{eq:red2eq7_rep}f^{\text{\tiny Red}}_{\uu}(\theta_{x,y,u_y})+\gamma_{u_y, x}f^{\text{\tiny Red}}_{\uu(x \rightarrow u_y)}(\theta_{x, y, u_y}) = 0.
\eea
Let $f_{\uu}^{\text{\tiny Red}}(x) = \sum\limits_{\ell=1}^{\mu} f_{\ell,\uu}^{\text{\tiny Red}}x^{\ell-1}$ and let $\uf^{\text{\tiny Red}}$ be a vector in $\mathbb{F}_q^{\mu p}$ such that $\uf^{\text{\tiny Red}} = (f_{\ell,\uv}^{\text{\tiny Red}} \mid \ell \in [\mu], \uv \in Q(E, \uz))^T$. \eqref{eq:red2eq7_rep} can be rewritten as:
\bean
\sum\limits_{\ell=1}^{\mu} f_{\ell,\uu}^{\text{\tiny Red}} H([\ell, \uu], [x,y,\uu]) + f_{\ell, \uu(x \rightarrow u_y)}^{\text{\tiny Red}} H([\ell, \uu(x \rightarrow u_y)], [x,y,\uu]) &=& 0 \text{ for all } \uu \in Q(E, \uz), \ (x,y) \in E_{2, \uu}\\
\sum\limits_{\ell=1}^{\mu} \sum\limits_{\uv \in Q(E, \uz)} f_{\ell,\uv}^{\text{\tiny Red}} H([\ell, \uv], [x,y,\uu]) &=& 0 \text{ for all } \uu \in Q(E, \uz), \ (x,y) \in E_{2, \uu} \\
{\uf^{\text{\tiny Red}}}^T H^{\text{\tiny Red}}_{E, {\tiny \uz}} &=& \underline{0} \text{ from the definition in } \eqref{eq:Hsubdefn2}.
\eean
If $H^{\text{\tiny Red}}_{E, {\tiny \uz}}$ is invertible, this would imply that $\uf^{\tiny \text{Red}} = \underline{0}$. From \eqref{eq:red2eq6_rep}, it is clear that $\uf=\underline{0}$ implying that
$H_{E, (x_0, y_0), {\tiny \uz}}$ is invertible.
\eprf

The Theorems~\ref{thm:reduction1_rep} and \ref{thm:reduction2_rep} together imply that it is enough to show invertibility of reduced matrix $H_{E, {\tiny \uz}}^{\tiny \text{Red}}$ to prove the optimal-access repair property. 

\section{Invertibility of Reduced Matrix \label{sec:inv_red}}
We will now prove that the matrix $H^{\text{\tiny Red}}_{E, {\tiny \uz}}$ is invertible. To do so we will first introduce some notation to prove this inductively. The induction will be over the number of $y$-columns over which there are non-zero erasures appearing in $E_{2, \uz}$.
Throughout the reminder of the paper we continue to work with set $E$ indicating erasure (aloof node) pattern and plane $\uz \in \Zst$ and ignore indicating them in the notations.

\bdefn\label{defn:erasvec}
Given an erasure (aloof node) pattern $E$, plane $\uz \in \Zst$ we define the erasure vector that indicates the number of erasures in $E_{2,\uz}$ with same $y$-value
\bean 
\underline{e} = (e_y \mid y \in \Zt) \text{ where }  e_y = |E_{2,\uz}(y)|, \ E_{2,\uz}(y) = \{ x' \mid (x',y') \in E_{2,\uz}, y'=y \}
\eean
\edefn

\begin{figure}[ht!]
	\bc
	\includegraphics[width=0.5\textwidth]{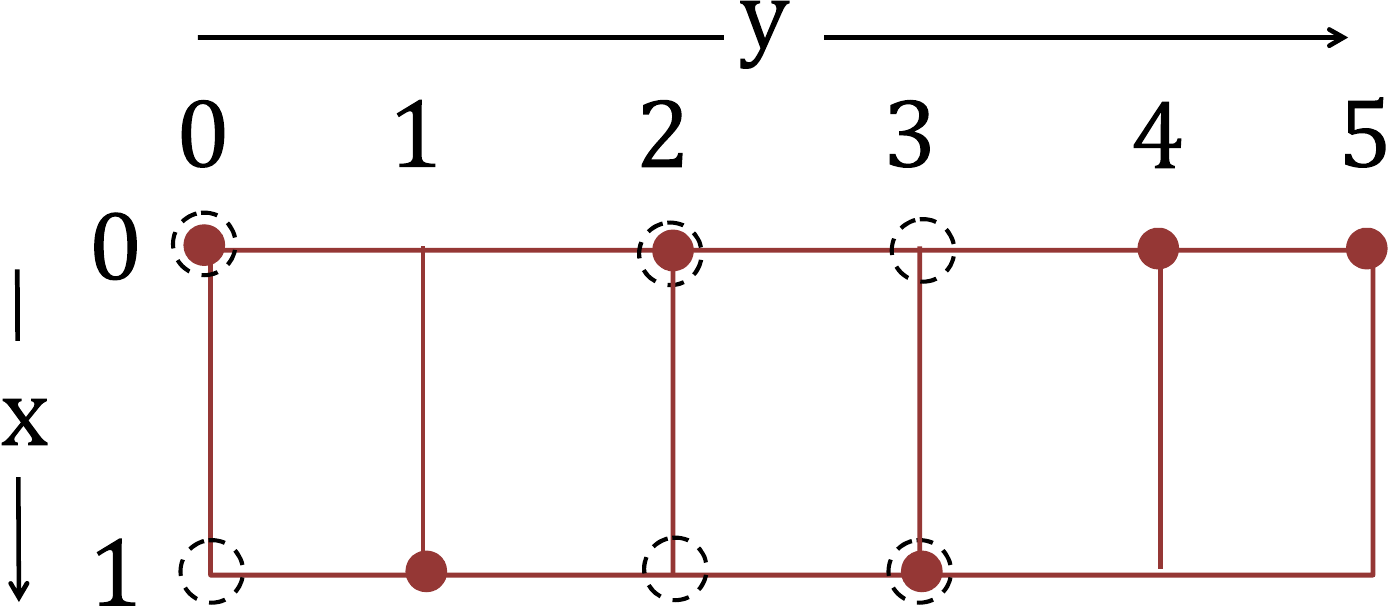}
	\caption{Illustration of an example erasure pattern for $E=\{ (0,0), (1,0), (0,2), (1,2), (0,3), (1, 3) \}$ indicated by dotted circles (holes) in plane $\uz = (0,1,0,1,0,0)$. $E_{2,\uz} = \{(1,0),(1,2),(0,3)\}$. \label{fig:erasvec}}
	\ec
\end{figure}

\bdefn\label{defn:M}
Let $Y = \text{Supp}(\underline{e})$ be the support set of erasure weight vector $\underline{e}$, $|Y| = m$, $Y = \{y_1, \cdots, y_m\}$. We define a subset of planes $Q^j(E, \uz) \subseteq Q(E, \uz)$ for any $j \in [m]$ as 
\bea
\label{eq:planesinM}Q^j(E, \uz) &=& S_0 \times S_1 \times \cdots \times S_{y_j} \times \{z_{y_j+1}\} \times \cdots \times \{z_{t-1}\}
\eea
where $S_y$ for $y \in \Zt$ is defined as shown in equation \eqref{eq:qezcomp}.
\edefn

\blem
The number of planes in $Q^j(E, \uz)$ is given by $p_j = \prod\limits_{i=1}^j (e_{y_i}+1)$ for all $j \in [m]$.
\elem
\bprf
From the Lemma~\ref{lem:decplane_prop}, $S_{y} = E_{2,\uz}(y) \cup \{(z_y,y)\}$. Therefore $|S_y| = (e_y+1)$ and 
\bean
 p_j = |Q^j(E,\uz)| = \prod\limits_{y=0}^{y_j} |S_y| =\prod\limits_{i=1}^j (e_{y_i}+1).
\eean
\eprf

For the example erasure pattern and plane shown in the Fig~\ref{fig:erasvec}, $\underline{e} = (1,0,1,1,0,0)$, $m = |Supp(\underline{e})|=3$ and $S_0=S_2=S_3=\{0,1\}$, $S_1 = \{1\}$, $S_4=S_5=\{0\}$. The number of planes in the sets $Q^1(E,\uz), Q^2(E,\uz)$ and $Q^3(E,\uz)$ are $2, 4$ and $8$ respectively. We can now define the sub-matrix of $H_{E,{\tiny \uz}}^{\tiny \text{Red}}$ that will be used to prove its invertibility by induction.

\bdefn[Induction Matrix]\label{defn:indmatr}
Let $j \in [m]$, $d \le \mu$  we define a matrix $M_{j, d}$ of size $(d p_j \times \mu_j p_j)$ where $\mu_j = \sum\limits_{i=1}^j e_{y_i}$, $p_j = \prod\limits_{i=1}^j (e_{y_i}+1)$ as below:
\bean
M_{j,d}([\ell,\uv],[x,y,\uu]) &=& H([\ell,\uv],[x,y,\uu])
\eean
where $\ell \in [d]$ and $\uu, \uv \in Q^j(E, \uz)$ and $(x,y) \in E_{2, \uu}$ such that $y \in \{y_1, y_2, \cdots, y_j \}$.
\edefn

\begin{note}
	It follows from the Definition~\ref{defn:M} that $H^{\text{\tiny Red}}_{E, {\tiny \uz}} = M_{m, \mu_m}$ and $p_m = p = |Q(E, \uz)| $, $\mu_m= \mu = |E_{2,\uz}|$.
\end{note}

\begin{note}
	$e_{y_i} \le s-1$ for all $i \in [m]$ and therefore in the proofs that follow, we consider cases of $e_{y_i} \in \{1,2,3\}$ as $s \in \{2, 3, 4\}$.
\end{note}

We will show in Lemma~\ref{lem:indMinv_initcond} that $M_{1, \mu_1}$ is invertible and then shown in Lemma~\ref{lem:indMinv} that $M_{i+1, \mu_{i+1}}$ is invertible given $M_{i, \mu_i}$ is invertible for any $i \in [m-1]$. This will imply that $H_{E, {\tiny \uz}}^{\text{\tiny Red}} = M_{m, \mu_m}$ is invertible.

\blem\label{lem:indMinv_initcond}
For any plane $\uz \in \Zst$ and any erasure pattern $E$ such that $|E| \le r$, the matrix $M_{1, \mu_1}$ is invertible.
\elem
\bprf It is clear that $\uz \in Q^1(E, \uz)$ and  the planes in $Q^1(E, \uz)$ are given by $\{ \uz(x \rightarrow z_y) | \  x \in S_{y_1} \}$. As $s \in \{2, 3, 4\}$ there are three possibilities for the $|S_{y_1}| = e_{y_1} + 1 \in \{2, 3, 4\}$. We will look at the invertibility of $M_{1, \mu_i}$ for each of these cases separately. Let 
\bea
\label{eq:vij}V_{i,j}^d = \left[ \begin{array}{cc}
\underline{\gamma \lambda_{i,y_j}}^d & \underline{\lambda_{i,y_j}}^d
\end{array} \right] \text{ be a } (d \times 2)\text{ matrix where } \underline{\theta}^d=\left[\begin{array}{cccc}
1 & \theta & \cdots & \theta^{d-1}
\end{array}\right]^T \text{ and } \Gamma = \left[\begin{array}{cc}
\gamma & 0\\
0 & 1
\end{array}\right].
\eea

\paragraph*{Case 1: $e_{y_1} = 1$, $\mu_1=1$, $p_1 = 2$}  Let $S_{y_1}= \{x_0,x_1\}$ and $x_0 < x_1$, then $M_{1, d}$ is an $(2d \times 2)$ matrix given by:
\bean
M_{1,d} &=& \left[ \begin{array}{c|c}
  \underline{\theta_{x_1, y_1, x_0}}^d & \underline{\theta_{x_0, y_1, x_1}}^d \\ 
  \gamma \underline{\theta_{x_1, y_1, x_0}}^d & \underline{\theta_{x_0, y_1, x_1}}^d\\
  \multicolumn{1}{c}{${\upbracefill}$} & \multicolumn{1}{c}{${\upbracefill}$}\\[-1ex]
  \multicolumn{1}{c}{u_{y_1}=x_0} & \multicolumn{1}{c}{u_{y_1}=x_1}\\
\end{array} \right]
\begin{array}{l}
	\MyLBrace{1ex}{$v_{y_1}=x_0$} \\
	\MyLBrace{1ex}{$v_{y_1}=x_1$,} \\
	\ \\
\end{array}
\eean
where the first $d$ p-c equations correspond to plane $\uv = \uz(x_0 \rightarrow z_{y_1})$, the last $d$ p-c equations correspond to plane $\uv = \uz(x_1 \rightarrow z_{y_1})$, and the first, second columns correspond to the symbols $(x,y,\uu) = (x_1, y_1, \uz(x_0 \rightarrow z_{y_1}))$ and $(x,y,\uu) = (x_0, y_1, \uz(x_1 \rightarrow z_{y_1}))$ respectively.
\bean
\ \implies
M_{1,d} &=& \left[\begin{array}{c} 
	V_{i1}^d \\ \hline
	V_{i1}^d \Gamma 
\end{array}\right]
\begin{array}{l}
	\MyLBrace{1ex}{$v_{y_1}=x_0$} \\
	\MyLBrace{1ex}{$v_{y_1}=x_1$} \\
\end{array} \ \text{ where } 
i = \begin{cases}
	1 & S_{y_1} = \{0,1\} \text{ or } \{2,3\}\\
	2 & S_{y_1} = \{0,2\} \text{ or } \{1,3\}\\
	3 & S_{y_1} = \{0,3\} \text{ or } \{1,2\}\\
\end{cases} (\text{due to } \eqref{eq:thetaassign}).
\eean Though we only needed the case when $d=1$, we provide the generic expression that will be used to recursively define $M_{j,d}$ in proof of Lemma \ref{lem:indMinv}. It is clear to see that the determinant of matrix $M_{1,1}$ is $(1-\gamma)$. As we have $\gamma \ne 1$ as defined in Section~\ref{subsec:pcmatrix}, the determinant is non-zero. Notice that in the example seen in Section~\ref{sec:example}, the matrix $M_{1,1}$ shown here appears as the reduced matrix $H_{E, {\tiny \uz}}^{\tiny \text{Red}}$ in the MDS property proof when two erasures appear in same-$y$ column. 
\paragraph*{Case 2: $e_{y_1} = 2$,  $\mu_1=2$, $p_1 = 3$} Let $S_{y_1}= \{x_0,x_1,x_2\}$ and $x_0 < x_1 < x_2$, then:
\bean
M_{1,d} &=& \left[\begin{array}{cc|cc|cc}
	\underline{\theta_{x_1, y_1, x_0}}^d & \underline{\theta_{x_2, y_1, x_0}}^d & \underline{\theta_{x_0, y_1, x_1}}^d &  & \underline{\theta_{x_0, y_1, x_2}}^d & \\ 
	\gamma \underline{\theta_{x_1, y_1, x_0}}^d &  & \underline{\theta_{x_0, y_1, x_1}}^d &  \underline{\theta_{x_2, y_1, x_1}}^d &  & \underline{\theta_{x_1, y_1, x_2}}^d\\ 
	& \gamma \underline{\theta_{x_2, y_1, x_0}}^d & & \gamma \underline{ \theta_{x_2, y_1, x_1}}^d & \underline{\theta_{x_0, y_1, x_2}}^d & \underline{\theta_{x_1, y_1, x_2}}^d\\ 
	\multicolumn{2}{c}{${\upbracefill}$} & \multicolumn{2}{c}{${\upbracefill}$}& \multicolumn{2}{c}{${\upbracefill}$}\\[-1ex]
	\multicolumn{2}{c}{u_{y_1}=x_0} & \multicolumn{3}{c}{u_{y_1}=x_1}& \multicolumn{1}{c}{u_{y_1}=x_2}\\
\end{array} \right]
\begin{array}{l}
	\MyLBrace{1ex}{$v_{y_1}=x_0$} \\
	\MyLBrace{1ex}{$v_{y_1}=x_1$} \\
	\MyLBrace{1ex}{$v_{y_1}=x_2$,} \\
	\ \\
\end{array}
\eean
is a $(3d \times 6)$ matrix. In $M_{1,d}$, first $d$ p-c equations correspond to plane $\uv = \uz(x_0 \rightarrow z_{y_1})$, the next $d$ p-c equations correspond to plane $\uv = \uz(x_1 \rightarrow z_{y_1})$ and the last $d$ p-c equations correspond to plane $\uv = \uz(x_2 \rightarrow z_{y_1})$ whereas the first $2$ columns correspond to erasures in plane $\uu = \uz(x_0 \rightarrow z_{y_1})$, next $2$ columns correspond to erasures in plane $\uu = \uz(x_1 \rightarrow z_{y_1})$ and the last $2$ columns correspond to erasures in plane $\uu=\uz(x_2 \rightarrow z_{y_1})$. After permuting the columns, $M_{1,d}$ can be written as:
\bean
M_{1,d} &=& \left[\begin{array}{cc|cc|cc}
	\underline{\theta_{x_1, y_1, x_0}}^d & \underline{\theta_{x_0, y_1, x_1}}^d & \underline{\theta_{x_2, y_1, x_0}}^d & \underline{\theta_{x_0, y_1, x_2}}^d & \\ 
	\gamma \underline{\theta_{x_1, y_1, x_0}}^d & \underline{\theta_{x_0, y_1, x_1}}^d & & & \underline{\theta_{x_2, y_1, x_1}}^d & \underline{\theta_{x_1, y_1, x_2}}^d\\ 
	& & \gamma \underline{\theta_{x_2, y_1, x_0}}^d & \underline{\theta_{x_0, y_1, x_2}}^d & \gamma \underline{\theta_{x_2, y_1, x_1}}^d  & \underline{\theta_{x_1, y_1, x_2}}^d\\ 
\end{array}\right] \\
&=& \left[\begin{array}{c|c|c}
	V_{i_11}^d & V_{i_21}^d & \\ \hline
	V_{i_11}^d \Gamma & & V_{i_31}^d\\ \hline
	& V_{i_21}^d \Gamma & V_{i_31}^d \Gamma
\end{array}\right]
\text{ where } (i_1, i_2, i_3) = \begin{cases}
(1, 2, 3 ) & S_{y_1} = \{0,1,2\}\\
(1, 3, 2 ) & S_{y_1} = \{0,1,3\}\\
(2, 3, 1 ) & S_{y_1} = \{0,2,3\}\\
(3, 2, 1 ) & S_{y_1} = \{1,2,3\}.
\end{cases}
\eean
The assignment of $(i_1, i_2, i_3)$ follows from the $\theta$ to $\lambda$ assignment defined in \eqref{eq:thetaassign}.
\bean
|M_{1,2}| &=& \left|\begin{array}{cc|cc|cc}
1 & 1 & 1 & 1 & & \\
\gamma \lambda_{i_1,y_1} & \lambda_{i_1,y_1} & \gamma \lambda_{i_2,y_1} &  \lambda_{i_2,y_1} & \\ \hline 
\gamma & 1 & & & 1 & 1 \\
\gamma^2 \lambda_{i_1,y_1} &  \lambda_{i_1,y_1} & & & \gamma \lambda_{i_3,y_1} & \lambda_{i_3,y_1} \\ \hline 
& & \gamma & 1 & \gamma & 1 \\
& & \gamma^2 \lambda_{i_2,y_1} &  \lambda_{i_2,y_1} & \gamma^2 \lambda_{i_3,y_1} &  \lambda_{i_3,y_1} \\
\end{array}\right|
\eean
The determinant of matrix $M_{1,2}$ can be computed to be equal to $\gamma (1-\gamma)^4 \lambda_{i_1,y_1} (\lambda_{i_2,y_1} - \lambda_{i_1,y_1})(\lambda_{i_2,y_1}-\lambda_{i_3,y_1})$. This is non-zero by the distinctness of $\lambda$'s as defined in Section~\ref{subsec:pcmatrix}. Note that the field $\fq$ being of characteristic $2$ is used in the determinant computation.
\paragraph*{Case 3: $e_{y_1} = 3$,  $\mu_1=3$, $p_1 = 4$} $S_{y_1} = \{0,1,2,3\}$ as $s \le 4$ and the $(4d \times 12)$ matrix $M_{1,d}$ is given by:
\bean
\scalebox{0.8}{$
\left[\begin{array}{ccc|ccc|ccc|ccc}
	\underline{\theta_{1, y_1, 0}}^d & \underline{\theta_{2, y_1, 0}}^d &
    \underline{\theta_{3, y_1, 0}}^d &
	 \underline{\theta_{0, y_1, 1}}^d & & & \underline{\theta_{0, y_1, 2}}^d & & & 
	 \underline{\theta_{0, y_1, 3}}^d & &\\ 
	\gamma \underline{\theta_{1, y_1, 0}}^d &  & & \underline{\theta_{0, y_1, 1}}^d & \underline{\theta_{2, y_1,1}}^d &  \underline{\theta_{3, y_1,1}}^d & & \underline{\theta_{1, y_1,2}}^d & & & \underline{\theta_{1, y_1,3}}^d &\\ 
	& \gamma \underline{\theta_{2, y_1,0}}^d & & & \gamma \underline{\theta_{2, y_1,1}}^d & & \underline{\theta_{0, y_1, 2}}^d & \underline{\theta_{1, y_1, 2}}^d & \underline{\theta_{3, y_1, 2}}^d & & & \underline{\theta_{2, y_1,3}}^d\\ 
	& & \gamma \underline{\theta_{3, y_1,0}}^d & & & \gamma \underline{\theta_{3, y_1,1}}^d & & & \gamma \underline{\theta_{3, y_1,2}}^d & \underline{\theta_{0, y_1, 3}}^d & \underline{\theta_{1, y_1, 3}}^d & \underline{\theta_{2, y_1, 3}}^d\\
	\multicolumn{3}{c}{${\upbracefill}$} & \multicolumn{3}{c}{${\upbracefill}$}& \multicolumn{3}{c}{${\upbracefill}$}&
	\multicolumn{3}{c}{${\upbracefill}$}\\[-1ex]
	\multicolumn{3}{c}{u_{y_1}=0} & \multicolumn{3}{c}{u_{y_1}=1}& \multicolumn{3}{c}{u_{y_1}=2}&
	\multicolumn{3}{c}{u_{y_1}=3}\\
\end{array} \right]
\begin{array}{l}
	\MyLBrace{2ex}{$v_{y_1}=0$} \\
	\MyLBrace{2ex}{$v_{y_1}=1$} \\
	\MyLBrace{2ex}{$v_{y_1}=2$} \\
	\MyLBrace{2ex}{$v_{y_1}=3.$} \\
	\ \\
\end{array}$}
\eean
In $M_{1,d}$, first $d$ p-c equations correspond to plane $\uv = \uz(x_0 \rightarrow z_{y_1})$, the next $d$ p-c equations correspond to plane $\uv = \uz(x_1 \rightarrow z_{y_1})$ followed by $d$ p-c equations corresponding to plane $\uv = \uz(x_2 \rightarrow z_{y_1})$ and the last $d$ p-c equations corresponding to plane $\uv = \uz(x_3 \rightarrow z_{y_1})$. Similarly, the first $3$ columns correspond to erasures in plane $\uu = \uz(x_0 \rightarrow z_{y_1})$, next $3$ columns correspond to erasures in plane $\uu = \uz(x_1 \rightarrow z_{y_1})$ followed by $3$ erased symbols from plane $\uu = \uz(x_2 \rightarrow z_{y_1})$ and the last $3$ erasures in plane $\uu=\uz(x_3 \rightarrow z_{y_1})$. After permuting columns $M_{1,d}$ can be written as:
\bean
\scalebox{0.8}{$
\begin{array}{cc|cc|cc|cc|cc|cc}
	\underline{\theta_{1, y_1, 0}}^d & 
	\underline{\theta_{0, y_1, 1}}^d & \underline{\theta_{2, y_1, 0}}^d & \underline{\theta_{0, y_1, 2}}^d & 
	\underline{\theta_{3, y_1, 0}}^d & 
	\underline{\theta_{0, y_1, 3}}^d & & &  & & &\\ 
	\gamma \underline{\theta_{1, y_1, 0}}^d &  \underline{\theta_{0, y_1, 1}}^d & & & & & & & \underline{\theta_{3, y_1,1}}^d &   \underline{\theta_{1, y_1,3}}^d & \underline{\theta_{2, y_1,1}}^d & \underline{\theta_{1, y_1,2}}^d\\ 
	& & \gamma \underline{\theta_{2, y_1,0}}^d & \underline{\theta_{0, y_1, 2}}^d & & & \underline{\theta_{3, y_1, 2}}^d &  \underline{\theta_{2, y_1,3}}^d & & &
	\gamma \underline{\theta_{2, y_1,1}}^d &  \underline{\theta_{1, y_1, 2}}^d \\ 
	& & & & \gamma \underline{\theta_{3, y_1,0}}^d& \underline{\theta_{0, y_1, 3}}^d & 
	\gamma \underline{\theta_{3, y_1,2}}^d& \underline{\theta_{2, y_1, 3}}^d & \gamma \underline{\theta_{3, y_1,1}}^d & \underline{\theta_{1, y_1, 3}}^d & &
\end{array}
\begin{array}{l}
	\MyLBrace{1ex}{$v_{y_1}=0$} \\
	\MyLBrace{1ex}{$v_{y_1}=1$} \\
	\MyLBrace{1ex}{$v_{y_1}=2$} \\
	\MyLBrace{1ex}{$v_{y_1}=3$} 
\end{array}$}\\
= \left[\begin{array}{c|c|c|c|c|c}
	V_{11}^d & V_{21}^d & V_{31}^d &  & &\\  \hline
	V_{11}^d \Gamma & & & & V_{21}^d & V_{31}^d\\  \hline
	& V_{21}^d \Gamma & & V_{11}^d & & V_{31}^d \Gamma\\ \hline
	& & V_{31}^d \Gamma & V_{11}^d \Gamma & V_{21}^d \Gamma &\\ 
\end{array}\right]
\begin{array}{l}
	\MyLBrace{1ex}{$v_{y_1}=0$} \\
	\MyLBrace{1ex}{$v_{y_1}=1$} \\
	\MyLBrace{1ex}{$v_{y_1}=2$} \\
	\MyLBrace{1ex}{$v_{y_1}=3$.} \\
\end{array}
\eean
\bean
|M_{1,3}| &=& \left|\begin{array}{cc|cc|cc|cc|cc|cc}
	1 & 1 & 1 & 1 & 1 & 1 & & & & & & \\
	\gamma \lambda_{1,y_1} & \lambda_{1,y_1} & \gamma \lambda_{2,y_1} &  \lambda_{2,y_1} & \gamma \lambda_{3,y_1} &  \lambda_{3,y_1} & & & & & \\
	\gamma^2 \lambda_{1,y_1}^2 & \lambda_{1,y_1}^2 & \gamma^2 \lambda_{2,y_1}^2 &  \lambda_{2,y_1}^2 & \gamma^2 \lambda_{3,y_1}^2 &  \lambda_{3,y_1}^2 & & & & & \\ \hline 
	\gamma & 1 & & &  & & & & 1 & 1 & 1 & 1 \\
	\gamma^2 \lambda_{1,y_1} &  \lambda_{1,y_1} & & & & & & & \gamma \lambda_{2,y_1} &  \lambda_{2,y_1} & \gamma \lambda_{3,y_1} &  \lambda_{3,y_1} \\
	\gamma^3 \lambda_{1,y_1}^2 &  \lambda_{1,y_1}^2 & & & & & & & \gamma^2 \lambda_{2,y_1}^2 &  \lambda_{2,y_1}^2 & \gamma^2 \lambda_{3,y_1}^2 &  \lambda_{3,y_1}^2 \\ \hline 
	& & \gamma & 1 & & & 1 & 1 & & & \gamma & 1 \\
	& & \gamma^2 \lambda_{2,y_1} & \lambda_{2,y_1} & & & \gamma \lambda_{1,y_1} & \lambda_{1,y_1}& & & \gamma^2 \lambda_{3,y_1} &  \lambda_{3,y_1} \\
	& & \gamma^3 \lambda_{2,y_1}^2 & \lambda_{2,y_1}^2 & & & \gamma^2 \lambda_{1,y_1}^2 & \lambda_{1,y_1}^2& & & \gamma^3 \lambda_{3,y_1}^2 &  \lambda_{3,y_1}^2 \\ \hline
	& & & & \gamma & 1 & \gamma & 1 & \gamma & 1 & &\\
	& & & & \gamma^2 \lambda_{3,y_1} & \lambda_{3,y_1} & \gamma^2 \lambda_{1,y_1} &  \lambda_{1,y_1} & \gamma^2 \lambda_{2,y_1} & \lambda_{2,y_1} & &\\
	& & & & \gamma^3 \lambda_{3,y_1}^2 & \lambda_{3,y_1}^2 & \gamma^3 \lambda_{1,y_1}^2 &  \lambda_{1,y_1}^2 & \gamma^3 \lambda_{2,y_1}^2 & \lambda_{2,y_1}^2 & &
\end{array}\right|.
\eean
The determinant of matrix $M_{1,3}$ from Lemma \ref{lem:dets4} is $\gamma^4 (1-\gamma)^6 (\lambda_{1,y_1}-\lambda_{2,y_1})^2(\lambda_{1,y_1}-\lambda_{3,y_1})^2(\lambda_{2,y_1}-\lambda_{3,y_1})^4(\lambda_{1,y_1}-\gamma\lambda_{3,y_1})(\gamma\lambda_{1,y_1}-\lambda_{3,y_1})(\lambda_{1,y_1}-\gamma\lambda_{2,y_1})(\gamma\lambda_{1,y_1}-\lambda_{2,y_1})$. This determinant is non-zero due to the $\lambda$ assignment conditions presented in Section~\ref{subsec:pcmatrix}. Hence we proved that $M_{1,\mu_1}$ is invertible. 
\eprf

\blem\label{lem:indMinv}
For any $i \in [2:m-1]$, $M_{i, \mu_i}$ is invertible given $M_{i-1, \mu_{i-1}}$ is invertible.
\elem
\bprf
Proof is provided in Appendix \ref{appendix}.
\eprf

\bcor\label{cor:hinv}
For any $E$ such that $|E|\le r$ and any plane $\uz \in \Zst$, $H^{\text{\tiny Red}}_{E, {\tiny \uz}}$ is invertible.
\ecor
\bprf
$H^{\text{\tiny Red}}_{E, {\tiny \uz}} = M_{m, \mu_m}$. Therefore the proof follows from Lemma \ref{lem:indMinv_initcond} and Lemma \ref{lem:indMinv}.
\eprf
\ \\ \ \\
We will now show that \smalld\ code is an MSR code with field size $q=O(n)$.
\bthm\label{thm:main}
\smalld\ coode is an optimal-access MSR code with parameters of the form $(n=st, k, d=k+s-1)$ and $\alpha = s^t$ where $s \in \{2,3,4\}$ over field $\fq$ that is an extension of binary field $q=2^w$, such that $q = O(n)$.
\ethm
\bprf
The MDS property of \smalld\ code follows from Theorems~\ref{thm:reduction1},\ref{thm:reduction2} and Corollary \ref{cor:hinv}. The optimal-access property follows from Theorems~\ref{thm:reduction1_rep},\ref{thm:reduction2_rep} and Corollary \ref{cor:hinv}.

We show here a way to do the $\lambda$ assignment that satisfies the conditions presented in Section~\ref{subsec:pcmatrix} with a field $\fq$ of size $q = O(n)$. Consider a multiplicative sub-group $G$ of $\fq \setminus \{0\}$ and cosets $\gamma G$,$\gamma^2 G$. Let $m_0$ be the number of distinct $\lambda$'s in $\Lambda_{s,y}$ matrix defined in \eqref{eq:thetaassign} excluding $\lambda_{0,y}$. It is clear to see that $m_0=1$ for $s=2$ and $m_0=3$ for $s=3,4$. We pick coefficients for $\{\lambda_{i,y} | \ y \in \Zt, i \in [m_0]\}$ from $G$ and the corresponding $\gamma$ multiples can be picked from $\gamma G$ and the remaining $t$,  $\lambda$'s corresponding to $\{\lambda_{0,y} \mid y \in [0:t-1]\}$ are picked from $\gamma^2G$. When $w$ is even $3|2^w-1$ and define $G = \{\psi^{3i}: 0 \leq i\leq \frac{2^w-1}{3}-1\}$, where $\psi$ is primitive element of $\mathbb{F}_{2^w}$ and set $\gamma=\psi$. Therefore, by choosing field size such that $|G|>m_0t$, we get $q \ge 3m_0t+1$. If the smallest possible field size $q$ that satisfies $2^w = q \ge 3m_0t+1$ results in a odd $w$, we just take double the field size. Therefore field size $q \ge 6t+2$ for $s=2$ and $q \ge 18t+2$ for $s \in 3,4$. 
\eprf

\bibliographystyle{IEEEtran}
\bibliography{msrjournal}

\begin{appendices} \label{appendix}
\section{Proof of Lemma~\ref{lem:indMinv}}
We will now introduce notation and prove few lemmas on factors of polynomials that will be used to prove Lemma~\ref{lem:indMinv}.

\blem\label{lem:poly1}
Let $g(x)$ be a polynomial in $\fq [x]$ and $f(x_1, \cdots, x_m)$ be a multivariate polynomial in $m$ variables i.e.,  $f \in \fq [x_1, \cdots, x_{m}]$ such that:
\bean
g(x) &=& f(x_1=x, x_2=x, \cdots, x_{m}=x) \\
f(x_1, \cdots, x_{i-1}, x_i = a, x_{i+1}, \cdots, x_m) &=& 0 \text{ for any } i \in [n] \text{ where } n \le m.
\eean
where $a$ is an element in $\fq$. Then $(x-a)^{n}$ divides $g(x)$. 
\elem
\bprf
It is clear to see that $\prod_{i=1}^n (x_i - a) \ | \ f(x_1, x_2, \cdots, x_m)$. Therefore $(x-a)^n$ divides $g(x)$.
\eprf

\blem\label{lem:poly2}
Let $g(x)$ be a polynomial in $\fq [x]$ and let $f(x_1, \cdots, x_{2n})$ be a multivariate polynomial in $2n$ variables i.e, $f \in \fq [x_1, \cdots, x_{2n}]$ such that:
\bean
g(x) &=& f(x_1=x, x_2=x, \cdots, x_{2n}=x) \\
f(x_1, \cdots, x_{2n}) \mid_{\text{on setting any } n+1 \text{ variables to } a} &=& 0 
\eean
where $a \in \fq$. Then $(x-a)^{n}$ divides $g(x)$. 
\elem
\bprf
We divide $f(x_1, \cdots, x_{2n})$ by $(x_1-a)$ to get:
\bea
\label{eq:poly1}
f(x_1, \cdots, x_{2n}) = (x_1-a) f_1(x_1, \cdots, x_{2n}) + R_1(x_2, \cdots, x_{2n}).
\eea
We now recursively define $R_i(x_{i+1}, \cdots, x_{2n})$ to be the reminder obtained on dividing $R_{i-1}(x_i, \cdots, x_{2n})$ by $(x_i-a)$ for $2 \le i \le (n+1)$. Let:
\bea
\label{eq:poly2}
R_{i-1}(x_i, \cdots, x_{2n}) = (x_i-a) f_i(x_1, \cdots, x_{2n}) + R_i(x_{i+1}, \cdots, x_{2n}).
\eea
From equations~\eqref{eq:poly1} and \eqref{eq:poly2} we get:
\bea
\label{eq:poly3} f(x_1, \cdots, x_{2n}) = (x_1-a) f_1(x_1, \cdots, x_{2n}) +\cdots + (x_{n+1}-a) f_{n+1}(x_1, \cdots, x_{2n}) +  R_{n+1}(x_{n+2}, \cdots, x_{2n}).
\eea
By setting $x_1 =x_2=\cdots=x_{n+1} = 0$ in equation~\eqref{eq:poly3} we get:
\bean
R_{n+1}(x_{n+2}, \cdots, x_{2n}) = 0.
\eean
Therefore:
\bea
\label{eq:poly4} f(x_1, \cdots, x_{2n}) = (x_1-a) f_1(x_1, \cdots, x_{2n}) +\cdots + (x_{n+1}-a) f_{n+1}(x_1, \cdots, x_{2n}).
\eea
On similarly expanding $f_{i_1}(x_1, \cdots, x_{2n})$ for $i_1 \in [n+1]$ by recursively dividing it by $(x_1-a), (x_2-a) \cdots, (x_{i_1-1}-a), (x_{i_1+1}-a), \cdots, (x_{n+2}-a)$ we get:
\bea
\label{eq:poly5}
f_{i_1}(x_1, \cdots x_{2n}) = \sum\limits_{\substack{i_2 = 1\\ i_2 \ne i_1}}^{n+2}(x_{i_2}-a)f_{i_1,i_2}(x_1, \cdots, x_{2n}) + R(x_i, x_{n+3}, \cdots, x_{2n}).
\eea
On setting $x_1 = x_2 =\cdots = x_{i_1-1}=x_{i_1+1}=\cdots=x_{n+1}=x_{n+2} = a$ in equation~\eqref{eq:poly4} we get:
\bea
\nonumber 0 &=& (x_{i_1} - a) f_{i_1}(x_1, \cdots, x_{2n}) |_{x_1=x_2=\cdots=x_{i_1-1} = x_{i_1+1} = \cdots = x_{n+2}= a}\\
\nonumber 0 &=& f_{i_1}(x_1, \cdots, x_{2n}) |_{x_1=x_2=\cdots=x_{i_1-1} = x_{i_1+1} = \cdots = x_{n+2}= a} \text{ as it is true for } x_{i_1} \ne a\\
\label{eq:poly6} &=& R(x_{i_1}, x_{n+3}, \cdots, x_{2n}) \text{ follows from equation~\eqref{eq:poly5}}
\eea
From equations~\eqref{eq:poly4}, \eqref{eq:poly5}, \eqref{eq:poly6} we get:
\bean
f(x_1, \cdots x_{2n}) &=& \sum\limits_{\substack{i_1 = 1}}^{n+1} \sum\limits_{\substack{i_2 = 1\\ i_2 \ne i_1}}^{n+2} (x_{i_1}-a)(x_{i_2}-a)f_{i_1,i_2}(x_1, \cdots, x_{2n})\\
&=& \sum\limits_{\substack{i_1 = 1}}^{n+1} \sum\limits_{i_2 > i_1}^{n+2} (x_{i_1}-a)(x_{i_2}-a)g_{i_1,i_2}(x_1, \cdots, x_{2n})
\eean
where $g_{i_1,i_2} = f_{i_1, i_2} + f_{i_2, i_1}$. Assuming that we have an expansion of the form:
\bea
\label{eq:poly7}
f(x_1, \cdots x_{2n}) = \sum\limits_{\substack{i_1 = 1}}^{n+1} \sum\limits_{i_2 > i_1}^{n+2} \cdots \sum\limits_{i_k > i_{k-1}}^{n+k} \left( \prod_{j=1}^k (x_{i_j}-a) \right) g_{i_1,i_2, \cdots, i_k}(x_1, \cdots, x_{2n}).
\eea
for $2 \le k \le n-1$. Expanding $g_{i_1,i_2, \cdots, i_k}(x_1, \cdots, x_{2n})$ similar using factors $(x_{i_{k+1}} - a)$ for $i_{k+1} \in [1, n+k+1] \setminus \{i_1, \cdots, i_k\}$ we get:
\bea
\label{eq:poly8}g_{i_1,i_2, \cdots, i_k}(x_1, \cdots, x_{2n}) = \sum\limits_{\substack{i_{k+1}=1 \\ i_{k+1} \notin \{i_1, \cdots, i_k\} }}^{n+k+1} (x_{i_{k+1}} - a) f_{i_1,\cdots, i_{k+1}}(x_1, \cdots, x_{2n}) + R(x_{i_1}, \cdots, x_{i_k}, x_{n+k+2}, \cdots, x_{2n}).
\eea
On setting $x_j = a$ for all $j \in [1, n+k+1] \setminus \{i_1, \cdots, i_k\}$ in equation~\eqref{eq:poly7} we get:
\bea
\nonumber 0 &=&  \left( \prod_{j=1}^k (x_{i_j}-a) \right) g_{i_1,i_2, \cdots, i_k}(x_1, \cdots, x_{2n})|_{x_j = a \text{ for all } j \in [1:n+k+1] \setminus \{i_1, \cdots, i_k\}}\\
\nonumber 0 &=& g_{i_1,i_2, \cdots, i_k}(x_1, \cdots, x_{2n})|_{x_j = a \text{ for all } j \in [1:n+k+1] \setminus \{i_1, \cdots, i_k\}}\\
\label{eq:poly9} &=& R(x_{i_1}, \cdots, x_{i_k}, x_{n+k+2}, x_{2n}) \text{ from equation~\eqref{eq:poly8}}
\eea
Equations~\eqref{eq:poly7} \eqref{eq:poly8}, \eqref{eq:poly9} imply that:
\bea
\nonumber f(x_1, \cdots x_{2n}) &=& \sum\limits_{\substack{i_1 = 1}}^{n+1} \sum\limits_{i_2 > i_1}^{n+2} \cdots \sum\limits_{i_k > i_{k-1}}^{n+k} \sum\limits_{\substack{i_{k+1} = 1\\ i_{k+1} \notin \{i_1, i_2, \cdots i_k\}}}^{n+k+1} \left( \prod_{j=1}^{k+1} (x_{i_j}-a) \right) f_{i_1,i_2, \cdots, i_{k+1}}(x_1, \cdots, x_{2n})\\
\label{eq:poly10} &=& \sum\limits_{\substack{i_1 = 1}}^{n+1} \sum\limits_{i_2 > i_1}^{n+2} \cdots  \sum\limits_{i_{k+1} > i_k}^{n+k+1} \left( \prod_{j=1}^{k+1} (x_{i_j}-a) \right) g_{i_1,i_2, \cdots, i_{k+1}}(x_1, \cdots, x_{2n}),
\eea
where $g_{i_1,i_2, \cdots, i_{k+1}} = f_{i_{k+1},i_1,i_2, \cdots, i_k}+f_{i_1,i_{k+1},i_2, \cdots, i_k}+\cdots+f_{i_1,i_2, \cdots, i_k, i_{k+1}}$.
Setting $k = n-1$ in equation~\eqref{eq:poly10} we get:
\bean
f(x_1, \cdots x_{2n}) = \sum\limits_{i_1 = 1}^{n+1} \sum\limits_{i_2 > i_1}^{n+2} \cdots \sum\limits_{i_n > i_{n-1}}^{2n} \left(\prod_{j=1}^n (x_{i_j}-a) \right) g_{i_1,i_2, \cdots, i_n}(x_1, \cdots, x_{2n}).
\eean
It now clearly follows that $(x-a)^n | g(x)$ where $g(x)=f(x_1=x, x_2=x, \cdots, x_{2n}=x)$.
\eprf

\blem\label{lem:dets4}
The determinant of matrix:
\bean
\scalebox{0.85}{$M = \left[ \begin{array}{cc|cc|cc|cc|cc|cc}
	1 & 1 & 1 & 1 & 1 & 1 & & & & & & \\
	\gamma \lambda_{1} & \lambda_{1} & \gamma \lambda_{2} &  \lambda_{2} & \gamma \lambda_{3} &  \lambda_{3} & & & & & \\
	\gamma^2 \lambda_{1}^2 & \lambda_{1}^2 & \gamma^2 \lambda_{2}^2 &  \lambda_{2}^2 & \gamma^2 \lambda_{3}^2 &  \lambda_{3}^2 & & & & & \\ \hline 
	\gamma & 1 & & &  & & & & 1 & 1 & 1 & 1 \\
	\gamma^2 \lambda_{1} &  \lambda_{1} & & & & & & & \gamma \lambda_{2} &  \lambda_{2} & \gamma \lambda_{3} &  \lambda_{3} \\
	\gamma^3 \lambda_{1}^2 &  \lambda_{1}^2 & & & & & & & \gamma^2 \lambda_{2}^2 &  \lambda_{2}^2 & \gamma^2 \lambda_{3}^2 &  \lambda_{3}^2 \\ \hline 
	& & \gamma & 1 & & & 1 & 1 & & & \gamma & 1 \\
	& & \gamma^2 \lambda_{2} & \lambda_{2} & & & \gamma \lambda_{1} & \lambda_{1}& & & \gamma^2 \lambda_{3} &  \lambda_{3} \\
	& & \gamma^3 \lambda_{2}^2 & \lambda_{2}^2 & & & \gamma^2 \lambda_{1}^2 & \lambda_{1}^2& & & \gamma^3 \lambda_{3}^2 &  \lambda_{3}^2 \\ \hline
	& & & & \gamma & 1 & \gamma & 1 & \gamma & 1 & &\\
	& & & & \gamma^2 \lambda_{3} & \lambda_{3} & \gamma^2 \lambda_{1} &  \lambda_{1} & \gamma^2 \lambda_{2} & \lambda_{2} & &\\
	& & & & \gamma^3 \lambda_{3}^2 & \lambda_{3}^2 & \gamma^3 \lambda_{1}^2 &  \lambda_{1}^2 & \gamma^3 \lambda_{2}^2 & \lambda_{2}^2 & &
	\end{array}\right]$}
\eean
is $\gamma^4 (1-\gamma)^6 (\lambda_{1}-\lambda_{2})^2(\lambda_{1}-\lambda_{3})^2(\lambda_{2}-\lambda_{3})^4(\lambda_{1}-\gamma\lambda_{3})(\gamma\lambda_{1}-\lambda_{3})(\lambda_{1}-\gamma\lambda_{2})(\gamma\lambda_{1}-\lambda_{2})$. 
\elem
\bprf
Let the determinant of matrix $M$ be a polynomial $f(\lambda_1)$. It can be observed the degree of this polynomial is atmost $8$. Let
\bean
M(\nu_1, \nu_2)&=&\left[ \begin{array}{cc|cc|cc|cc|cc|cc}
	1 & 1 & 1 & 1 & 1 & 1 & & & & & & \\
	\nu_{1} & \nu_{2} & \gamma \lambda_{2} &  \lambda_{2} & \gamma \lambda_{3} &  \lambda_{3} & & & & & \\
	\nu_{1}^2 & \nu_{2}^2 & \gamma^2 \lambda_{2}^2 &  \lambda_{2}^2 & \gamma^2 \lambda_{3}^2 &  \lambda_{3}^2 & & & & & \\ \hline 
	\gamma & 1 & & &  & & & & 1 & 1 & 1 & 1 \\
	\gamma \nu_{1} &  \nu_{2} & & & & & & & \gamma \lambda_{2} &  \lambda_{2} & \gamma \lambda_{3} &  \lambda_{3} \\
	\gamma \nu_{1}^2 &  \nu_{2}^2 & & & & & & & \gamma^2 \lambda_{2}^2 &  \lambda_{2}^2 & \gamma^2 \lambda_{3}^2 &  \lambda_{3}^2 \\ \hline 
	& & \gamma & 1 & & & 1 & 1 & & & \gamma & 1 \\
	& & \gamma^2 \lambda_{2} & \lambda_{2} & & & \nu_1 & \nu_2& & & \gamma^2 \lambda_{3} &  \lambda_{3} \\
	& & \gamma^3 \lambda_{2}^2 & \lambda_{2}^2 & & & \nu_{1}^2 & \nu_{2}^2& & & \gamma^3 \lambda_{3}^2 &  \lambda_{3}^2 \\ \hline
	& & & & \gamma & 1 & \gamma & 1 & \gamma & 1 & &\\
	& & & & \gamma^2 \lambda_{3} & \lambda_{3} & \gamma \nu_{1} &  \nu_{2} & \gamma^2 \lambda_{2} & \lambda_{2} & &\\
	& & & & \gamma^3 \lambda_{3}^2 & \lambda_{3}^2 & \gamma \nu_{1}^2 &  \nu_{2}^2 & \gamma^3 \lambda_{2}^2 & \lambda_{2}^2 & &
\end{array}\right],
\eean
and $g(\nu_1, \nu_2) = \text{det}(M(\nu_1, \nu_2))$. Then $f(\lambda_1)= g(\gamma \lambda_1, \lambda_1)$. We will now show that $g(v_1, v_2)$ is divisible by $(v_1-\lambda_2)(v_1-\lambda_3)(v_1-\gamma\lambda_2)(v_1-\gamma\lambda_3)(v_2-\lambda_2)(v_2-\lambda_3)(v_2-\gamma \lambda_2)(v_2-\gamma \lambda_3)$. It can be seen that:
\bean
g(\gamma \lambda_2, \nu_2) = 0 \text{ as } \underline{c}_{1} + \underline{c}_{3} + \gamma \underline{c}_7 + \gamma \underline{c}_9 = 0\\
g( \lambda_2, \nu_2) = 0 \text{ as } \underline{c}_{1} + \underline{c}_{4} + \underline{c}_7 + \gamma \underline{c}_{10} = 0\\
g( \nu_1, \lambda_2) = 0 \text{ as } \underline{c}_{2} + \underline{c}_{4} + \underline{c}_8 + \underline{c}_{10} = 0\\
g( \nu_1, \gamma \lambda_2) = 0 \text{ as } \underline{c}_{2} + \underline{c}_{3} + \gamma \underline{c}_8 + \underline{c}_{9} = 0\\
g( \nu_1, \lambda_3) = 0 \text{ as } \underline{c}_{2} + \underline{c}_{6} + \underline{c}_8 + \underline{c}_{12} = 0\\
g( \nu_1, \gamma \lambda_3) = 0 \text{ as } \underline{c}_{2} + \underline{c}_{5} + \gamma \underline{c}_8 + \underline{c}_{11} = 0
\eean
where $\underline{c}_i$ is the $i$-th column of the matrix $M(\nu_1, \nu_2)$. We will now show that the matrix $M(\lambda_3, \nu_2)$ has determinant $0$. This is implied if there exists a non-zero vector $\underline{\mu}$ such that $\underline{\mu}^T M(\lambda_3, \nu_2) = 0$. Let $\underline{\mu}^T = \left[\gamma\underline{\mu_1}^T \ \gamma \underline{\mu_1}^T \  \underline{\mu_1}^T \ \underline{\mu_1}^T\right]$ and $\underline{\mu}_1^T = [m_0, m_1, m_2]$ such that the polynomial $m(x)=m_0+m_1x+m_2x^2 = (x- \lambda_3)(x-\lambda_2)$. It can be seen that $\underline{\mu}^T M(\lambda_3, \nu_2) = \underline{0}^T$. \\
We will now similarly show $M(\gamma \lambda_3, \nu_2)$ has determinant zero. Let $\underline{\mu}^T = \left[\underline{\mu_1}^T \ \underline{\mu_1}^T \  \underline{\mu_1}^T \ \underline{\mu_1}^T\right]$ and $\underline{\mu}_1^T = [m_0, m_1, m_2]$ such that the polynomial $m(x)=m_0+m_1x+m_2x^2 = (x- \gamma \lambda_3)(x- \gamma \lambda_2)$. It can be seen that $\underline{\mu}^T M(\gamma \lambda_3, \nu_2) =\underline{0}^T$. \\
This implies that:
\bean
g(\nu_1, \nu_2) &=& g' \prod\limits_{i=1}^2 (\nu_i - \lambda_2) (\nu_i - \lambda_3) (\nu_i - \gamma \lambda_2) (\nu_i - \gamma \lambda_3),
\eean
where $g'$ is a constant dependant only on $\lambda_2$, $\lambda_3$ as $g$ is of degree $4$ in variables $\nu_1$, $\nu_2$ . Therefore:
\bean
f(\lambda_1) = g(\gamma \lambda_1, \lambda_1) &=&  g' \gamma^2 (\lambda_1 - \lambda_2)^2 (\lambda_1 - \lambda_3)^2 (\lambda_1 - \gamma \lambda_2) (\lambda_1 - \gamma \lambda_3) (\gamma \lambda_1 - \lambda_2) (\gamma \lambda_1 - \lambda_3)
\eean
We will now determine $g'$ by computing $f(0) = \text{det} (M(0,0))$.
\bean
M(0, 0)&=&\left[ \begin{array}{cc|cc|cc|cc|cc|cc}
	1 & 1 & 1 & 1 & 1 & 1 & & & & & & \\
	&  & \gamma \lambda_{2} &  \lambda_{2} & \gamma \lambda_{3} &  \lambda_{3} & & & & & \\
	& & \gamma^2 \lambda_{2}^2 &  \lambda_{2}^2 & \gamma^2 \lambda_{3}^2 &  \lambda_{3}^2 & & & & & \\ \hline 
	\gamma & 1 & & &  & & & & 1 & 1 & 1 & 1 \\
	& & & & & & & & \gamma \lambda_{2} &  \lambda_{2} & \gamma \lambda_{3} &  \lambda_{3} \\
	& & & & & & & & \gamma^2 \lambda_{2}^2 &  \lambda_{2}^2 & \gamma^2 \lambda_{3}^2 &  \lambda_{3}^2 \\ \hline 
	& & \gamma & 1 & & & 1 & 1 & & & \gamma & 1 \\
	& & \gamma^2 \lambda_{2} & \lambda_{2} & & & && & & \gamma^2 \lambda_{3} &  \lambda_{3} \\
	& & \gamma^3 \lambda_{2}^2 & \lambda_{2}^2 & & & && & & \gamma^3 \lambda_{3}^2 &  \lambda_{3}^2 \\ \hline
	& & & & \gamma & 1 & \gamma & 1 & \gamma & 1 & &\\
	& & & & \gamma^2 \lambda_{3} & \lambda_{3} & & & \gamma^2 \lambda_{2} & \lambda_{2} & &\\
	& & & & \gamma^3 \lambda_{3}^2 & \lambda_{3}^2 & & & \gamma^3 \lambda_{2}^2 & \lambda_{2}^2 & &
\end{array}\right]
\eean
\bean
\text{det} (M(0,0)) &=& (1-\gamma)^2 \gamma^4 \lambda_2^4 \lambda_3^4 \left| \begin{array}{cc|cc|cc|cc}
	1 & 1 & 1 & 1 & & & &\\
	\gamma \lambda_2 & \lambda_2 & \gamma \lambda_3 & \lambda_3 & & & &\\ \hline
	& & & & 1 & 1 & 1 & 1\\
	& & & & \gamma \lambda_2 & \lambda_2 & \gamma \lambda_3 & \lambda_3\\ \hline
	\gamma & 1 & & & & & \gamma & 1\\
	\gamma^2 \lambda_2 & \lambda_2 & & & & & \gamma^2 \lambda_3 & \lambda_3\\ \hline
	& & \gamma & 1 & \gamma & 1 & &\\
	& & \gamma^2 \lambda_3 & \lambda_3 & \gamma^2 \lambda_2 & \lambda_2 & &
\end{array}\right|\\
&=& (1-\gamma)^6 \gamma^6 \lambda_2^4 \lambda_3^4 (\lambda_2-\lambda_3)^4 \text{ by computing the determinant of the $8 \times 8$ matrix above}\\
&=& f(0) = g' \gamma^4 \lambda_2^4 \lambda_3^4 \\
g' &=& (1-\gamma)^6 \gamma^2 (\lambda_2 - \lambda_3)^4
\eean
\bean
f(\lambda_1) &=&  (1-\gamma)^6 \gamma^4 (\lambda_2 - \lambda_3)^4 \prod_{i=2}^3 (\lambda_1 - \lambda_i)^2  (\lambda_1 - \gamma \lambda_i) (\gamma \lambda_1 - \lambda_i) 
\eean
\eprf

\subsection{Recursive definition of $M_{i,d}$}
 \label{remark:recursionM} 
 
 The notation introduced in Definitions~\ref{defn:erasvec},\ref{defn:M} and \ref{defn:indmatr} within Section~\ref{sec:inv_red} will be used throughout this appendix. Note that the induction matrix $M_{i, \mu_i}$ introduced in Definition~\ref{defn:indmatr} is a sub matrix of $H_{E, {\tiny \uz}}^{ \text{\tiny Red}}$ where $E$ is an erasure (aloof node) pattern and $\uz$ is plane in $\Zst$. Since we restrict our attention to a particular $E$, $\uz$ we do not mention it explicitly in the notation and this holds for any $|E| \le r$ and $\uz \in \Zst$.
 
 We note here the recursive definition of $M_{i, d}$ in terms of $M_{i-1, d}$. The recursive description is dependent on the value of $e_{y_i} \in \{1, 2, 3\}$.
	
Recall that from the Definition~\ref{defn:indmatr}, the inductive matrix $M_{i, d}$ is a $(d p_i \times \mu_i p_i)$ sub-matrix of the parity-check matrix $H$ where $p_i = |Q^i(E, \uz)|$. The rows are indexed by $[\ell, \uv]$ and columns by  $[x,y,\uu]$ where $\ell \in [d]$, $\uu, \uv \in Q^i(E, \uz)$, $(x,y) \in E_{2,\uu}$ such that $y \in \{y_1, y_2, \cdots, y_i \}$. Let,
	\bean
	\scalebox{0.84}{$
		D_{j,i}^d = \underbrace{\left[\begin{array}{ccc}
			V_{j,i}^d & & \\
			& \ddots & \\
			& & V_{j,i}^d
			\end{array}\right]}_{(d p_{i-1} \times 2 p_{i-1} )} \ \ \Psi = \underbrace{\left[\begin{array}{ccc}
			\Gamma & & \\
			& \ddots &\\
			& & \Gamma
			\end{array}\right]}_{(2 p_{i-1} \times 2 p_{i-1})} 
		\text{ where } V_{j, i}^d = \left[\begin{array}{cc}
		1 & 1\\
		\gamma \lambda_{j,y_i} & \lambda_{j, y_i}\\
		\vdots & \vdots\\
		(\gamma \lambda_{j,y_i})^{d-1} & \lambda_{j, y_i}^{d-1}\\
		\end{array}\right] \text{ and } \Gamma = \left[\begin{array}{cc}
		\gamma & \\
		& 1
		\end{array}\right].$}
	\eean
	\subsubsection{$e_{y_i}=1$} For this case $|S_{y_i}| = (e_{y_i}+1) = 2$. Let $S_{y_i} = \{x_0, x_1\}$ where $x_0 < x_1$ then it can be seen that:
	\bean
	M_{i,d}  = \left[\begin{array}{ccc}
		M_{i-1,d}  & & D_{i_1,i}^{d}\\
		& M_{i-1,d}  & D_{i_1,i}^{d} \Psi\\
		\multicolumn{1}{c}{${\upbracefill}$} &
		\multicolumn{1}{c}{${\upbracefill}$} & \multicolumn{1}{c}{${\upbracefill}$}\\
		\multicolumn{1}{c}{\substack{y\ne y_i \\ u_{y_i}=x_0}} &
		\multicolumn{1}{c}{\substack{y\ne y_i \\  u_{y_i}=x_1}} & \multicolumn{1}{c}{\substack{y=y_i \\ x,u_{y_i} \in \{x_0, x_1\}}}\\
	\end{array}\right]
	\begin{array}{l}
		\MyLBrace{2ex}{$v_{y_i}=x_0$} \\
		\MyLBrace{2ex}{$v_{y_i}=x_1$} \\
		\ \\ \ \\
	\end{array} \ \text{ where } i_1 = \scalebox{0.9}{$\begin{cases}
		1 & S_{y_i} = \{0,1\} \text{ or } \{2,3\}\\
		2 & S_{y_i} = \{0,2\} \text{ or } \{1,3\}\\
		3 & S_{y_i} = \{0,3\} \text{ or } \{1,2\}\\
		\end{cases} $}
	\eean
	due to the $\theta$ to $\lambda$ assignment shown in equation \eqref{eq:thetaassign}. It can be noted that number of rows in $M_{i, d}$ is two times the rows of $M_{i-1, d} =  2dp_{i-1} = d p_i$. Number of columns of $M_{i, d}$ is equal to $2\mu_{i-1}p_{i-1} + 2 p_{i-1} = (\mu_{i-1}+1) p_i = \mu_i p_i$.\\
	
	\subsubsection{$e_{y_i}=2$} For this case $|S_{y_i}| = (e_{y_i}+1) = 3$.  Let $S_{y_i} = \{x_0, x_1, x_2\}$ where $x_0 < x_1 < x_2$ then it can be seen that:
	\bean
	M_{i,d} = \left[\begin{array}{cccccc}
		M_{i-1,d}  & & & D_{i_1,i}^{d} & D_{i_2,i}^{d} &\\
		& M_{i-1,d}  & & D_{i_1,i}^{d}\Psi & & D_{i_3,i}^{d}\\
		& & M_{i-1,d}  & & D_{i_2,i}^{d}\Psi  & D_{i_3,i}^{d} \Psi\\
		\multicolumn{1}{c}{${\upbracefill}$} &
		\multicolumn{1}{c}{${\upbracefill}$} & \multicolumn{1}{c}{${\upbracefill}$} &
		\multicolumn{1}{c}{${\upbracefill}$} &
		\multicolumn{1}{c}{${\upbracefill}$} &
		\multicolumn{1}{c}{${\upbracefill}$}\\
		\multicolumn{1}{c}{\substack{y\ne y_i \\u_{y_i}=x_0}}&
		\multicolumn{1}{c}{\substack{y\ne y_i \\u_{y_i}=x_1}}&
		\multicolumn{1}{c}{\substack{y\ne y_i \\u_{y_i}=x_2}}&
		\multicolumn{1}{c}{\substack{y=y_i \\ x,u_{y_i} \in \{x_0, x_1\}}}&
		\multicolumn{1}{c}{\substack{y=y_i \\ x,u_{y_i} \in \{x_0, x_2\}}}&
		\multicolumn{1}{c}{\substack{y=y_i \\ x,u_{y_i} \in \{x_1, x_2\}}}\\
	\end{array}\right]
	\begin{array}{l}
		\MyLBrace{2ex}{$v_{y_i}=x_0$} \\
		\MyLBrace{2ex}{$v_{y_i}=x_1$} \\
		\MyLBrace{2ex}{$v_{y_i}=x_2$} \\
		\ \\ \ \\
	\end{array}
	\eean
	where $(i_1, i_2, i_3) = \begin{cases}
	(1, 2, 3 ) & S_{y_1} = \{0,1,2\}\\
	(1, 3, 2 ) & S_{y_1} = \{0,1,3\}\\
	(2, 3, 1 ) & S_{y_1} = \{0,2,3\}\\
	(3, 2, 1 ) & S_{y_1} = \{1,2,3\}
	\end{cases}$ due to \eqref{eq:thetaassign}.\\
	Number of rows in $M_{i, d}$ is equal to $3dp_{i-1} = dp_i$ and number of columns in $M_{i, d}$ is equal to $3 \mu_{i-1}p_{i-1} + 6p_{i-1} = 3p_{i-1}(\mu_{i-1}+2) = p_i \mu_i$.\\
	
	\subsubsection{$e_{y_i}=3$} For this case $|S_{y_i}| = (e_{y_i}+1) = 4$.  $S_{y_i} = \{0,1,2,3\}$ and the recursion for $M_{i, d}$ is given by:
	\bean
	\scalebox{0.8}{$
		\left[\begin{array}{cccccccccc}
		M_{i-1,d} &  & & & D_{1,i}^{d} & D_{2,i}^{d} & D_{3,i}^{d} & & &\\ 
		& M_{i-1,d} & & & D_{1,i}^{d} \Psi & & & & D_{2,i}^{d} & D_{3,i}^{d} \\ 
		&  & M_{i-1,d} &  & & D_{2,i}^{d} \Psi & & D_{1,i}^{d} & & D_{3,i}^{d} \Psi\\ 
		&  &  & M_{i-1,d} & &  & D_{3,i}^{d} \Psi & D_{1,i}^{d} \Psi & D_{2,i}^{d} \Psi &\\
		\multicolumn{1}{c}{${\upbracefill}$} &
		\multicolumn{1}{c}{${\upbracefill}$} & \multicolumn{1}{c}{${\upbracefill}$} &
		\multicolumn{1}{c}{${\upbracefill}$} &
		\multicolumn{1}{c}{${\upbracefill}$} &
		\multicolumn{1}{c}{${\upbracefill}$} &
		\multicolumn{1}{c}{${\upbracefill}$} &
		\multicolumn{1}{c}{${\upbracefill}$}&
		\multicolumn{1}{c}{${\upbracefill}$}&
		\multicolumn{1}{c}{${\upbracefill}$}\\
		\multicolumn{1}{c}{\substack{y\ne y_i \\u_{y_i}=0}}&
		\multicolumn{1}{c}{\substack{y\ne y_i \\u_{y_i}=1}}&
		\multicolumn{1}{c}{\substack{y\ne y_i \\u_{y_i}=2}}&
		\multicolumn{1}{c}{\substack{y\ne y_i \\u_{y_i}=3}}&
		\multicolumn{1}{c}{\substack{y=y_i \\ x,u_{y_i} \in \{0, 1\}}}&
		\multicolumn{1}{c}{\substack{y=y_i \\ x,u_{y_i} \in \{0, 2\}}}&
		\multicolumn{1}{c}{\substack{y=y_i \\ x,u_{y_i} \in \{0, 3\}}}&
		\multicolumn{1}{c}{\substack{y=y_i \\ x,u_{y_i} \in \{2, 3\}}}&
		\multicolumn{1}{c}{\substack{y=y_i \\ x,u_{y_i} \in \{1, 3\}}}&
		\multicolumn{1}{c}{\substack{y=y_i \\ x,u_{y_i} \in \{1, 2\}}}
		\end{array}\right]
		\begin{array}{l}
		\MyLBrace{2ex}{$v_{y_i}=0$} \\
		\MyLBrace{2ex}{$v_{y_i}=1$} \\
		\MyLBrace{2ex}{$v_{y_i}=2$} \\
		\MyLBrace{2ex}{$v_{y_i}=3$} \\
		\ \\ \ \\
		\end{array}$}
	\eean
	Number of rows in $M_{i, d}$ is equal to $4dp_{i-1} = dp_i$ and number of columns in $M_{i, d}$ is equal to $4 \mu_{i-1}p_{i-1} + 12p_{i-1} = 4p_{i-1}(\mu_{i-1}+3) = p_i \mu_i$.

We prove the Lemma~\ref{lem:indMinv} case by case depending on the value of $e_{y_i} \in \{1, 2, 3\}$. 
\subsection{Proof of Lemma ~\ref{lem:indMinv} for $e_{y_i}=1$}\label{subsec:app_1}
We will first prove that the determinant of $M_{i, \mu_i}$ has certain factors in Lemma~\ref{lem:coup_poly}. This particular lemma holds for $e_{y_i} \in \{1, 2, 3\}$. We use this to prove that $M_{i, \mu_i}$ is invertible given $M_{i-1, \mu_{i-1}}$ is invertible when $e_{y_i}=1$. Proving  this implies that $H_{E, {\tiny \uz}}^{\tiny \text{Red}}$ is invertible for $s=2$ case completing the MSR property proof for the case when $s=2$ i.e., $d=k+1$.

\blem\label{lem:coup_poly}
$f_{coup}(\theta)$ divides $|M_{i,\mu_i}|$ for all $\theta \in \{\theta_{x_i, y_i, x_i'} \ | \ x_i,x_i' \in S_{y_i}, x_i \ne x_i' \}$ ($S_{y_i}$ is defined in equation \eqref{eq:qezcomp} and $M_{i, \mu_i}$ defined in Definition~\ref{defn:M}) where 
\bean
f_{coup}(\theta) = \prod\limits_{j=1}^{i-1} \prod\limits_{\substack{x_j, x_j' \in S_{y_j} \\ x_j \ne x_j'}} \left(\theta-\theta_{x_j,y_j,x_j'} \right)^{\frac{p_{i-1}}{e_{y_j}+1}}.
\eean
\elem
\bprf
We use the Lemma~\ref{lem:poly1} to prove the current lemma. Determinant of matrix $M_{i, \mu_i}$ can be expressed as a polynomial in $\theta_{x_i, y_i, x_i'}$ where $x_i, x_i' \in S_{y_i}$. $\theta_{x_i, y_i, x_i'}$ appears in columns of $M_{i, \mu_i}$ indexed by:
\bean
\{ [x_i, y_i, \uu] \mid \uu \in Q^i(E, \uz), u_{y_i} = x_i' \}.
\eean
The number of columns where $\theta_{x_i, y_i, x_i'}$ appears is equal to $\frac{p_i}{|S_{y_i}|} = p_{i-1}$. Let $j \in [i-1]$, $x_j, x_j' \in S_{y_j}$ with $x_j \ne x_j'$. To show that $(\theta_{x_i, y_i, x_i'} - \theta_{x_j, y_j, x_j'})$ is a factor with multiplicity $\frac{p_{i-1}}{(e_{y_j}+1)}$, we consider $\theta_{x_i, y_i, x_i'}$ in columns 
$[x_i, y_i, \uu], [x_i, y_i, \uu(x_j \rightarrow u_{y_j})] $ to be equal to $\phi_{\uu}$ for all $\uu \in C$ where $C = \{ \uw \in Q^i(E, \uz) \mid w_{y_i} = x_i', w_{y_j} = x_j' \}$. We will now show that on substituting $\phi_{\uu} = \theta_{x_j, y_j, x_j'}$ in columns $[x_i, y_i, \uu], [x_i, y_i, \uu(x_j \rightarrow u_{y_j})]$ for any $\uu \in C$, we get the determinant of the matrix to be $0$. From Lemma~\ref{lem:poly1} it follows that $(\theta_{x_i, y_i, x_i'} - \theta_{x_j, y_j, x_j'})$ is a factor with multiplicity $|C| = \frac{p_i}{|S_{y_i}| |S_{y_j}|} = \frac{p_{i-1}}{(e_{y_j}+1)}$ implying the lemma.

This columns indexed by $[x_i,y_i, \uu],[x_i,y_i, \uu(x_j \rightarrow u_{y_j})], [x_j, y_j, \uu],   [x_j, y_j, \uu(x_i \rightarrow u_{y_i})]$ have support only in rows indexed by $[\ell, \uv]$ where $\uv \in \{\uu, \uu(x_i \rightarrow u_{y_i}), \uu(x_j \rightarrow u_{y_j}), (\uu(x_j \rightarrow u_{y_j}))(x_i \rightarrow u_{y_i}) \}$, $\ell \in [\mu_i]$ as shown below:
\bean
\scalebox{0.85}{$
	\left[\begin{array}{cccc}
	\underline{\phi_{\uu}}^{\mu_i} & \underline{\theta_{x_j, y_j, x_j'}}^{\mu_i} & &\\ 
	\gamma_{x_i',x_i}\underline{\phi_{\uu}}^{\mu_i} & & & \underline{\theta_{x_j, y_j, x_j'}}^{\mu_i}\\
	& \gamma_{x_j',x_j} \underline{\theta_{x_j, y_j, x_j'}}^{\mu_i} &  \underline{\phi_{\uu}}^{\mu_i} &\\
	& & \gamma_{x_i',x_i} \underline{\phi_{\uu}}^{\mu_i} & \gamma_{x_j',x_j}\underline{\theta_{x_j, y_j, x_j'}}^{\mu_i} \\
	\multicolumn{1}{c}{${\upbracefill}$} & \multicolumn{1}{c}{${\upbracefill}$}& \multicolumn{1}{c}{${\upbracefill}$}&
	\multicolumn{1}{c}{${\upbracefill}$}\\[-1ex]
	\multicolumn{1}{c}{[x_i,y_i, \uu]} & \multicolumn{1}{c}{[x_j,y_j, \uu]}& \multicolumn{1}{c}{[x_i,y_i,\uu(x_j \rightarrow u_{y_j})]}&
	\multicolumn{1}{c}{[x_j,y_j,\uu(x_i \rightarrow u_{y_i})]}\\
	\end{array}\right]
	\begin{array}{l}
	\MyLBrace{2.2ex}{$\uv=\uu$} \\
	\MyLBrace{2.2ex}{$\uv=\uu( x_i \rightarrow u_{y_i})$} \\
	\MyLBrace{2.2ex}{$\uv=\uu(x_j \rightarrow u_{y_j})$} \\
	\MyLBrace{2.2ex}{$\uv=(\uu(x_i \rightarrow u_{y_i}))( x_j \rightarrow u_{y_j})$} \\
	\ \\ \ \\
	\end{array}$}
\eean
where $\underline{\theta}^d = [\begin{array}{cccc}
1 & \theta & \cdots & \theta^{d-1}
\end{array}]$. It can be observed that $c_1 + c_2 + \gamma_{x_j',x_j}c_3 + \gamma_{x_i',x_i} c_4 = 0$ on substituting $\phi_{\uu} = \theta_{x_j, y_j, x_j'}$ where $c_{\ell}$ is $\ell$-th column in the matrix shown above. Therefore the determinant is zero on substitution $\phi_{\uu} = \theta_{x_j, y_j, x_j'}$ for any $\uu \in C$.
\eprf

\ \\
\emph{\textbf{Proof of Lemma ~\ref{lem:indMinv} for $e_{y_i}=1$: }}
Let $S_{y_i} = \{x_0, x_1\}$. From the recursive definition shown in Section \ref{remark:recursionM} we know that:
\bean
M_{i,\mu_i}  &=& \left[\begin{array}{ccc}
	M_{i-1,\mu_i}  & & D_{i_1,i}^{\mu_i}\\
	& M_{i-1,\mu_i}  & D_{i_1,i}^{\mu_i} \Psi\\
	\multicolumn{1}{c}{${\upbracefill}$} &
	\multicolumn{1}{c}{${\upbracefill}$} & \multicolumn{1}{c}{${\upbracefill}$}\\
	\multicolumn{1}{c}{\substack{y\ne y_i \\ u_{y_i}=x_0}} &
	\multicolumn{1}{c}{\substack{y\ne y_i \\  u_{y_i}=x_1}} & \multicolumn{1}{c}{\substack{y=y_i \\ x,u_{y_i} \in \{x_0, x_1\}}}\\
\end{array}\right]
\begin{array}{l}
	\MyLBrace{2ex}{$v_{y_i}=x_0$} \\
	\MyLBrace{2ex}{$v_{y_i}=x_1$} \\
	\ \\ \ \\
\end{array} \ \\ && \ \ \ \text{ where } i_1 = \begin{cases}
	1 & S_{y_i} = \{0,1\} \text{ or } \{2,3\}\\
	2 & S_{y_i} = \{0,2\} \text{ or } \{1,3\}\\
	3 & S_{y_i} = \{0,3\} \text{ or } \{1,2\}\\
\end{cases} (\text{due to } \eqref{eq:thetaassign})
\eean
We will prove factors of determinant of $M_{i,\mu_i}$ (considering this determinant as polynomial in $\lambda$'s) corresponding to $\lambda_{i_1,y_i}$, that can match up to maximum degree $2(\mu_i-1)p_{i-1}$ it can take. Hence determinant of $M_{i,\mu_i}$ can be written as product of these factors involving $\lambda_{i_1,y_i}$ and  polynomial in rest of the $\lambda$'s. Now we substitute $\lambda_{i_1,y_i}=0$ in $M_{i,\mu_i}$ and show that resulting determinant is non-zero given $|M_{i-1,\mu_{i-1}}| \ne 0$  which implies that the factor of determinant of $M_{i,\mu_i}$ corresponding to polynomial in rest of the $\lambda$'s is non zero. First, we show factors that can account to a degree of $2p_{i-1}(\mu_i - 1)$. 

From Lemma \ref{lem:coup_poly} for $\theta \in \{\theta_{x_0, y_i, x_1}, \theta_{x_1, y_i, x_0}\} = \{ \lambda_{i_1, y_i}, \gamma \lambda_{i_1, y_i} \}$, $f_{coup}(\theta)$ divides $|M_{i, \mu_i}|$. Therefore, \bean
f_{coup}(\lambda_{i_1,y_i})f_{coup}(\gamma \lambda_{i_1,y_i}) \text{ divides } |M_{i, \mu_i}|, \text{ where } 
f_{coup}(\theta) = \prod\limits_{j=1}^{i-1} \prod\limits_{\substack{x_j, x_j' \in S_{y_j} \\ x_j \ne x_j'}} \left(\theta-\theta_{x_j,y_j,x_j'} \right)^{\frac{p_{i-1}}{e_{y_j}+1}}.
\eean
This amounts to degree equal to $2\sum\limits_{j=1}^{i-1} e_{y_j} (e_{y_j}+1) \frac{p_{i-1}}{e_{y_j}+1} = 2p_{i-1}\mu_{i-1} = 2p_{i-1}(\mu_i - e_{y_i})=2p_{i-1}(\mu_i - 1)$. We have all the factors of $\lambda_{i_1,y_i}$ that accounts to the degree $2p_{i-1}(\mu_i-1)$. Therefore, 
\bean
|M_{i,\mu_i}| = f_{coup}(\lambda_{i_1,y_i})f_{coup}(\gamma \lambda_{i_1,y_i})c,
\eean 
$c$ is not a function of $\lambda_{i_1,y_i}$.
We will now show that $|M_{i,\mu_i}|$ when evaluated at $\lambda_{i_1,y_i}=0$ is non-zero. This will imply that:
\bean
|M_{i,\mu_i}|_{\{\lambda_{i_1,y_i} = 0\}} &=& cf_{coup}^2(0) \ne 0,
\eean
indicating that $c \ne 0$ as $f_{coup}(0) \ne 0$ by choice of $\lambda$'s and therefore $|M_{i, \mu_i}| \ne 0$. Recall that:
\bean
M_{i,\mu_i} = \left[\begin{array}{ccc}
	M_{i-1,\mu_i} & & D_{i_1,i}^{\mu_i}\\
	& M_{i-1,\mu_i} & D_{i_1,i}^{\mu_i} \Psi
\end{array}\right]
\eean
On substitution of $\lambda_{i_1,y_i} = 0$:
\bean 
D_{i_1,i}^{\mu_i}  = \underbrace{\left[\begin{array}{ccccc}
		v_1 & v_1 & & &  \\
		& & \ddots & & \\
		& & &  v_1 & v_1\\
	\end{array}\right]}_{(\mu_i p_{i-1} \times 2p_{i-1})} \text{ where } v_1 = \underbrace{\left[\begin{array}{c}
		1 \\ 0 \\ \vdots \\ 0
	\end{array}\right]}_{(\mu_i \times 1)}.
\eean
By doing column operations on the matrix $M_{i,\mu_i}$, we can remove the rows corresponding to non-zero entries of $\left[\begin{array}{c}
D_{i_1,i}^{\mu_i}\\
D_{i_1,i}^{\mu_i}\Psi
\end{array}\right]$ while calculating the determinant and the $2 p_{i-1}$ columns of $\left[\begin{array}{c}
D_{i_1,i}^{\mu_i}\\
D_{i_1,i}^{\mu_i}\Psi
\end{array}\right]$ with an effect of the factor $(1-\gamma)^{p_{i-1}}$. In the resultant matrix,  any column $[x,y,\uu]$ has $\theta_{x,y,u_y}$ as a factor as we removed the first row corresponding to $\underline{\theta_{x,y,u_y}}^{\mu_i}$. We therefore get:
\bean
|M_{i,\mu_i}|_{\{\lambda_{i_1,y_i}=0\}} &=& (1-\gamma)^{p_{i-1}} \left( \prod\limits_{\substack{\uu \in Q^{i-1}(E, {\tiny \uz}),\\ (x,y) \in E_{2, {\tiny \uu}}, y \le y_{i-1}}} \theta_{x,y,u_y}\right)^2 \left|\begin{array}{cc}
	M_{i-1,\mu_{i-1}} & \\
	& M_{i-1,\mu_{i-1}}
\end{array}\right|\\ 
&=&(1-\gamma)^{p_{i-1}} \left( \prod\limits_{j=1}^{i-1}\prod\limits_{\substack{x,x' \in S_{y_j}\\ x \ne x'}}\left(\theta_{x,y_j,x'}\right)^{\frac{2p_{i-1}}{e_{y_j}+1}} \right) |M_{i-1, \mu_{i-1}}|^2\\
&=& (1-\gamma)^{p_{i-1}} f_{coup}^2(0)|M_{i-1,\mu_{i-1}}|^2 = c f_{coup}^2(0).
\eean

Therefore, $|M_{i,\mu_i}| = |M_{i-1,\mu_{i-1}}|^2 (1-\gamma)^{p_{i-1}} f_{coup}(\lambda_{i_1,y_i})f_{coup}(\gamma \lambda_{i_1,y_i}) \ne 0$ and hence invertible.
\eprf

\subsection{Proof of Lemma ~\ref{lem:indMinv} for $e_{y_i}=2$}\label{subsec:app_2}
Before proving that $M_{i,\mu_i}$ is invertible given $M_{i-1, \mu_{i-1}}$ is invertible, we first show certain factors of $|M_{i,\mu_i}|$ in the following lemma for the case when $e_{y_i} = 2$. We will use this along with Lemma~\ref{lem:coup_poly} to prove the invertibility of $M_{i,\mu_i}$ for the case when $e_{y_i}=2$. Note that this along with the proof for the case when $e_{y_i} = 1$ imply that $H_{E, {\tiny \uz}}^{\tiny \text{Red}}$ is invertible for $s = 3$ proving the MSR property for $s=3$ i.e., $d=k+2$.
 
\blem\label{lem:q3factors} For the case when $e_{y_i} = 2$, $f_{base}(\lambda_{i_2,y_i})$ divides $det(M_{i,\mu_i})= |M_{i,\mu_i}|$ where
\bean
f_{base}(\theta) =  \left( (\theta - \lambda_{i_1,y_i})  (\theta - \lambda_{i_3,y_i}) \right)^{p_{i-1}} \text{ and } (i_1, i_2, i_3) = \begin{cases}
	(1, 2, 3 ) & S_{y_i} = \{0,1,2\}\\
	(1, 3, 2 ) & S_{y_i} = \{0,1,3\}\\
	(2, 3, 1 ) & S_{y_i} = \{0,2,3\}\\
	(3, 2, 1 ) & S_{y_i} = \{1,2,3\}
\end{cases}
\eean
\elem
\bprf
The proof shown here in similar to Lemma~\ref{lem:coup_poly}. Let $S_{y_i} = \{x_0, x_1, x_2\}$ and $x_0 < x_1 < x_2$. We will first show $(\lambda_{i_2,y_i} -\lambda_{i_1,y_i})^{p_{i-1}}$ divides $M_{i,\mu_i}$. To do this we substitute $\lambda_{i_2,y_i} = \lambda_{i_1,y_i}$ in any $p_{i-1}+1$ columns out of the $2p_{i-1}$ columns where $\lambda_{i_2, y_i}$ appears. The set of columns where $\lambda_{i_2, y_i}$ appears is given by $C_2 = \{(x_0,y_i;\uu(x_2 \rightarrow u_{y_i})), (x_2,y_i;\uu(x_0 \rightarrow u_{y_i})) \mid \uu \in Q^{i-1}(E,\uz)\}$ by Section~\ref{remark:recursionM}. Let the set of the selected $p_{i-1}+1$ columns be ${\cal P} \subset C_2$. We will show that upon substitution the determinant $|M_{i,\mu_i}|$ is zero by showing that a non-zero vector exists in the null space of $M_{i, \mu_i}$. $M_{i, \mu_i}$ can be recursively expressed in the following form when $e_{y_i} = 2$.
\bean
M_{i,\mu_i} = \left[
\begin{array}{cccccc}
	M_{i-1,\mu_i} & & & D_{i_1,i}^{\mu_i} & \underbrace{D_{i_2,i}^{\mu_i}}_{(\mu_i p_{i-1} \times 2 p_{i-1})} &\\
	& M_{i-1,\mu_i} & & \underbrace{D_{i_1,i}^{\mu_i}\Psi}_{(\mu_i p_{i-1} \times 2 p_{i-1})} & & D_{i_3,i}^{\mu_i}\\
	& & \underbrace{M_{i-1,\mu_i}}_{( \mu_i p_{i-1} \times \mu_{i-1} p_{i-1})} & & D_{i_2,i}^{\mu_i}\Psi & \underbrace{D_{i_3,i}^{\mu_i} \Psi}_{(\mu_i p_{i-1} \times 2 p_{i-1})}\\
	& & & \multicolumn{1}{c}{${\upbracefill}$} &
	\multicolumn{1}{c}{${\upbracefill}$} &\\
	& & & \multicolumn{1}{c}{C_1} & \multicolumn{1}{c}{C_2} & 
\end{array}\right]
\eean
Now we add these (${\cal P}$) columns to corresponding columns in $C_1$ resulting in an equivalent matrix $M'$ (same determinant). Let $M(:, [x,y,{\uu}])$ be the column at index $[x,y,\uu]$  in matrix $M_{i, \mu_i}$ and let $M'(:, [x,y,{\uu}])$ be a column in matrix $M'$. The column operations to result in $M'$ are given by:
\bean
\scalebox{0.95}{$M'(:, [x,y,{\uu}]) = \begin{cases}
	M(:, [x,y,\uu]) + M(:, [x_0,y,\uu(x_1 \rightarrow u_y)])
	& [x,y,\uu] \in {\cal P}, u_{y_i} = x_2, x =x_0, y=y_i\\
	M(:, [x, y, \uu]) + M(:, [x_1,y,\uu(x_0 \rightarrow u_y)])
	& [x,y,\uu] \in {\cal P}, u_{y_i} = x_0, x =x_2, y=y_i\\
	M(:, [x, y, \uu]) & \text{otherwise}
	\end{cases}$}
\eean

This implies that after setting $\lambda_{i_2, y_i} = \lambda_{i_1, y_i}$ in columns given by ${\cal P}$ we get:
\bean
|M'|_{\lambda_{i_2,y_i} = \lambda_{i_1,y_i} \text{ in columns given by } {\cal P}} = \left| \begin{array}{ccccccc}
	M_{i-1,\mu_i} & & & & &  D_{i_1,i}^{\mu_i} & D_2' \\
	&  M_{i-1,\mu_i} & & D_1'\Psi' & D_{i_3,i}^{\mu_i} & D_{i_1,i}^{\mu_i}\Psi &\\
	&  & M_{i-1,\mu_i} & \underbrace{D_1'\Psi'}_{\cal P} & D_{i_3,i}^{\mu_i}\Psi & & \underbrace{D_2'\Psi''}_{C_2 \setminus {\cal P}}
\end{array}\right|
\eean
where 
$\left[\begin{array}{c}
D_2'\\
\\
D_2'\Psi''
\end{array}
\right]$ is a submatrix of $\left[\begin{array}{c}
D_{i_2,i}^{\mu_i}\\
\\
D_{i_2,i}^{\mu_i}\Psi
\end{array}
\right]$ that contains columns from $C_2 \setminus {\cal P}$ and the submatrix $\left[\begin{array}{c}
\\
D_1'\Psi'\\
D_1'\Psi'
\end{array}
\right]$ is the set of $p_{i-1}+1$ columns $\cal P$ after the column operations.

Consider a vector $(\uf_1, \uf, \uf)$ where, $\uf_1, \uf \in \fq^{\mu_i p_{i-1}}$ are any vectors in the left null space of $M_{i-1,\mu_i}$ such that $\uf$ vector satisfies the conditions: 
\bea
\label{eq:fcond}
\uf &=& (f_{\ell,\uu} \mid \ell \in [\mu_i], \uu \in Q^{i-1}(E, \uz)), \ f_{\uu}(\gamma \lambda_{3,y_i}) = 0 \\ && \ \ \ \ \text{ for all } \uu \in Q^{i-1}(E,\uz), \text{ where } f_{\uu}(x) = \sum\limits_{\ell=1}^{\mu_i} f_{\ell,\uu}x^{\ell-1} \notag
\eea
With these conditions  $(\uf_1, \uf, \uf)$ is a vector in null space of:
\bean
\left[\begin{array}{cccccc}
	M_{i-1,\mu_i} & & & & &   \\
	&  M_{i-1,\mu_i} & & D_1'\Psi' & D_{i_3, i}^{\mu_i} \\
	&  & M_{i-1,\mu_i} & D_1'\Psi' & D_{i_3, i}^{\mu_i}\Psi 
\end{array}\right]
\eean
The dimension of matrix $M_{i-1,\mu_i}$ is $(\mu_i p_{i-1} \times \mu_{i-1} p_{i-1})$. Therefore the dimension of its left null space is atleast $(\mu_i - \mu_{i-1})p_{i-1}=e_ip_{i-1}=2p_{i-1}$. But, by additional conditions that need to be satisfied by $\uf$ shown in equation \eqref{eq:fcond}, the dimension of vector space in which $\uf$ can take values reduces to $p_{i-1}$. However, $\uf_1$ can takes values in a vector space of dimension $2p_{i-1}$. Now we will show that there should be at least one non-zero vector $(\uf_1, \uf, \uf)$ that is also in the left null space of matrix:
\bean
D = \underbrace{\left[\begin{array}{cc}
		D_{i_1, i}^{\mu_i} & D_2'\\
		D_{i_1, i}^{\mu_i}\Psi & \\
		& D_2'\Psi''
	\end{array}\right]}_{(3\mu_i p_{i-1} \times 3p_{i-1}-1)}
\eean
The dimension of left null space of $D$, NS$(D)$ is atleast $3(\mu_i-1)p_{i-1}+1$. Let the space of vectors $(\uf_1,\uf,\uf)$ be ${\cal F}$. dim${\cal F} \ge 3p_{i-1}$ and dim$({\cal F} + \text{NS}(D)) \le 3\mu_i p_{i-1}$ as the vector $(\uf_1, \uf, \uf) \in \fq^{3\mu_i p_{i-1}}$. Therefore,
\bean
\text{dim}({\cal F} \cap \text{NS}(D)) &=& \text{dim}{\cal F} + \text{dim}(\text{NS}(D)) - \text{dim}({\cal F} + \text{NS}(D))\\
&\ge& 1
\eean
implying that ${\cal F} \cap \text{NS(D)} \ne \phi$ and there is non zero vector in the null space of $M_{i,\mu_i}$.


This shows that on substituting $\lambda_{i_2,y_i}=\lambda_{i_1,y_i}$ in any $p_{i-1}+1$ columns out of $2p_{i-1}$ columns, the determinant is zero. This results in $(\lambda_{i_2,y_i}-\lambda_{i_1,y_i})^{p_{i-1}}$ being a factor of $|M_{i,\mu_i}|$ by Lemma~\ref{lem:poly2}. The proof for $(\lambda_{i_2,y_i}-\lambda_{i_3,y_i})^{p_{i-1}}$ being a factor of $|M_{i,\mu_i}|$ follows in the same lines, in that case, we set $\lambda_{i_2,y_i} = \lambda_{i_3,y_i}$ in $p_{i-1}+1$ columns. This proves that $f_{base}(\lambda_{i_2,y_i})$ divides $|M_{i, \mu_i}|$.
\eprf
\ \\

\emph{\textbf{Proof of Lemma ~\ref{lem:indMinv} for $e_{y_i}=2$: }}
Let $S_{y_i} = \{x_0, x_1, x_2\}$. From the recursion defined in Section \ref{remark:recursionM}:
\bean
M_{i,\mu_i} = \left[\begin{array}{cccccc}
	M_{i-1, \mu_i}  & & & D_{i_1,i}^{\mu_i} & D_{i_2,i}^{\mu_i} &\\
	& M_{i-1, \mu_i}  & & D_{i_1,i}^{\mu_i}\Psi & & D_{i_3,i}^{\mu_i}\\
	& & M_{i-1, \mu_i}  & & D_{i_2,i}^{\mu_i}\Psi  & D_{i_3,i}^{\mu_i} \Psi\\
	\multicolumn{1}{c}{${\upbracefill}$} &
	\multicolumn{1}{c}{${\upbracefill}$} & \multicolumn{1}{c}{${\upbracefill}$} &
	\multicolumn{1}{c}{${\upbracefill}$} &
	\multicolumn{1}{c}{${\upbracefill}$} &
	\multicolumn{1}{c}{${\upbracefill}$}\\
	\multicolumn{1}{c}{\substack{y\ne y_i \\u_{y_i}=x_0}}&
	\multicolumn{1}{c}{\substack{y\ne y_i \\u_{y_i}=x_1}}&
	\multicolumn{1}{c}{\substack{y\ne y_i \\u_{y_i}=x_2}}&
	\multicolumn{1}{c}{\substack{y=y_i \\ x,u_{y_i} \in \{x_0, x_1\}}}&
	\multicolumn{1}{c}{\substack{y=y_i \\ x,u_{y_i} \in \{x_0, x_2\}}}&
	\multicolumn{1}{c}{\substack{y=y_i \\ x,u_{y_i} \in \{x_1, x_2\}}}\\
\end{array}\right]
\begin{array}{l}
	\MyLBrace{2ex}{$v_{y_i}=x_0$} \\
	\MyLBrace{2ex}{$v_{y_i}=x_1$} \\
	\MyLBrace{2ex}{$v_{y_i}=x_2$} \\
	\ \\ \ \\
\end{array}
\eean
where $(i_1, i_2, i_3) = \begin{cases}
(1, 2, 3 ) & S_{y_i} = \{0,1,2\}\\
(1, 3, 2 ) & S_{y_i} = \{0,1,3\}\\
(2, 3, 1 ) & S_{y_i} = \{0,2,3\}\\
(3, 2, 1 ) & S_{y_i} = \{1,2,3\}
\end{cases}$ due to \eqref{eq:thetaassign}.

We will prove factors of determinant of $M_{i,\mu_i}$ corresponding to $\lambda_{i_2,y_i}$, that can match up to maximum degree $2p_{i-1}(\mu_i-1)$ it can take. Hence determinant of $M_{i,\mu_i}$ can be written as product of these factors involving $\lambda_{i_2,y_i}$ and  polynomial in rest of the $\lambda$'s. Now we substitute $\lambda_{i_2,y_i}=0$ in $M_{i,\mu_i}$ and show that resulting determinant is non-zero given $|M_{i-1,\mu_{i-1}}| \ne 0$  which implies that the factor of determinant of $M_{i,\mu_i}$ corresponding to polynomial in rest of the $\lambda$'s is non zero. 

From Lemma \ref{lem:coup_poly}, Lemma \ref{lem:q3factors} and the fact that the factors are coprime, $f_{coup}(\lambda_{i_2,y_i})f_{coup}(\gamma \lambda_{i_2,y_i})f_{base}(\lambda_{i_2,y_i})$  divides $|M_{i, \mu_i}|$. This amounts to degree equal to $2\sum\limits_{j=1}^{i-1} e_{y_j} (e_{y_j}+1) \frac{p_{i-1}}{e_{y_j}+1} + 2p_{i-1} = 2p_{i-1}(\mu_{i-1}+1) = 2p_{i-1}(\mu_i - e_{y_i}+1)=2p_{i-1}(\mu_i - 1)$. 

Hence we have all the factors involving $\lambda_{i_2,y_i}$ in the polynomial $|M_{i,\mu_i}|$. Therefore, $|M_{i,\mu_i}|$ can be written as:
\bea
|M_{i,\mu_i}| = f_{coup}(\lambda_{i_2,y_1})f_{coup}(\gamma\lambda_{i_2,y_1})f_{base}(\lambda_{i_2,y_1})c \label{Eq:det_2} 
\eea
where $c$ is a polynomial not involving $\lambda_{i_2,y_i}$. 
Now its enough to prove $c \neq 0$ for the chosen $\lambda$'s. To show that we set $\lambda_{i_2,y_i}=0$ in \eqref{Eq:det_2} and prove that the polynomial $|M_{i,\mu_i}|$ is not equal to zero for the chosen $\lambda$'s when $\lambda_{i_2,y_i}=0$. 
We will therefore, prove that $|M_{i,\mu_i}|_{\{\lambda_{i_2,y_i} = 0\}} \neq 0$. Recall that
\bean 
M_{i,\mu_i} = \left[\begin{array}{c|c|c|c|c|c}
	M_{i-1,\mu_i} &  & & D_{i_1,i}^{\mu_i} & D_{i_2,i}^{\mu_i} &\\ \hline
	& M_{i-1,\mu_i} &  & D_{i_1,i}^{\mu_i} \Psi & & D_{i_3,i}^{\mu_i} \\ \hline
	&  & M_{i-1,\mu_i} &  & D_{i_2,i}^{\mu_i} \Psi & D_{i_3,i}^{\mu_i} \Psi
\end{array}\right].
\eean
On substituting, $\lambda_{i_2, y_i} = 0$ to prove that $M_{i, \mu_i}$ is invertible, it is enough to show that the only vector in the left null space of $M_{i,\mu_i}$ is zero. On substituting $\lambda_{i_2,y_i}=0$ we have:
\bean 
D_{i_2,i}^{\mu_i}  = \underbrace{\left[\begin{array}{ccccc}
		v_1 & v_1 & & &  \\
		& & \ddots & & \\
		& & &  v_1 & v_1\\
	\end{array}\right]}_{(\mu_i p_{i-1} \times 2p_{i-1})} \text{ where } v_1 = \underbrace{\left[\begin{array}{c}
		1 \\ 0 \\ \vdots \\ 0
	\end{array}\right]}_{(\mu_i \times 1)}.
\eean
Hence by doing columnn operations, we can remove all rows corresponding to non-zero entries in $D_{i_2,i}^{\mu_i}$, $D_{i_2,i}^{\mu_i}\Psi$ from $M_{i, \mu_i}$ and all columns corresponding to $D_{i_2,i}^{\mu_i}, D_{i_2,i}^{\mu_i}\Psi$ from $M_{i,\mu_i} $ without affecting the invertibility of determinant of $M_{i,\mu_i}$ as $\gamma \neq 0,1$. This can be seen as follows:
\bean 
|M_{i,\mu_i}|_{\{\lambda_{i_2,y_i} = 0\}} &=& (1-\gamma)^{ p_{i-1}} \times \left( \prod\limits_{\substack{\uu \in Q^{i-1}(E, \uz)\\ (x, y) \in E_{2, {\tiny \uu}}, y \le y_{i-1}}} \theta_{x, y, u_y} \right)^2 \times |M'| \\ 
\text{where } M' &=& \left[\begin{array}{c|c|c|c|c}
	M_{i-1,\mu_i-1} &  & & \lambda_{i_1,y_i} D_{i_1,i}^{\mu_i-1} \Psi &\\ \hline
	& M_{i-1,\mu_i} &  &  D_{i_1,i}^{\mu_i} \Psi & D_{i_3,i}^{\mu_i} \\ \hline
	&  & M_{i-1,\mu_i-1} &  & \lambda_{i_3,y_i} D_{i_3,i}^{\mu_i-1} \Psi^2
\end{array}\right].
\eean
\bean
|M_{i,\mu_i}|_{\{\lambda_{i_2,y_i} = 0\}} &=& (1-\gamma)^{ p_{i-1}} \times \left( \prod\limits_{j=1}^{i-1} \prod\limits_{\substack{x,x' \in S_{y_j}\\ x \ne x'}} \left(\theta_{x, y, x'}\right)^{\frac{2p_i}{e_{y_j}+1}} \right) \times |M'|\\
&=& (1-\gamma)^{p_{i-1}} \times f_{coup}(0)^2 |M'|\\
&=& f_{coup}^2(0) f_{base}(0) c \ \ \text{ from equation \eqref{Eq:det_2}}\\
c &=& \frac{1}{f_{base}(0)}(1-\gamma)^{p_{i-1}} |M'|.
\eean
Hence its enough to show that the left null space of matrix $M'$ is zero to show invertibility of $M'$ and hence invertibility of $M_{i, \mu_i}$.
Let the vector in left null space of matrix $M'$ be of the form $F=[F_1, F_2, F_3]$ where $F_1=(f_{1,\ell,\uv} \mid \ell \in [\mu_i-1], \uv \in Q^{i-1}(E, \uz))$, $F_2=(f_{2,\ell,\uv}  \mid \ell \in [\mu_i], \uv \in Q^{i-1}(E, \uz))$, $F_3=(f_{3,\ell,\uv} \mid \ell \in [\mu_i-1], \uv \in Q^{i-1}(E, \uz))$ and $f_{1,\uv}(x)=\sum\limits_{\ell=1}^{\mu_i-1}f_{1,\ell,\uv}x^{\ell-1}$, $f_{2,\uv}(x)=\sum\limits_{\ell=1}^{\mu_i}f_{2,\ell,\uv}x^{\ell-1}$ and $f_{3,\uv}(x)=\sum\limits_{\ell=1}^{\mu_i-1}f_{3,\ell,\uv}x^{\ell-1}$. 
Any vector $F=[F_1,F_2,F_3]$ in the left null space of $M'$ must be such that $F'=[F_1,F_2]$ is in left null space of :
\bean
M'' = \left[\begin{array}{c|c|c}
	M_{i-1,\mu_i-1} &  & \lambda_{i_1,y_i} D_{i_1,i}^{\mu_i-1} \Psi \\ \hline
	& M_{i-1, \mu_i} &   D_{i_1,i}^{\mu_i} \Psi  \\ 
\end{array}\right].
\eean
The above matrix is a $((2\mu_i-1)p_{i-1} \times 2(\mu_i-1)p_{i-1})$ matrix. $M''$ can be shown to be of rank $2(\mu_i-1)p_{i-1}$ by showing that the matrix $M'''$ created by appending $p_{i-1}$ columns is of full rank. Let:
\bean
M'''= \left[\begin{array}{c|c|c|c|c|c}
	M_{i-1,\mu_i-1} & 0 & \lambda_{i_1,y_i} D_{i_1,i}^{\mu_i-1} \Psi & 0 & 0 & 0 \\ \hline
	0 & M_{i-1,\mu_i} &   D_{i_1,i}^{\mu_i} \Psi &  v_1 & \hdots & v_{p_{i-1}}  \\ 
\end{array}\right],
\eean
where $v_j$ is a $\mu_ip_{i-1} \times 1$ vector with $1$ at $((j-1)\mu_{i}+1)^{th}$ component and with $0$ at other components. As $v_j$'s are columns with single non-zero element, we can remove the rows corresponding to these non-zero elements and bring out the factors that are common to the columns without affecting the determinant.
\bean
|M'''|&=& (\gamma \lambda_{i_1,y_i}^2)^{p_{i-1}} \prod\limits_{j=1}^{i-1}\prod_{\substack{x,x' \in S_{y_j} \\ x \ne x'}} (\theta_{x,y_j,x'})^{\frac{p_{i-1}}{e_j+1}} \left|\begin{array}{c|c|c}
	M_{i-1,\mu_i-1} & 0 &  D_{i_1,i}^{\mu_i-1}  \\ \hline
	0 & M_{i-1,\mu_i-1} &   D_{i_1,i}^{\mu_i-1} \Psi \\ 
\end{array}\right|\\
&=& (\gamma \lambda_{i_1,y_i}^2)^{p_{i-1}} f_{coup}(0) |M_{i-1, \mu_i-2}|^2 f_{coup}(\lambda_{i_1,y_i})f_{coup}(\gamma\lambda_{i_1,y_i})  (1-\gamma)^{p_{i-1}} \ne 0
\eean
The last equality follows from the derivation for the case when $e_{y_i}=1$ presented in Section~\ref{subsec:app_1}. Hence if we produce a set of vectors {$F'=[F_1,F_2]$} forming $p_{i-1}$ dimensional left null space for $M''$ then it is the exact full left null space of $M''$ as rank of $M''$ is exactly $2(\mu_i-1)p_{i-1}$.
Let $F_1$ be vector in left null space of $M_{i-1,\mu_i-1}$. Define $F_2$, such that $f_{2,\ell,\uv} = \lambda_{i_1,y_i}f_{1,\ell,\uv}$ for all $\ell \in [\mu_i-1]$, $\uv \in Q^{i-1}(E, \uz)$ and $f_{2,\mu_i, \uv}=0$ for all $\uv \in Q^{i-1}(E, \uz)$. Now it can be seen that $[F_1,F_2]$ is in the left null space of $M''$ and we can produce such a vector $F'=[F_1, F_2]$ in left null space of $M''$ for every vector $F_1$ in left null space of $M_{i-1,\mu_i-1}$. By invertibility of $M_{i-1, \mu_{i-1}}$, the left null space of $M_{i-1, \mu_i - 1}$ has dimension $(\mu_i - 1)p_{i-1} - \mu_{i-1} p_{i-1} = (\mu_i - \mu_{i-1}-1)p_{i-1} = p_{i-1}$ and 
we have produced a $p_{i-1}$ dimensional left null space for $M''$ which is the exactly full left null space of $M''$. Now any vector in left null space of $M'$ must be of the form $[F_1,F_2,F_3]$ where $F_1,F_2$ is as described just now in the left null space of $M''$.
As $[F_1 \ F_2 \ F_3] M' = 0^T$. By looking at the last $2p_{i-1}$ columns of $M'$ we get : 
\bea
f_{2,\uv}(\gamma \lambda_{i_3,y_i}) &=& \gamma^2\lambda_{i_3,y_i} f_{3,\uv}(\gamma\lambda_{3,y_i}) \text{ for all } \uv \in Q^{i-1}(E, \uz) \label{root2} \\
f_{2,\uv}(\lambda_{i_3,y_i}) &=& \lambda_{i_3,y_i} f_{3,\uv}(\lambda_{i_3,y_i}) \label{root1} \text{ for all } \uv \in Q^{i-1}(E, \uz)
\eea
We now define for any $\uv \in Q^{i-1}(E, \uz)$ $g_{\uv}(x) = f_{2,\uv}(x)-\lambda_{i_3,y_i}f_{3,\uv}(x) = \sum\limits_{\ell=1}^{\mu_i-1}g_{\ell,\uv}x^{\ell-1}$. From equation \eqref{root1}, $g_{\uv}(x)$ has root at  $\lambda_{i_3,y_i}$.

From the definition of $F_2,F_3$, it is implied that $G=(g_{\ell,\uv} \mid \ell \in [\mu_i-1], \uv \in Q^{i-1}(E, \uz))$ is in left null space of $M_{i-1,\mu_i-1}$. We will show that the condition that $GM_{i-1,\mu_i-1}=0$ together with constraints that $g_{\uv}(\lambda_{i_3,y_i}) = 0$ will force $G$ to be a null vector. Let $g_{\uv}(x) = (x-\lambda_{i_3, y_i})g_{\uv}'(x)$ where $g_{\uv}'(x)$ is of degree $\mu_i-3$ and let $g_{\uv}'(x) = \sum\limits_{\ell=1}^{\mu_i-2} g_{\ell,\uv}'x^{\ell-1}$. $GM_{i-1,\mu_i-1}$ implies that for any $\uu \in Q^{i-1}(E, \uz)$ and $(x,y) \in E_{2,\uu}$, $y \le y_{i-1}$:
\bean
g_{\uu}(\theta_{x,y,u_y}) + \gamma_{u_y, x} g_{\uu(x \rightarrow u_{y})}(\theta_{x,y,u_y}) &=& 0\\
\implies (\theta_{x,y,u_y}-\lambda_{i_3,y_i}) (g_{\uu}'(\theta_{x,y,u_y}) + \gamma_{u_y, x} g_{\uu(x \rightarrow u_{y})}'(\theta_{x,y,u_y})) &=&  0 
\eean
By setting $G'=(g_{\ell,\uv}' \mid \ell \in [\mu_i-2], \uv \in Q^{i-1}(E, \uz))$ it is implied that:
\bea
G'M_{i-1, \mu_i-2} &=& 0 \label{eq:smallermatrixred}\\
G' &=& 0  \scalebox{0.9}{$ \ \text{as } M_{i-1, \mu_{i-1}} \text{ is invertible and } \mu_i = \mu_{i-1} + 2 \notag$}.
\eea
Therefore $G=0$ and $f_{2,\uv}(x) = \lambda_{i_3, y_i} f_{3,\uv}(x)$ for all $\uv \in Q^{i-1}(E, \uz)$. Substituting this in equation \eqref{root2} we get:
\bea
\lambda_{i_3,y_i} f_{3,\uv}(\gamma \lambda_{i_3,y_i}) &=& \gamma^2 \lambda_{i_3,y_i} f_{3,\uv}(\gamma\lambda_{i_3,y_i}) \notag\\
(1-\gamma^2)f_{3,\uv}(\gamma \lambda_{i_3,y_i})&=& 0 \label{imposs1}
\eea
Let $f_{3,\uv}(x) = (x-\gamma \lambda_{i_3,y_i})f_{3,\uv}'(x)$ for any $\uv \in Q^{i-1}(E, \uz)$ where $f_{3,\uv}'(x)$ is of degree $\mu_i-3$. $F_3$ is null space of 
$M_{i-1,\mu_i-1}$. Therefore for any $\uu \in Q^{i-1}(E, \uz)$, $(x,y) \in E_{2,\uu}$, $y \le y_{i-1}$:
\bean
f_{3,\uu}(\theta_{x,y,u_y}) + \gamma_{u_y, x} f_{3,\uu( x \rightarrow u_{y})}(\theta_{x,y,u_y}) &=& 0\\
(\theta_{x,y,u_y}-\gamma \lambda_{i_3,y_i}) (f_{3,\uu}'(\theta_{x,y,u_y}) + \gamma_{u_y, x} f_{3,\uu( x \rightarrow u_y)}'(\theta_{x,y,u_y})) &=& 0
\eean
By setting $F_3'=(f_{3,\ell,\uv}' \mid \ell \in [\mu_i-2], \uv \in Q^{i-1}(E, \uz))$ it is implied that:
\bean
F_3'M_{i-1, \mu_i-2} &=& 0 \\
F_3' &=& 0 \text{ as } M_{i-1, \mu_{i-1}} \text{ is invertible}.
\eean
This implies that $F_3 = 0$ and therefore $F_2=0$ and $F_1=0$. Hence the left null space of $M'$ has only zero vector and hence is invertible. It follows that $M_{i, \mu_i}$ is invertible.
\eprf

\subsection{Proof of Lemma ~\ref{lem:indMinv} for $e_{y_i}=3$}

Similar to the proof for the case where $e_{y_i}=2$, before proving that $M_{i,\mu_i}$ is invertible given $M_{i-1, \mu_{i-1}}$ is invertible, we first show certain factors of $|M_{i,\mu_i}|$ in the following lemma for the case when $e_{y_i} = 3$. We will use this along with Lemma~\ref{lem:coup_poly} to prove the invertibility of $M_{i,\mu_i}$ for the case when $e_{y_i}=3$. Note that this along with the proof for the case when $e_{y_i} = 1, 2$ imply that $H_{E, {\tiny \uz}}^{\tiny \text{Red}}$ is invertible for $s \le 4$.

\blem\label{lem:q4factors}
For the case when $e_{y_i} = 3$, $f_{base,2}(\lambda_{1,y_i})f_{base,3}(\lambda_{1,y_i})$ divides $|M_{i,\mu_i}|$, where
\bean
f_{base,j}(\theta)=((\theta-\lambda_{j,y_i})^2(\theta-\gamma\lambda_{j,y_i})(\gamma\theta-\lambda_{j,y_i}))^{p_{i-1}}. 
\eean
\elem
\bprf
\paragraph*{Case 1: $f_{base, 2}(\lambda_{1,y_i})$ divides $ |M_{i, \mu_i}|$} Set $\lambda_{1,y_i} = \lambda_{2,y_i}$ in columns given by $\{(0,y_i,\uu(1 \rightarrow u_{y_i})), (2,y_i,\uu(3 \rightarrow u_{y_i}))\}$ for any $\uu \in Q^{i-1}(E, \uz)$. This is because $\theta_{0,y_i, 1} = \theta_{2,y_i, 3} = \lambda_{1,y_i}$ from the $\theta$ to $\lambda$ assignment shown in equation~\eqref{eq:thetaassign}. Then we will show that the columns: 
\bean \{(0,y_i,\uu(1 \rightarrow u_{y_i})), (2,y_i,\uu(3 \rightarrow u_{y_i})), (0,y_i,\uu(2 \rightarrow u_{y_i})), (1,y_i,\uu(3 \rightarrow u_{y_i}))\},
\eean 
are linearly dependent. It can be seen that the rows where they have non zero elements is indexed by $[\ell, \uv], \ell \in [\mu_i], \uv \in \{\uu(0 \rightarrow u_{y_i}), \uu(1 \rightarrow u_{y_i}), \uu(2 \rightarrow u_{y_i}), \uu(3 \rightarrow u_{y_i}) \}$ and restricted to these rows the columns are as shown below:
\bean
\scalebox{0.9}{$
	\left[\begin{array}{c|c|c|c}
	\gamma_{1,0} \underline{\theta_{0,y_i, 1}}^{\mu_i} & &  \gamma_{2,0}\underline{\theta_{0,y_i, 2}}^{\mu_i} &\\ 
	\underline{\theta_{0,y_i, 1}}^{\mu_i} & & & \gamma_{3,1}\underline{\theta_{1,y_i, 3}}^{\mu_i}\\ 
	& \gamma_{3,2}\underline{\theta_{2,y_i,3}}^{\mu_i} & \underline{\theta_{0, y_i, 2}}^{\mu_i} &\\ 
	& \underline{\theta_{2,y_i,3}}^{\mu_i} & & \underline{\theta_{1,y_i,3}}^{\mu_i} \\
	\multicolumn{1}{c}{${\upbracefill}$} &
	\multicolumn{1}{c}{${\upbracefill}$} & \multicolumn{1}{c}{${\upbracefill}$} &
	\multicolumn{1}{c}{${\upbracefill}$} \\
	\multicolumn{1}{c}{ (0, y_i, \uu(1 \rightarrow u_{y_i}))}&
	\multicolumn{1}{c}{ (2, y_i, \uu(3 \rightarrow u_{y_i}))}&
	\multicolumn{1}{c}{(0, y_i, \uu(2 \rightarrow u_{y_i}))}&
	\multicolumn{1}{c}{ (1, y_i, \uu(3 \rightarrow u_{y_i}))}
	\end{array}\right]
	\begin{array}{l}
	\MyLBrace{2ex}{$v_{y_i}=0$} \\
	\MyLBrace{2ex}{$v_{y_i}=1$} \\
	\MyLBrace{2ex}{$v_{y_i}=2$} \\
	\MyLBrace{2ex}{$v_{y_i}=3$} \\
	\ \\ \ \\
	\end{array}$}\\  \scalebox{0.9}{$= \left[\begin{array}{c|c|c|c}
	\underline{\lambda_{1,y_i}}^{\mu_i} & &  \underline{\lambda_{2,y_i}}^{\mu_i} &\\ 
	\underline{\lambda_{1,y_i}}^{\mu_i} & & & \underline{\lambda_{2,y_i}}^{\mu_i}\\ 
	& \underline{\lambda_{1,y_i}}^{\mu_i} & \underline{\lambda_{2, y_i}}^{\mu_i} &\\ 
	& \underline{\lambda_{1,y_i}}^{\mu_i} & & \underline{\lambda_{2,y_i}}^{\mu_i}
	\end{array} \right] \text{(by } \theta \text{ to } \lambda \text{ assignment from equation \eqref{eq:thetaassign})}$}
\eean
It is clear to see that on setting $\lambda_{1, y_i} = \lambda_{2, y_i}$ the columns in matrix shown above sum to zero. This is true for any $\uu \in Q^{i-1}(E, \uz)$ resulting in $(\lambda_{1,y_i} - \lambda_{2,y_i})^{p_{i-1}}$ being a factor of $|M_{i, \mu_i}|$. 

Now set $\lambda_{1,y_i} = \lambda_{2,y_i}$ in the columns given by $\{(1,y_i,\uu(0 \rightarrow u_{y_i})), (3,y_i,\uu(2 \rightarrow u_{y_i}))\}$ for a fixed $\uu \in Q^{i-1}(E, \uz)$. Then we show that the columns 
\bean
\{(1,y_i,\uu(0 \rightarrow u_{y_i})), (3,y_i,\uu(2 \rightarrow u_{y_i})), (2,y_i,\uu(0 \rightarrow u_{y_i})), (3,y_i,\uu(1 \rightarrow u_{y_i})) \} 
\eean are linearly dependent. It can be seen that the rows where they have non zero elements is indexed by $[\ell, \uv], \ell \in [\mu_i], \uv \in \{\uu(0 \rightarrow u_{y_i}), \uu(1 \rightarrow u_{y_i}), \uu(2 \rightarrow u_{y_i}),\uu(3 \rightarrow u_{y_i}) \}$ and restricted to these rows the columns are as shown below:
\bean
\scalebox{0.9}{$\left[\begin{array}{c|c|c|c}
	\underline{\gamma\lambda_{1,y_i}}^{\mu_i} & &  \underline{\gamma \lambda_{2,y_i}}^{\mu_i} &\\ 
	\gamma \underline{\gamma\lambda_{1,y_i}}^{\mu_i} & & & \underline{\gamma\lambda_{2,y_i}}^{\mu_i}\\ 
	& \underline{\gamma\lambda_{1,y_i}}^{\mu_i} & \gamma \underline{\gamma\lambda_{2, y_i}}^{\mu_i} &\\ 
	& \gamma \underline{\gamma\lambda_{1,y_i}}^{\mu_i} & & \gamma
	\underline{\gamma\lambda_{2,y_i}}^{\mu_i} 
	\end{array} \right]$}
\eean
It is clear to see that on setting $\lambda_{1, y_i}=\lambda_{2, y_i}$, $c_1 + \gamma c_2 + c_3 + \gamma c_4=0$ where $c_i$ is $i$-th column of the matrix shown above. This is true for any $\uu \in Q^{i-1}(E, \uz)$ resulting in a factor of  $(\lambda_{1,y_i} - \lambda_{2,y_i})^{p_{i-1}}$. Therefore $(\lambda_{1,y_i} - \lambda_{2,y_i})^{2p_{i-1}}$ divides $|M_{i,\mu_i}|$.

We now set $\lambda_{1,y_i}=\gamma \lambda_{2,y_i}$ in columns indexed by $(0,y_i,\uu(1 \rightarrow u_{y_i})), (2,y_i,\uu(3 \rightarrow u_{y_i}))$ for some $\uu \in Q^{i-1}(E, \uz)$. Then we show that the columns $\{(0,y_i,\uu(1 \rightarrow u_{y_i})), (2,y_i,\uu(3 \rightarrow u_{y_i})), (2,y_i,\uu(0 \rightarrow u_{y_i})), (3,y_i,\uu(1 \rightarrow u_{y_i}))\}$ are linearly dependent. It can be seen that the rows where they have non zero elements is indexed by $[\ell, \uv], \ell \in [\mu_i], \uv \in \{\uu(x \rightarrow u_{y_i}) \ | \ x \in [0, 3]  \}$ and restricted to these rows the columns are as shown below:
\bean
\scalebox{0.9}{$ \left[\begin{array}{c|c|c|c}
	\underline{\lambda_{1,y_i}}^{\mu_i} & &  \underline{\gamma\lambda_{2,y_i}}^{\mu_i} &\\ 
	\underline{\lambda_{1,y_i}}^{\mu_i} & & & \underline{\gamma\lambda_{2,y_i}}^{\mu_i}\\ 
	& \underline{\lambda_{1,y_i}}^{\mu_i} & \gamma  \underline{\gamma \lambda_{2, y_i}}^{\mu_i} &\\ 
	& \underline{\lambda_{1,y_i}}^{\mu_i} & & \gamma \underline{\gamma \lambda_{2,y_i}}^{\mu_i} 
	\end{array} \right]$}
\eean
It is clear to see that $c_1 + \gamma c_2 + c_3 + c_4=0$ where $c_i$ is $i$-th column of the matrix shown above. This is true for any $\uu \in Q^{i-1}(E, \uz)$ resulting in a factor of  $(\lambda_{1,y_i} - \gamma \lambda_{2,y_i})^{p_{i-1}}$.

We similarly set $\lambda_{2,y_i} = \gamma \lambda_{1,y_i}$ in columns $(0,y_i,\uu(2 \rightarrow u_{y_i})), (1,y_i,\uu(3 \rightarrow u_{y_i}))$. Then we can show that the columns $(0,y_i,\uu(2 \rightarrow u_{y_i})), (1,y_i,\uu(3 \rightarrow u_{y_i})), (1,y_i,\uu(0 \rightarrow u_{y_i})), (3,y_i,\uu(2 \rightarrow u_{y_i}))$ are linearly dependent. The rows where these columns have non zero elements is indexed by $[\ell, \uv], \ell \in [\mu_i], \uv \in \{\uu(x \rightarrow u_{y_i}) \ | \ x \in [0, 3]  \}$ and restricted to these rows the columns are as shown below:
\bean
\scalebox{0.9}{$ \left[\begin{array}{c|c|c|c}
	\underline{\lambda_{2,y_i}}^{\mu_i} & &  \underline{\gamma\lambda_{1,y_i}}^{\mu_i} &\\ 
	&  \underline{\lambda_{2,y_i}}^{\mu_i} & \gamma \underline{\gamma\lambda_{1,y_i}}^{\mu_i} & \\ 
	\underline{\lambda_{2,y_i}}^{\mu_i} & &  & \underline{\gamma \lambda_{1, y_i}}^{\mu_i} \\ 
	& \underline{\lambda_{2,y_i}}^{\mu_i} & & \gamma \underline{\gamma \lambda_{1,y_i}}^{\mu_i} 
	\end{array} \right]$}
\eean
It is clear to see that $c_1 + \gamma c_2 + c_3 + c_4=0$ where $c_i$ is $i$-th column of the matrix shown above. This is true for any $\uu \in Q^{i-1}(E, \uz)$ resulting in a factor of  $(\lambda_{2,y_i} - \gamma \lambda_{1,y_i})^{p_{i-1}}$.
Therefore $f_{base, 2}(\lambda_{1,y_i})$ divides $|M_{i,\mu_i}|$.

\paragraph*{Case 2: $f_{base, 3}(\lambda_{1,y_i})$ divides $ |M_{i, \mu_i}|$} Set $\lambda_{1,y_i} = \lambda_{3,y_i}$ in columns given by $\{(0,y_i,\uu(1 \rightarrow u_{y_i})), (2,y_i,\uu(3 \rightarrow u_{y_i}))\}$ for any $\uu \in Q^{i-1}(E, \uz)$. Then we will show that the columns 
\bean
(0,y_i,\uu(1 \rightarrow u_{y_i})), (2,y_i,\uu(3 \rightarrow u_{y_i})), (0,y_i,\uu(3 \rightarrow u_{y_i})), (1,y_i, \uu(2 \rightarrow u_{y_i}))
\eean
are linearly dependent. It can be seen that the rows where they have non zero elements is indexed by $[\ell, \uv], \ell \in [\mu_i], \uv \in \{\uu(x \rightarrow u_{y_i}) \ | \ x \in [0, 3]  \}$ and restricted to these rows the columns are as shown below:
\bean
\scalebox{0.8}{$ \left[\begin{array}{c|c|c|c}
	\underline{\lambda_{1,y_i}}^{\mu_i} & &  \underline{\lambda_{3,y_i}}^{\mu_i} &\\ 
	\underline{\lambda_{1,y_i}}^{\mu_i} & & & \underline{\lambda_{3,y_i}}^{\mu_i}\\ 
	& \underline{\lambda_{1,y_i}}^{\mu_i} & & \underline{ \lambda_{3, y_i}}^{\mu_i} \\ 
	& \underline{\lambda_{1,y_i}}^{\mu_i} & \underline{ \lambda_{3,y_i}}^{\mu_i}  &
	\end{array} \right]$}
\eean
It is clear to see that $c_1 +  c_2 + c_3 + c_4=0$ where $c_i$ is $i$-th column of the matrix shown above. This is true for any $\uu \in Q^{i-1}(E, \uz)$ resulting in a factor of  $(\lambda_{1,y_i} - \gamma \lambda_{3,y_i})^{p_{i-1}}$.
%
Set $\lambda_{1,y_i} = \lambda_{3,y_i}$ in columns given by $\{(1,y_i,\uu(0 \rightarrow u_{y_i})), (3,y_i,\uu(2 \rightarrow u_{y_i}))\}$ for some fixed $\uu \in Q^{i-1}(E,\uz)$. We will show that there exists a non zero vector in the null space of the matrix.
From the recursion in Section \ref{remark:recursionM} the matrix $M_{i,\mu_i}$ can be expressed as following:
\bean 
\left[\begin{array}{c|c|c|c|c|c|c|c|c|c}
	M_{i-1, \mu_i} &  & & & D_{1,i}^{\mu_i} & D_{2,i}^{\mu_i} & D_{3,i}^{\mu_i} & & &\\ \hline
	& M_{i-1, \mu_i} & & & D_{1,i}^{\mu_i} \Psi & & & & D_{2,i}^{\mu_i} & D_{3,i}^{\mu_i} \\ \hline
	&  & M_{i-1, \mu_i} &  & & D_{2,i}^{\mu_i} \Psi & & D_{1,i}^{\mu_i} & & D_{3,i}^{\mu_i} \Psi\\ \hline
	&  &  & M_{i-1, \mu_i} & &  & D_{3,i}^{\mu_i} \Psi & D_{1,i}^{\mu_i} \Psi & D_{2,i}^{\mu_i} \Psi &
\end{array}\right]
\eean
The matrix $M_{i-1,\mu_i}$ is of dimension $(\mu_ip_{i-1} \times \mu_{i-1}p_{i-1})$ and therefore has a left null space of dimension atleast $(\mu_i-\mu_{i-1})p_{i-1} = 3p_{i-1}$. Let $\underline{f} = (f_{\ell,\uv})_{\ell \in [\mu_i], \uv \in Q^{i-1}(E, {\tiny \uz})}$ be a vector in null space of $M_{i-1,\mu_i}$. We use notation $f_{\uv}(x) = \sum\limits_{j=1}^{\mu_i} f_{\ell,\uv}x^{\ell-1}$ and introduce some more conditions on the vector $\underline{f}$ such that the vector $\left[\begin{array}{cccc}
\underline{ f} & \underline{f} & \underline{f} & \underline{f}\end{array}\right]$ is in null space of $M_{i,\mu_i}$. Let $f_{\uv}(\gamma \lambda_{3,y_i}) = f_{\uv}(\gamma \lambda_{2,y_i}) = 0$ for all $\uv \in Q^{i-1}(E, \uz)$ and $f_{\uv}(\gamma\lambda_{1,y_i})=0$ for all $\uv \in Q^{i-1}(E, \uz) \setminus \{\uu\}$. These conditions will ensure that  $\left[\begin{array}{cccc}
\underline{f} & \underline{f} & \underline{f} & \underline{f}\end{array}\right]$ is in null space of $M_{i,\mu_i}$ as we are setting $\lambda_{1,y_i} = \lambda_{3,y_i}$ in columns given by $\{(1,y_i,\uu(0 \rightarrow u_{y_i})), (3,y_i,\uu(2 \rightarrow u_{y_i}))\}$. This introduces $3p_{i-1}-1$ extra conditions on $\underline{f}$ apart from the constraint that $\underline{f}$ is in null space of $M_{i-1,\mu_i}$. Therefore, the null space is of dimension atleast $1$ implying that determinant $|M_{i,\mu_i}|$ is $0$ on setting $\lambda_{1,y_i}=\lambda_{3,y_i}$. This is true for any $\uu \in Q^{i-1}(E, \uz)$. Therefore it follows from Lemma~\ref{lem:poly1} that  $(\lambda_{1,y_i} -\lambda_{3,y_i})^{p_{i-1}}$ divides $|M_{i,\mu_i}|$. In total we have $(\lambda_{1,y_i}-\lambda_{3,y_i})^{2p_{i-1}}$ divides $M_{i,\mu_i}$. This is because we used $4 p_{i-1}$ distinct columns containing $\lambda_{1, y_i}$ while substituting it in two columns to be equal to $\lambda_{3, y_i}$ at a time.

We now set $\lambda_{3,y_i} = \gamma \lambda_{1,y_i}$ in columns given by $\{(0,y_i,\uu(3 \rightarrow u_{y_i})), (1,y_i,\uu(2 \rightarrow u_{y_i}))\}$ for some fixed $\uu \in Q^{i-1}(E, \uz)$. We will show that there exists a non zero vector in the null space of the matrix.
We use similar ideas as before. The matrix $M_{i-1,\mu_i}$ has a left null space of dimension $(\mu_i-\mu_{i-1})p_{i-1} = 3p_{i-1}$. Let $\underline{f} = (f_{\ell,\uv} \mid \ell \in [\mu_i], \uv \in Q^{i-1}(E, \uz))$ be a vector in null space of $M_{i-1,\mu_i}$. We introduce some more conditions on the vector $\underline{f}$ such that the vector $\left[\begin{array}{cccc}
\uf & \uf & \gamma^{-1} \uf &  \gamma^{-1} \uf\end{array}\right]$ is in null space of $M_{i,\mu_i}$. Let $f_{\uv}(\gamma \lambda_{1,y_i}) = f_{\uv}(\lambda_{2,y_i}) = 0$ for all $\uv \in Q^{i-1}(E,\uz)$ and $f_{\uv}(\lambda_{3,y_i})=0$ for all $\uv \in Q^{i-1}(E, \uz) \setminus \{\uu\}$. This introduces $3p_{i-1}-1$ extra conditions on $\underline{f}$ and with these conditions there exists a non zero vector of the form $\left[\begin{array}{cccc}
\uf & \uf & \gamma^{-1} \uf &  \gamma^{-1} \uf\end{array}\right]$ is in null space of $M_{i, \mu_i}$ obtained after substitution. This implies that $(\lambda_{3,y_i} - \gamma \lambda_{1,y_i})$ divides $|M_{i,\mu_i}|$. This is true for any $\uu \in Q^{i-1}(E, \uz)$. Hence, $(\gamma\lambda_{1,y_i}-\lambda_{3,y_i})^{p_{i-1}}$ is a factor of $|M_{i,\mu_i}|$.

Set $\lambda_{1,y_i} = \gamma \lambda_{3,y_i}$ in columns given by $\{(0,y_i,\uu(1 \rightarrow u_{y_i})), (2,y_i,\uu(3 \rightarrow u_{y_i}))\}$ for some fixed $\uu \in Q^{i-1}(E,\uz)$. Then we will show that the columns $(0,y_i,\uu(1 \rightarrow u_{y_i})), (2,y_i,\uu(3 \rightarrow u_{y_i})), (3,y_i,\uu(0 \rightarrow u_{y_i})), (2,y_i, \uu(1 \rightarrow u_{y_i}))$ are linearly dependent. It can be seen that the rows where they have non zero elements is indexed by $[\ell, \uv], \ell \in [\mu_i], \uv \in \{\uu(x \rightarrow u_{y_i}) \ | \ x \in [0, 3]  \}$ and restricted to these rows the columns are as shown below:
\bean
\scalebox{0.9}{$ \left[\begin{array}{c|c|c|c}
	\underline{\lambda_{1,y_i}}^{\mu_i} & &   \underline{\gamma\lambda_{3,y_i}}^{\mu_i} &\\ 
	\underline{\lambda_{1,y_i}}^{\mu_i} & & &  \underline{\gamma \lambda_{3,y_i}}^{\mu_i}\\ 
	&  \underline{\lambda_{1,y_i}}^{\mu_i} & & \gamma  \underline{ \gamma \lambda_{3, y_i}}^{\mu_i} \\ 
	& \underline{\lambda_{1,y_i}}^{\mu_i} & \gamma  \underline{ \gamma \lambda_{3,y_i}}^{\mu_i}  &
	\end{array} \right]$}
\eean
It is clear to see that $c_1 + \gamma c_2 + c_3 + c_4=0$ where $c_i$ is $i$-th column of the matrix shown above. This is true for any $\uu \in Q^{i-1}(E, \uz)$ resulting in a factor of  $(\lambda_{1,y_i} - \gamma \lambda_{3,y_i})^{p_{i-1}}$.

Hence $f_{base,3}(\lambda_{1,y_i})$ divides $|M_{i,\mu_i}|$.
\eprf

\ \\

\emph{\textbf{Proof of Lemma ~\ref{lem:indMinv} for $e_{y_i}=3$: }}
From Lemma \ref{lem:coup_poly} and Lemma \ref{lem:q4factors} \bean
f_{coup}(\lambda_{1,y_i})^2f_{coup}(\gamma \lambda_{1,y_i})^2f_{base,2}(\lambda_{1,y_i})f_{base,3}(\lambda_{1,y_i}) \text{ divides } |M_{i, \mu_i}|,
\eean 
as $\theta_{0, y_i, 1} = \theta_{2, y_i, 3} = \lambda_{1, y_i}$ and $\theta_{1,y_i,0} = \theta_{3, y_i, 2} = \gamma \lambda_{1, y_i}$. This accounts to degree $4\sum\limits_{j=1}^{i-1} e_{y_j}(e_{y_j}+1) \frac{p_{i-1}}{e_{y_j}+1}+8p_{i-1} = 4(\mu_{i-1}+2)p_{i-1}=4(\mu_i-1)p_{i-1}$.
Hence we have all the factors involving $\lambda_{1,y_i}$ in the polynomial $|M_{i,\mu_i}|$. Therefore, $|M_{i,\mu_i}|$ can be written as:
\bea
|M_{i,\mu_i}|&= & f^2_{coup}(\lambda_{1,y_i})f^2_{coup}(\gamma\lambda_{1,y_i})f_{base,2}(\lambda_{1,y_i}) f_{base,3}(\lambda_{1,y_i})c
\label{Eq:det_24}
\eea
where $c$ is a polynomial not involving $\lambda_{1,y_i}$. 
We will now show that $|M_{i,\mu_i}|$ when evaluated at $\lambda_{1,y_i}=0$ is invertible for the chosen $\lambda$'s. Given that it is true:
\bean
|M_{i,\mu_i}|_{\{\lambda_{1,y_i} = 0\}} = f^4_{coup}(0)f_{base,2}(0) f_{base,3}(0)c \neq 0\\ 
\text{We know that } f^4_{coup}(0) f_{base,2}(0) f_{base,3}(0) \neq 0\\ 
\text{Hence : } c \neq 0 \\
\text{From equation \eqref{Eq:det_24} } |M_{i,\mu_i}| \neq 0.
\eean
Therefore, we prove that $|M_{i,\mu_i}|_{\{\lambda_{1,y_i} = 0\}} \neq 0$. Recall,
\bean 
\scalebox{0.85}{$M_{i,\mu_i} = \left[\begin{array}{c|c|c|c|c|c|c|c|c|c}
	M_{i-1,\mu_i} &  & & & D_{1,i}^{\mu_i} & D_{2,i}^{\mu_i} & D_{3,i}^{\mu_i} & & &\\ \hline
	& M_{i-1,\mu_i} & & & D_{1,i}^{\mu_i} \Psi & & & & D_{2,i}^{\mu_i} & D_{3,i}^{\mu_i} \\ \hline
	&  & M_{i-1,\mu_i} &  & & D_{2,i}^{\mu_i} \Psi & & D_{1,i}^{\mu_i} & & D_{3,i}^{\mu_i} \Psi\\ \hline
	&  &  & M_{i-1,\mu_i} & &  & D_{3,i}^{\mu_i} \Psi & D_{1,i}^{\mu_i} \Psi & D_{2,i}^{\mu_i} \Psi &
	\end{array}\right]$}
\eean
It is enough to prove that the only vector in the left null space of the above matrix is zero vector on substituting  $\lambda_{1,y_i}=0$.
We have that on substituting $\lambda_{1,y_i}=0$:
\bean 
D_{1,i}^{\mu_i}  = \underbrace{\left[\begin{array}{ccccc}
		v_1 & v_1 & & &  \\
		& & \ddots & & \\
		& & &  v_1 & v_1\\
	\end{array}\right]}_{(\mu_i p_{i-1} \times 2p_{i-1})} \text{ where } v_1 = \underbrace{\left[\begin{array}{c}
		1 \\ 0 \\ \vdots \\ 0
	\end{array}\right]}_{(\mu_i \times 1)}
\eean
where $v_1$ is a vector with $\mu_i$ components with $1$ at the first component and with $0$ at rest of the components. Hence by doing column operations, we can remove all rows corresponding to non-zero entries in $D_{1,i}^{\mu_i}$, $D_{1,i}^{\mu_i}\Psi$ from $M_{i,\mu_i}$ and all columns corresponding to $D_{1,i}^{\mu_i}, D_{1,i}^{\mu_i}\Psi$ from $M_{i,\mu_i} $ without affecting the invertibility of determinant of $M_{i, \mu_i}$ as $\gamma \neq 0,1$. This can be seen as follows:
\bean 
|M_{i,\mu_i}|_{\{\lambda_{1,y_i} = 0\}} = (1-\gamma)^{ 2p_{i-1}} \times \left( \prod\limits_{j=1}^{i-1} \prod\limits_{\substack{x, x' \in S_{y_j} \\ x \ne x'}} \left( \theta_{x,y_j;x'} \right)^{\frac{4p_{i-1}}{e_{y_j}+1}} \right) \times
(\lambda_{2,y_i}\lambda_{3,y_i})^{4p_{i-1}} \times \gamma^{4p_{i-1}} \times \\  \left|\begin{array}{c|c|c|c|c|c|c|c}
	M_{i-1,\mu_i-1} &  & & & D_{2,i}^{\mu_i-1} & D_{3,i}^{\mu_i-1} & & \\ \hline
	& M_{i-1,\mu_i-1} & & & & &  D_{2,i}^{\mu_i-1} & D_{3,i}^{\mu_i-1} \\ \hline
	&  & M_{i-1,\mu_i-1}  & & D_{2,i}^{\mu_i-1} \Psi & & & D_{3,i}^{\mu_i-1} \Psi\\ \hline
	&  &  & M_{i-1,\mu_i-1}  &  & D_{3,i}^{\mu_i-1} \Psi  & D_{2,i}^{\mu_i-1} \Psi
\end{array}\right| 
\eean
Hence its enough to show that the left null space of following matrix is zero in order to prove $|M_{i, \mu_i}|_{\{\lambda_{1, y_i} = 0\}}$.
\bean
\scalebox{0.85}{$M' =\left[\begin{array}{c|c|c|c|c|c|c|c}
	M_{i-1,\mu_i-1} &  & & & D_{2,i}^{\mu_i-1} & D_{3,i}^{\mu_i-1} & & \\ \hline
	& M_{i-1,\mu_i-1} & & & & &  D_{2,i}^{\mu_i-1} & D_{3,i}^{\mu_i-1} \\ \hline
	&  & M_{i-1,\mu_i-1}  & & D_{2,i}^{\mu_i-1} \Psi & & & D_{3,i}^{\mu_i-1} \Psi \\ \hline
	&  &  & M_{i-1,\mu_i-1}  & &  D_{3,i}^{\mu_i-1} \Psi  &  D_{2,i}^{\mu_i-1} \Psi & \\
	& & & &\multicolumn{1}{c}{${\upbracefill}$} &
	\multicolumn{1}{c}{${\upbracefill}$} & \multicolumn{1}{c}{${\upbracefill}$} &
	\multicolumn{1}{c}{${\upbracefill}$}\\
	& & & & C_1 & C_2 & C_3 & C_4
	\end{array}\right]$}
\eean
Let the vector in left null space of above matrix $M'$ be of the form $F=[F_1, F_2, F_3, F_4]$ where for $1 \leq  j \leq 4$, $F_{j}=(f_{j, \ell,\uv} \mid \ell \in [\mu_i-1], \uv \in Q^{i-1}(E, \uz))$, and $f_{j,\uv}$ a polynomial of degree $\mu_i-2$ given by $f_{j,\uv}(x) = \sum\limits_{\ell=1}^{\mu_i-1}f_{j,\ell,\uv}x^{\ell-1}$. 
\ben
\item Since $FM'=0$, we write the null space equations corresponding to $2p_{i-1}$ columns $C_1$.
\bea
f_{1,\uv}(\lambda_{2,y_i})&=& f_{3,\uv}(\lambda_{2,y_i}) \text{ for all } \uv \in Q^{i-1}(E, \uz)\label{eq:q41} \\
f_{1,\uv}(\gamma \lambda_{2,y_i})&=&\gamma f_{3,\uv}(\gamma \lambda_{2,y_i}) \text{ for all } \uv \in Q^{i-1}(E, \uz).\label{eq:q42}
\eea  Equation \eqref{eq:q41} implies that $f_{1,\uv}(x)-f_{3,\uv}(x)$ have $\lambda_{2,y_i}$ as root. Let:
\bea
g_{1,\uv}(x) &=& f_{3,\uv}(x)-f_{1,\uv}(x) \text{ for all } \uv \in Q^{i-1}(E, \uz) \label{eq:q411}\\
\nonumber &=& \sum\limits_{\ell=1}^{\mu_i-1} g_{1, \ell, \uv} x^{\ell-1}
\eea where $g_{1,\uv}(x)$ has root at $\lambda_{2,y_i}$. Let $G_1=(g_{1,\ell,\uv} \mid \ell \in [\mu_i-1], \uv \in Q^{i-1}(E, \uz))$. Hence $F_3=F_1+G_1$.
By equation \eqref{eq:q42} and \eqref{eq:q411}:
\bea
f_{1,\uv}(\gamma \lambda_{2,y_i})&=&\gamma f_{1,\uv}(\gamma \lambda_{2,y_i})+ \gamma g_{1,\uv}(\gamma \lambda_{2,y_i}) \notag \\
g_{1,\uv}(\gamma \lambda_{2,y_i}) &=& \gamma^{-1} (1-\gamma)f_{1,\uv}(\gamma \lambda_{2,y_i}) \label{eq:q412}
\eea
Since $F_3=F_1+G_1$ and $F_3,F_1$ are in left null space of $M_{i-1,\mu_i-1}$, we have that $G_1$ is also in left null space of $M_{i-1,\mu_i-1}$. 
\item Since $FM'=0$, we write the $2p_{i-1}$ null space equations corresponding to $C_2$. For any $\uv \in Q^{i-1}(E, \uz)$:
\bea
f_{1,\uv}(\lambda_{3,y_i})=f_{4,\uv}(\lambda_{3,y_i}) \label{eq:q415} \\
f_{1,\uv}(\gamma \lambda_{3,y_i})=\gamma f_{4,\uv}(\gamma \lambda_{3,y_i}) \label{eq:q425}
\eea
Equation \eqref{eq:q415} implies that $f_{1,\uv}(x)-f_{4,\uv}(x)$ has $\lambda_{3,y_i}$ as root. Let:
\bea
g_{2,\uv}(x) = f_{4,\uv}(x)-f_{1,\uv}(x) = \sum\limits_{\ell=1}^{\mu_i-1} g_{2, \ell, \uv} x^{\ell-1} \label{eq:q4115}
\eea where $g_{2,\uv}$ has root at $\lambda_{3,y_i}$. Let $G_2=(g_{2,\ell,\uv} \mid \ell \in [\mu_i-1], \uv \in Q^{i-1}(E, \uz))$. Hence $F_4=F_1+G_2$. By same argument that led to equation \eqref{eq:q412}, we have that:
\bea
g_{2,\uv}(\gamma \lambda_{3,y_i}) &=& \gamma^{-1} (1-\gamma)f_{1,\uv}(\gamma \lambda_{3,y_i}) \label{eq:q4125}
\eea
Since $F_4=F_1+G_2$ and $F_4,F_1$ are in left null space of $M_{i-1,\mu_i-1}$, we have that $G_2$ is also in left null space of $M_{i-1,\mu_i-1}$.
\item Since $FM'=0$, we write the null space equations corresponding to $2p_{i-1}$ columns $C_3$. For any $\uv \in Q^{i-1}(E, \uz)$:
\bea
f_{2,\uv}(\lambda_{2,y_i})=f_{4,\uv}(\lambda_{2,y_i}) \label{eq:q4156} \\
f_{2,\uv}(\gamma \lambda_{2,y_i})=\gamma f_{4,\uv}(\gamma \lambda_{2,y_i}) \label{eq:q4256}
\eea
Equation \eqref{eq:q4156} implies that $f_{2,\uv}(x)-f_{4,\uv}(x)$ has $\lambda_{2,y_i}$ as root. Let:
\bean
g_{3,\uv}(x) = f_{2,\uv}(x)-f_{4,\uv}(x) = \sum\limits_{\ell=1}^{\mu_i-1} g_{3, \ell, \uv} x^{\ell-1}
\eean
By above equation and equation \eqref{eq:q4115}:
\bea
f_{2,\uv}(x)=f_{1,\uv}(x)+g_{2,\uv}(x)+g_{3,\uv}(x) \label{eq:q41156}
\eea where $g_{3,\uv}(x)$ has root at $\lambda_{2,y_i}$. Let $G_3=(g_{3,\ell,\uv} \mid \ell \in [\mu_i-1], \uv \in Q^{i-1}(E, \uz))$. Hence $F_2=F_1+G_2+G_3$ and $G_3$ is in left null space of $M_{i-1, \mu_i-1}$.
\item Since $FM'=0$, we write the null space equations corresponding to $2p_{i-1}$ columns $C_4$. For any $\uv \in Q^{i-1}(E, \uz)$:
\bea
f_{2,\uv}(\lambda_{3,y_i})=f_{3,\uv}(\lambda_{3,y_i}) \label{eq:q41567} \\
f_{2,\uv}(\gamma \lambda_{3,y_i})=\gamma f_{3,\uv}(\gamma \lambda_{3,y_i}) \label{eq:q42567}
\eea
Equation \eqref{eq:q41567} implies that $f_{2,\uv}(x)-f_{3,\uv}(x)$ has $\lambda_{3,y_i}$ as root. Let
\bean
g_{4,\uv}(x) = f_{2,\uv}(x)-f_{3,\uv}(x)
\eean
By above equation and equation \eqref{eq:q411}:
\bea
f_{2,\uv}(x)=f_{1,\uv}(x)+g_{1,\uv}(x)+g_{4,\uv}(x) \label{eq:q411567}
\eea 
where $g_{4,\uv}(x)$ has root at $\lambda_{3,y_i}$. Let $G_4=(g_{4,\ell,\uv} \mid \ell \in [\mu_i-1], \uv \in Q^{i-1}(E, \uz))$. Hence $F_2=F_1+G_1+G_4$ and $G_4$ is in left null space of $M_{i-1, \mu_i-1}$.
\item From equations, \eqref{eq:q41156} and \eqref{eq:q411567}, we have that:
\bea
f_{1,\uv}(x)+g_{1,\uv}(x)+g_{4,\uv}(x) = f_{1,\uv}(x)+g_{2,\uv}(x)+g_{3,\uv}(x) \notag \\
g_{1,\uv}(x)+g_{4,\uv}(x) = g_{2,\uv}(x)+g_{3,\uv}(x) \notag \\
g_{1,\uv}(x)-g_{3,\uv}(x)=g_{2,\uv}(x)-g_{4,\uv}(x) \label{eq:equality1}
\eea
We know that $g_{1, \uv}$ and $g_{3, \uv}$ have $\lambda_{2, y_i}$ as root and $g_{2, \uv}$ and $g_{4, \uv}$ have $\lambda_{3, y_i}$ as root. By equation \eqref{eq:equality1} it follows that $g_{1,\uv}-g_{3,\uv}$ has root at both $\lambda_{2,y_i},\lambda_{3,y_i}$, this constraint when combined with the condition that $(G_1-G_3)M_{i-1,\mu_i-1}=0$ as $G_1, G_3$ are in left null space of $M_{i-1,\mu_i-1}$. This implies that for any $\uu \in Q^{i-1}(E, \uz)$, $(x, y) \in E_{2, \uu}$ such that $y \le y_{i-1}$:
\bea
\label{eq:equality1_1} (g_{1,\uu}-g_{3,\uu})(\theta_{x,y,u_y}) - (g_{1,\uu(x \rightarrow u_y)}+g_{3,\uu(x \rightarrow u_y)})(\theta_{x,y,u_y}) = 0.
\eea
As $g_{1,\uv}-g_{3,\uv}$ has roots at $\lambda_{2,y_i}$, $\lambda_{3,y_i}$ for every $\uv \in Q^{i-1}(E, \uz)$, we can write $(g_{1,\uv}-g_{3,\uv})(x) = (x-\lambda_{2,y_i})(x-\lambda_{3,y_i})g_{13,\uv}'(x)$ where $g_{13,\uv}'(x) = \sum\limits_{\ell=1}^{\mu_i-3}g_{13,\ell,\uv}'x^{\ell-1}$ is a polynomial of degree $\mu_i-4$. Substituting this in equation~\eqref{eq:equality1_1} we get that for any $\uu \in Q^{i-1}(E, \uz)$, $(x, y) \in E_{2, \uu}$ such that $y \le y_{i-1}$:
\bea
\label{eq:equality1_2} g_{13,\uu}'(\theta_{x,y,u_y}) + g_{13,\uu(x \rightarrow u_y)}'(\theta_{x,y,u_y}) = 0.
\eea
By setting $G_{13}'=(g_{13,\ell,\uv}' \mid \ell \in [\mu_i-3], \uv \in Q^{i-1}(E, \uz))$ equation~\eqref{eq:equality1_2} implies that $(G_{13}')M_{i-1, \mu_i-3} = 0$. But since $\mu_i-3 = \mu_{i-1}$ and $M_{i-1, \mu_{i-1}}$ is invertible it follows that $G_{13}' = 0$. By the definition of $G_{13}'$ it follows that $G_1-G_3=0$, $G_2-G_4=0$.
Hence 
\bea
f_{2,\uv}(x)=f_{1,\uv}(x)+g_{1,\uv}(x)+g_{2,\uv}(x) \label{eq:final1}
\eea
\item By equation \eqref{eq:q4256},equation \eqref{eq:final1}, equation \eqref{eq:q4115}:
\bea
f_{2,\uv}(\gamma \lambda_{2,y_i})&=&\gamma f_{4,\uv}(\gamma \lambda_{2,y_i}) \notag \\
(f_{1,\uv}+g_{1,\uv}+g_{2,\uv})(\gamma \lambda_{2,y_i})&=&\gamma (f_{1,\uv}+g_{2,\uv})(\gamma \lambda_{2,y_i}) \notag \\
g_{1,\uv}(\gamma \lambda_{2,y_i})&=&-(1-\gamma)(f_{1,\uv}+g_{2,\uv})(\gamma \lambda_{2,y_i}) \label{f1}
\eea
From \eqref{f1} and equation \eqref{eq:q412} we have:\\
\bea
\gamma^{-1} (1-\gamma)f_{1,\uv}(\gamma \lambda_{2,y_i}) &=& -(1-\gamma)(f_{1,\uv}+g_{2,\uv})(\gamma \lambda_{2,y_i})  \notag \\
g_{2,\uv}(\gamma \lambda_{2,y_i})& =& \gamma^{-1}(1-\gamma) f_{1,\uv}(\gamma \lambda_{2,y_i})  \label{eq:equality2}
\eea
The constraint that $G_2M_{i-1,\mu_i-1}=0$ and the constraints given in equation \eqref{eq:equality2} and equation \eqref{eq:q4125} form a set of linear constraints as shown below:
\bea
\scalebox{0.8}{$G_2 \left[\begin{array}{ccc} \underbrace{M_{i-1,\mu_i-1}}_{((\mu_i-1) p_{i-1} \times \mu_{i-1} p_{i-1})} & \underbrace{V_2}_{((\mu_1-1)p_{i-1} \times p_{i-1})} & \underbrace{V_3}_{((\mu_1-1)p_{i-1} \times p_{i-1})}\end{array}\right] = \left[\begin{array}{ccc}
	\underbrace{0}_{1 \times \mu_{i-1}p_{i-1}} & \underbrace{F_{1,2}}_{1 \times p_{i-1}} & \underbrace{F_{1,3}}_{1 \times p_{i-1}}
	\end{array}\right]$} \label{eq:g2soln} \\
\scalebox{0.8}{$\text{ where } V_j = \left[\begin{array}{cccc}
	v_j & & &\\
	& \underbrace{v_j}_{((\mu_{i}-1) \times 1)} & &\\
	& & \ddots &\\
	& & & v_j
	\end{array}\right], \ \ {\uv}_{j} = \left[ \begin{array}{c}
	1 \\ 
	\gamma \lambda_{j,y_i}\\ \vdots \\ (\gamma \lambda_{j,y_i})^{\mu_i-2}
	\end{array}\right]$}, \notag
\eea
and $F_{1,j} = (\gamma^{-1}(1-\gamma) f_{1, \uv}(\gamma \lambda_{j, y_i}) \mid \uv \in Q^{i-1}(E, \uz))^T$ for $j \in \{2, 3\}$. 

We will now show that the appended matrix is invertible implying that there is a unique solution to $G_2$. We prove the invertibility by showing that only in its null space is the zero vector. Let $G_2'$ be in the null space of the extended matrix. Then:
\bean
G_2' \left[\begin{array}{ccc} M_{i-1,\mu_i-1} & V_2 & V_3\end{array}\right] &=& 0 \text{ where } G_2' = (g_{2,\ell,\uv}' \mid \ell \in [\mu_i-1], \uv \in Q^{i-1}(E, \uz)),\\
g_{2, \uv}'(\gamma \lambda_{2,y_i}) &=& g_{2,\uv}'(\gamma \lambda_{3,y_i}) = 0 \text{ where }  g_{2,\uv}' = \sum\limits_{\ell=1}^{\mu_i-1} g_{2,\ell,\uv}'x^{\ell-1}.
\eean
Let us now define $G_2'' = (g_{2,\ell, \uv}'' \mid \ell \in [\mu_i-3], \uv \in Q^{i-1}(E, \uz))$ where $g_{2,\uv}'(x) = (x-\gamma \lambda_{2,y_i}) (x - \lambda_{3, y_i}) g_{2,\uv}''(x)$, $g_{2,\uv}'' = \sum\limits_{\ell=1}^{\mu_i-3}g_{2,\ell, \uv}''x^{\ell-1}$. It can be shown that $G_2' M_{i-1, \mu_i-1}=0$ implies that $G_2'' M_{i-1, \mu_i-3}=0$. Since $ M_{i-1, \mu_i-3} =  M_{i-1, \mu_{i-1}}$ is invertible it follows that $G_2''=0$ and therefore $G_2'=0$.
Hence, the extended matrix is invertible and $G_2$ in equation \eqref{eq:g2soln} has unique solution. But is clear to see that $\gamma^{-1}(1-\gamma) F_1$ is a solution. Hence $G_2=\gamma^{-1}(1-\gamma) F_1$,
\bea
g_{2,\uv}(x)=\gamma^{-1}(1-\gamma) f_{1,\uv}(x). \label{eq:final2}
\eea
By similar argument it can be proven that $G_1=\gamma^{-1}(1-\gamma) F_1$ and therefore:
\bea
g_{1,\uv}(x)=\gamma^{-1}(1-\gamma) f_{1,\uv}(x)  \label{eq:final3}
\eea
\item  $F_1M_{i-1,\mu_i-1}=0$ and by equation  \eqref{eq:final2}, \eqref{eq:final3}, $f_{1,\uu}$ has roots at $\lambda_{2,y_i},\lambda_{3,y_i}$. From an argument similar to the one shown in step 5 to prove $G_1-G_3=0$ can be followed to show that $F_1=0$ and hence $G_1=G_2=0$ and hence $F_3=F_4=F_2=0$.
Therefore, the only vector in left null space of $M'$ is zero vector implying $M'$ is invertible and hence $M_{i, \mu_i}$ to be invertible.
\een

\end{appendices}
\end{document}